\documentclass[twocolumn]{aastex631}
\newcommand{\sph}{S_\mathrm{ph}}
\newcommand{\feh}{\left[\mbox{Fe/H}\right]}
\newcommand{\teff}{T_\mathrm{eff}}
\newcommand{\logg}{\log g}
\newcommand{\prot}{P_\mathrm{rot}}
\newcommand{\tauc}{\tau_c}

\shorttitle{Stellar magnetism evolution with Rossby number}
\shortauthors{Mathur et al.}
\graphicspath{{./}{figures/}}

\begin{document}


\title{Magnetic activity evolution of solar-like stars: \\
II. $S_{\rm ph}$-Ro evolution of {\it Kepler} main-sequence targets}

\correspondingauthor{Savita Mathur}
\email{smathur@iac.es}

\author[0000-0002-0129-0316]{Savita Mathur}
\affil{Instituto de Astrof\'isica de Canarias (IAC), E-38205 La Laguna, Tenerife, Spain}
\affil{Universidad de La Laguna (ULL), Departamento de Astrof\'isica, E-38206 La Laguna, Tenerife, Spain}

\author[0000-0001-7195-6542]{\^Angela R. G. Santos}
\affil{Instituto de Astrof\'isica e Ci\^encias do Espa\c{c}o, Universidade do Porto, CAUP, Rua das Estrelas, PT4150-762 Porto, Portugal}
\affil{Departamento de F\'isica e Astronomia, Faculdade de Ci\^encias, Universidade do Porto, Rua do Campo Alegre 687, PT4169-007 Porto, Portugal}

\author[0000-0002-9879-3904]{Zachary R. Claytor}
\affiliation{Space Telescope Science Institute, 3700 San Martin Drive, Baltimore, MD 21218, USA}
\affiliation{Department of Astronomy, University of Florida, 211 Bryant Space Science Center, Gainesville, FL 32611, USA}
\affiliation{Institute for Astronomy, University of Hawai‘i at Mānoa, 2680 Woodlawn Drive, Honolulu, HI 96822, USA}


\author[0000-0002-8854-3776]{Rafael A. Garc\'{i}a}
\affil{Universit\'e Paris-Saclay, Universit\'e Paris Cit\'e, CEA, CNRS, AIM, 91191, Gif-sur-Yvette, France}

\author[0000-0002-9630-6463]{Antoine Strugarek}
\affil{Universit\'e Paris-Saclay, Universit\'e Paris Cit\'e, CEA, CNRS, AIM, 91191, Gif-sur-Yvette, France}

\author[0000-0002-3020-9409]{Adam J. Finley}
\affil{Universit\'e Paris-Saclay, Universit\'e Paris Cit\'e, CEA, CNRS, AIM, 91191, Gif-sur-Yvette, France}

\author[0000-0002-7422-1127]{Quentin Noraz}
\affiliation{Rosseland Centre for Solar Physics, University of Oslo, P.O. Box 1029 Blindern, Oslo, NO-0315, Norway}

\author{Louis Amard}
\affil{Universit\'e Paris-Saclay, Universit\'e Paris Cit\'e, CEA, CNRS, AIM, 91191, Gif-sur-Yvette, France}

\author[0000-0000-0000-0000]{Paul G. Beck}
\affil{Instituto de Astrof\'isica de Canarias (IAC), E-38205 La Laguna, Tenerife, Spain}
\affil{Universidad de La Laguna (ULL), Departamento de Astrof\'isica, E-38206 La Laguna, Tenerife, Spain}

\author{Alfio Bonanno}
\affiliation{INAF -- Osservatorio Astrofisico di Catania, Via S. Sofia 78, I-95123, Italy}

\author[0000-0003-0377-0740]{Sylvain N. Breton}
\affiliation{INAF -- Osservatorio Astrofisico di Catania, Via S. Sofia 78, I-95123, Italy}

\author{Allan S. Brun}
\affil{Universit\'e Paris-Saclay, Universit\'e Paris Cit\'e, CEA, CNRS, AIM, 91191, Gif-sur-Yvette, France}

\author[0000-0002-8849-9816]{Lyra Cao}
\affiliation{Department of Physics and Astronomy, Vanderbilt University, 6301 Stevenson Science Center, Nashville, TN 37212, USA}

\author[0000-0001-8835-2075]{Enrico Corsaro}
\affiliation{INAF -- Osservatorio Astrofisico di Catania, Via S. Sofia 78, I-95123, Italy}

\author[0000-0003-4556-1277]{Diego Godoy-Rivera}
\affil{Instituto de Astrof\'isica de Canarias (IAC), E-38205 La Laguna, Tenerife, Spain}
\affil{Universidad de La Laguna (ULL), Departamento de Astrof\'isica, E-38206 La Laguna, Tenerife, Spain}

\author{St\'ephane Mathis}
\affil{Universit\'e Paris-Saclay, Universit\'e Paris Cit\'e, CEA, CNRS, AIM, 91191, Gif-sur-Yvette, France}

\author[0000-0002-6812-4443]{Dinil B. Palakkatharappil}
\affil{Universit\'e Paris-Saclay, Universit\'e Paris Cit\'e, CEA, CNRS, AIM, 91191, Gif-sur-Yvette, France}

\author{Marc H. Pinsonneault}
\affiliation{Ohio State University, McPherson Laboratory, 140 W 18th Ave, Columbus, OH 43210, USA}

\author{Jennifer van Saders}
\affiliation{Institute for Astronomy, University of Hawai‘i at Mānoa, 2680 Woodlawn Drive, Honolulu, HI 96822, USA}




\begin{abstract}


There is now a large sample of stars observed by the {\it Kepler} satellite with measured rotation periods and photometric activity index $\sph$. We use this data, in conjunction with stellar interiors models, to explore the interplay of magnetism, rotation, and convection. {  Stellar activity proxies other than $\sph$} are correlated with the Rossby number, $Ro$, or ratio of rotation period to convective overturn timescale. We compute the latter using the Yale Rotating Evolution Code stellar models. We observe different $\sph$-$Ro$ relationships for different stellar spectral types. Though the overall trend of decreasing magnetic activity versus $Ro$ is recovered, we find a localized dip in $\sph$ around $Ro/Ro_{\odot} \sim$\,0.3 for the G and K dwarfs. F dwarfs show little to no dependence of $\sph$ on $Ro$ due to their shallow convective zones; further accentuated as $\teff$ increases. 
The dip in activity for the G and K dwarfs corresponds to the intermediate rotation period gap, suggesting that the dip in $\sph$ could be associated with the redistribution of angular momentum between the core and convective envelope inside stars. For G-type stars, we observe enhanced magnetic activity above solar $Ro$. Compared to other Sun-like stars with similar effective temperature and metallicity, we find that the Sun's current level of magnetic activity is comparable to its peers and lies near the transition to increasing magnetic activity at high $Ro$. We confirm that metal-rich stars have a systematically larger $\sph$ level than metal-poor stars, which is likely a consequence of their deeper convective zones.


\end{abstract}

\keywords{stars: rotation -- stars: activity -- starspots -- techniques: photometric -- methods: data analysis -- catalogs}


\section{Introduction} \label{sec:intro}


The magnetic and rotational properties of solar-like stars are strongly coupled due to the interaction of rotation and convection in their outer envelopes, which fuels their magnetic dynamos \citep{2017LRSP...14....4B}. Magnetic activity cycles result from the interaction between rotation, convection, and magnetic field, processes included in dynamo models \citep[e.g.][]{2004ApJ...614.1073B,2008A&A...483..949J,2011ApJ...731...69B,2017LRSP...14....4B,2017ApJ...836..192B, 2022ApJ...931L..17K}. The level of magnetic activity hosted by a solar-like star is highly correlated with stellar rotation \citep[e.g.][]{2003ApJ...586L.145B,2014MNRAS.444.3517M, 2014MNRAS.441.2361V}, and powers the stellar wind \citep{2015ApJ...798..116R,2018ApJ...854...78F}, which governs their rotation period evolution throughout the main-sequence \citep{1972ApJ...171..565S,1988ApJ...333..236K, 1990ApJS...74..501P,2013A&A...556A..36G,2015ApJ...799L..23M}.

Current dynamo models allow us to explore the mechanisms responsible for maintaining magnetic activity in solar-like stars \citep{2017Sci...357..185S, 2018ApJ...863...35S, 2022ApJ...926...21B,2024A&A...684A..92F}, however observational constraints are limited. Zeeman broadening and Zeeman-Doppler imaging techniques have been used to recover the magnetic field strengths and topologies for a broad range of stars \citep[e.g.][]{2019ApJ...876..118S,2023MNRAS.525.2015D}, however this type of observation is limited in number. Our closest star, the Sun, is subject to a wide variety of observations at different spatial/temporal scales and wavelengths, with observations spanning decades (magnetograms, Extreme Ultra Violet, X-ray), and sometimes centuries (Ca II H\&K, sunspots). Models of the solar dynamo are subsequently more advanced, however predicting the magnetic activity level of the Sun remains challenging \citep[e.g.][]{2021SoPh..296...54N}. Dynamo models rely on the knowledge of many parameters, from the convective flows, internal rotation, and meridional circulation, to the magnetic field strength at the base of the convection zone and its emergence to the surface \citep[e.g.][]{2007A&A...474..239J,2011ApJ...731...69B}. 
Unfortunately, so far it has not been possible to constrain these processes and quantities for other stars and for the deep solar convective zone. One key parameter in these models is the Rossby number (hereafter $Ro$) that compares timescales between the Coriolis force and the advection. Typically, this is evaluated as the ratio between the rotation period, $\prot$, and the convective overturn timescale, $\tauc$, either at a given depth of the convective zone or more globally as an average value. Obtaining this stellar parameter is then of prime importance in order to quantify the rotation-convection interplay at the origin of magnetic activity and characterize its evolution over secular timescales.

From previous spectroscopic and X-ray luminosity surveys that provided proxies of the magnetic activity in many other stars, two regimes have been identified: 1) the saturated regime, at low $Ro$ values, where the magnetic activity is flat as a function of the Rossby number and 2) the unsaturated regime where the magnetic activity decreases as a function of $Ro$ \citep[e.g.][]{1984ApJ...279..763N,2011ApJ...743...48W}. Similar studies also investigated other proxies of magnetic activity such as the magnetic field strength from spectropolarimetric observations \citep{2014MNRAS.444.3517M} that probed the unsaturated regime. More recently, both regimes have been recovered with estimates of the flare energy of {\it Kepler} targets \citep{2019ApJS..241...29Y}, with the mid-frequency continuum in {\it Kepler}/K2 data, and by measuring the star-spot filling fraction obtained from spectroscopic observations \citep{2022MNRAS.517.2165C}. Note that another regime at higher $Ro$ may exist according to simulations \citep{2024A&A...684A.156N}. 

Regarding the {\it Kepler} observations \citep{2010Sci...327..977B}, several photometric magnetic activity proxies have been defined in addition to the flare energy. Some of them measure the range of stellar variability as defined by \citet{2010ApJ...713L.155B,2011AJ....141...20B,2013ApJ...769...37B}, however this is a broader definition of the stellar variability and depends on the length of the segments used to compute it and the time scale of the pulsations or granulation, making the proxy sensitive to those, and possibly biasing them if one is interested in the magnetic variability only. For this reason, other proxies are computed based on the knowledge of the surface rotation period measured via the presence of spots or active regions that produce a periodic modulation in the photometric light curves. This  ensures that the variability measured is linked to spots, and hence to magnetic activity. This can be measured in the time domain \citep[e.g.][]{2014JSWSC...4A..15M,2014ApJS..211...24M} or in the power spectrum density \citep{2021ApJ...916...66B}. Various studies of the relation between the magnetic proxies and the properties of the stars have been done, showing interesting features such as the effect of metallicity or stellar mass on the evolution of magnetic activity \citep[e.g.][]{2021ApJ...912..127S}. However these studies were based on the previous {\it Kepler} rotation catalog \citep{2014ApJS..211...24M}. In this work, we use a more recent one with 60\% more stars \citep{2019ApJS..244...21S,2021ApJS..255...17S}.


In this paper, following Paper I of this series \citep{2023ApJ...952..131M} that investigated magnetic activity-rotation-age relations of the {\it Kepler} targets, we take advantage of that larger catalog to study the evolution of their magnetic activity with the $Ro$ number. In Section~\ref{sec:Sample}, we describe the provenance of the stellar fundamental parameters from different catalogs as well as the rotation periods and the photometric activity indexes. We then explain how the Rossby number is computed with stellar models (Section~\ref{sec:model}). In Section~\ref{sec:sph_Ro}, we present different features in the magnetic activity-Rossby number diagram. We discuss their possible origins in Section~\ref{sec:discussion} and conclude in Section~\ref{sec:conclusion}.

\section{Stellar parameters of the sample}
\label{sec:Sample}
In this section, we describe the origin of the stellar parameters and explain how the rotation periods and magnetic activity proxies were obtained for that sample. We start with the sample of 55,232 {\it Kepler}\, targets from \citet[][]{2019ApJS..244...21S,2021ApJS..255...17S}, which represents one of the largest catalogs of rotation periods available for the {\it Kepler} field to date. 

\subsection{Stellar fundamental parameters}

We followed the same procedure as in \citet{2023ApJ...952..131M} that we briefly remind here. Several catalogs with stellar fundamental parameters have been published for the stars in the {\it Kepler} field: the {\it Kepler} Input Catalog \citep{2011AJ....142..112B}, the revised catalog for Quarters 1-16 \citep{2014ApJS..211....2H}, and finally the {\it Kepler} close-out catalog for the Data Release 25 \citep[][hereafter DR25]{2017ApJS..229...30M}. With the {\it Gaia} DR2 \citep{2018A&A...616A...1G}, an updated catalog that used the parallaxes of the aforementioned mission was built \citep[][hereafter B20]{2020AJ....159..280B}, providing improved stellar parameters for 186,301 entries compared to the 197,097 stars targeted during the {\it Kepler} nominal mission.  For our sample of stars with measured rotation periods, we utilize effective temperature, $\teff$, and metallicity, $\feh$, from five different catalogs. 

We prioritized the stellar atmospheric parameters from spectroscopic surveys in the following order: {\it Kepler} Community Follow-up Observation Program \citep[CFOP,][]{2018ApJ...861..149F}, DR16 of the APOGEE survey \citep[Apache Point Observatory for Galactic Evolution Experiment,][]{2020ApJS..249....3A}, and DR7 of the Large Sky Area Multi-ObjectFiber Spectroscopic Telescope \citep[LAMOST,][]{2012RAA....12..723Z, 2020ApJS..251...15Z}. For the remainder of the sample, we used in priority the B20 $\teff$ and $\feh$ values and finally, the DR25 values for the remaining stars. We end up with 402 stars with CFOP parameters, 2,313 stars with APOGEE atmospheric parameters, and 16,473 stars with LAMOST spectroscopic values. More than 36,000 stars have B20 or DR25 input values.

Finally, we used the luminosity, $L$, from {\it Gaia} DR2 as given in B20 when available and $\logg$ from that same catalog for the remaining cases. We also take the {\it Gaia} Renormalized
Unit Weight Error (RUWE) as we are interested in separating binary candidates. The RUWE is not available for all the targets so we end up with a sample of 52,150 stars. 

\subsection{Rotation and magnetic proxy} 
\label{sec:rot}

The rotation periods and magnetic activity proxies that are used in this study come from the analysis by \citet[][]{2019ApJS..244...21S,2021ApJS..255...17S}. We quickly remind here how those rotation periods were obtained. 

Three different methods were used with a time-period analysis based on wavelets techniques \citep{1998BAMS...79...61T,liu2007,2010A&A...511A..46M}, the auto-correlation function \citep[ACF,][]{2013MNRAS.432.1203M,2014A&A...572A..34G,2014ApJS..211...24M}, and the composite spectrum that is a combination of the first two methods \citep{2017A&A...605A.111C}. In addition to these analyses, a machine learning algorithm, Random fOrest Over STEllar Rotation \citep[ROOSTER,][]{2021A&A...647A.125B} was run to select the most likely rotation periods and leading to fewer visual inspections. 


The measurement of the surface rotation periods relies on the presence of active regions on the stellar surface. As a consequence, stars with measured rotation periods are magnetically active and we can define a proxy of magnetic activity based on photometric data. This proxy is produced by evaluating  the standard deviation of the photometric time series \citep{2010Sci...329.1032G}. But we can also use the knowledge of the rotation period of the star to measure the standard deviation on subseries of length $k\,\times P_{\rm rot}$ and then take the mean value \citep{2014JSWSC...4A..15M}. A value of 5 for $k$ proved to be a good trade-off to avoid smoothing too much the data while keeping enough signal from the active regions. This index has been shown to be a good proxy of the magnetic activity for both the Sun \citep{2017A&A...608A..87S} and other solar-like stars \citep{2016A&A...596A..31S} where $S_{\rm ph}$ was compared to other classical magnetic activity indexes such as chromospheric emission in Ca H\&K, $R'_{\rm HK}$ or the Mount Wilson S-index \citep{1978ApJ...226..379W}.


We note that, similarly to the chromospheric activity indexes \citep[e.g.][]{2021ApJ...914...21S}, the photometric magnetic proxy depends on the {  projected surface of the active regions on the line-of-sight, which is related to the combination between the} inclination angle of the rotation axis of the star {  and the active latitudes}. If the inclination angle is low (closer to pole-on), and assuming that the active regions form within $\pm \sim 30^\circ$ of latitude like the solar case, the $S_{\rm ph}$ measurement is lower than the real magnetic activity level of the star. {  We also note that the existence of active longitudes could affect the $\sph$. For instance, in the case where the active regions are separated by 180$^\circ$ in longitude, we would measure a smaller $\sph$ value compared to cases where the active regions are at a similar longitude. This was also mentioned by \citet{2018ApJ...863..190B} for lightcurves with double dips.} In the remainder of the paper, since the inclination angle of the rotation axis {  and the active longitudes are} unknown for most of the stars, we emphasize that the $S_{\rm ph}$ is a lower limit of the magnetic activity of the star. Besides, similarly to the Mount Wilson S-index \citep{1991ApJS...76..383D}, it depends on the moment of the magnetic cycle when the star was observed \citep{2023A&A...672A..56S}. 


\section{Rossby number from stellar models}
\label{sec:model}
The Rossby number, $Ro$, plays a crucial role in understanding and characterizing the magnetic activity of a star. This parameter measures the balance between the two primary ingredients at the origin of the large-scale dynamo: rotation and convection. Different ways can be used to compute it: semi-empirical relations \citep[e.g.][]{1984ApJ...279..763N,2011ApJ...741...54C,2011ApJ...743...48W,2021A&A...652L...2C}, stellar modeling \citep[e.g.][]{2014A&A...562A.124M,2016Natur.529..181V,2017A&A...605A.102C,2017ApJ...850..134S,2023MNRAS.519.5304L}, and analytical formula constrained by stellar evolution modeling combined with magneto-hydrodynamical simulations \citep[e.g.][]{2017ApJ...836..192B,2019ApJ...876...83A,2022A&A...667A..50N}. Caution should be taken when $Ro$ values from different approaches are compared. Indeed, while the definitions can be different, the Rossby number also depends on the stellar models used and on the location in the convective zone where it is evaluated. As a consequence, the absolute values of the $Ro$ numbers cannot always be compared directly as there could be a multiplicative factor between different methods \citep[see Appendix of][]{2024A&A...684A.156N}. For this reason, our results are presented with the $Ro$ number normalized by the solar value obtained with our approach.


 \begin{figure*}
    \centering
    \includegraphics[width=\hsize]{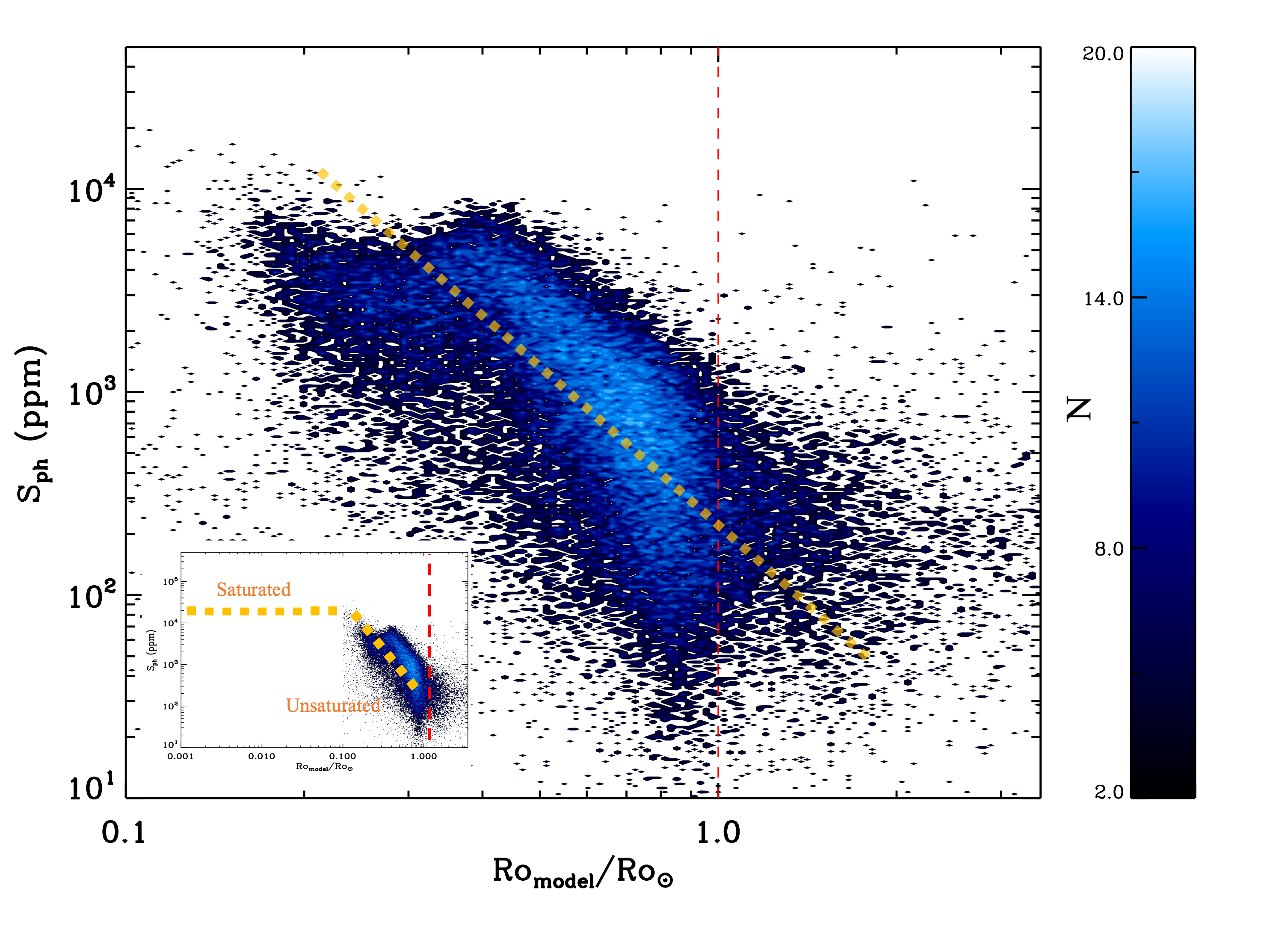}
    \caption{$S_\mathrm{ph}$ as a function of the model Rossby number normalized to the solar value of 2.16 (represented by the red dash line) for all the main-sequence stars without potential pollution from binary systems and color-coded with the number density of stars. The inset illustrates where our sample falls compared to the saturated and unsaturated regimes found in X-ray observations \citep{2018MNRAS.479.2351W} (yellow dotted lines), with the slope of the unsaturated regime taken from the bulk of our sample.}
    \label{Sph_Rom_all}
\end{figure*}

In this work, the Rossby number is computed as:

\begin{equation}\label{eq:Ro}
     Ro = \frac{P_{\rm rot}}{\tau_c},
 \end{equation}
where $P_{\rm rot}$ is the observed surface rotation period at the active latitudes  and $\tau_c$ is the convective overturn timescale. While the rotation period can be measured from photometric data as explained in Section~\ref{sec:rot}, the convective overturn timescale requires semi-empirical relations or models in order to be computed. A comparison between the different methods developed to measure this quantity is given in Appendix~\ref{app:tauc_comp}.

We derived the convective overturn timescale, $\tauc$, with the model grid interpolation tool named \texttt{kiauhoku} \citep{Claytor2020} to fit stellar evolution models to the observables. We used the models of \citet{vanSaders2013}, which were computed using the Yale Rotating Evolution Code \citep[YREC;][]{1989ApJ...338..424P,2001ApJ...555..990B, Demarque2008}. The convective overturn timescale used in this work is computed one pressure scale height above the base of the convective zone. 

We performed Markov Chain Monte Carlo (MCMC) using a $\chi^2$ log-likelihood of the form
\begin{equation} \nonumber
    \mathcal{L}_{\chi^2} = -\frac{1}{2} \sum_{i} \frac{\left(x_i - x_i'\right)^2}{\sigma_{x_i}^2}, \nonumber
\end{equation}
where $x_i$ and $\sigma_{x_i}$ are input parameters and uncertainties respectively, and $x_i'$ is the model-computed value. As input we used $\teff$, $\feh$, and $L$ (or $\logg$ when $L$ was not available). Compared to Paper I, we did not use the rotation period as input since the convective overturn timescale is sufficiently constrained by an isochrone fit. We required that the chains ran for at least fifty autocorrelation times (and in practice, they were usually longer), ensuring at least fifty independent samples in each chain and increasing the likelihood that the chains were converged. We also required that the posterior $\teff$, $\feh$, and $L$ (or $\logg$) were within the uncertainties of the input values. Using these criteria, 50,065 out of 52,150 stars were successfully fit.
Finally, the Rossby number is computed with Equation~\ref{eq:Ro}.

\section{The $S_{\rm ph}$-Ro diagram}
\label{sec:sph_Ro}

\subsection{Sample selection}

Similarly to what was done in \citet{2023ApJ...952..131M}, we want to ensure that none of the considered stars are binary candidates as they can have a different evolution than single stars. We used different metrics or published binary candidates to remove them: {\it Gaia} Renormalized Unit Weight Error (RUWE$<$1.2 to remove binary candidates), close-binary candidates \citep[CPCB1 from][]{2019ApJS..244...21S, 2021ApJS..255...17S}, targets from the Non-Single Stars catalog \citep{2022arXiv220605439H}, binary candidates from the literature \citep[][]{2020AJ....159..280B,2020ApJ...898...76S}, {  stars flagged as RS Canum Venaticorum variables \citep{2023A&A...674A..14R},} and photometric binaries from the Color-Magnitude Diagram \citep{2025arXiv250118719G}. The threshold on RUWE is based on the suggestion by \citet{2018A&A...616A...1G}. However, we note that lower values of RUWE are not a guarantee for a star to be single, as this metric favours significant positional changes. As shown by \citet{2024A&A...682A...7B} $\sim$40\% of all confirmed binaries in the SB9 catalogue \citep{Pourbaix2004} have RUWE\,$\leq$\,1.2. Since most of the missing binaries in {\it Gaia} are long-period systems, it will not affect the evolution of our targets and not bias our subsequent analysis.

Finally, in the remainder of the paper, we focus on the main-sequence stars that we select based on the Evolution Equivalent Phase \citep[EEP,][]{2008ApJS..178...89D} obtained with the models  with EEP\,$<$\,454. The EEP is a renormalization of the age such that the same value corresponds to roughly the same evolutionary stage for all models. The value 454 corresponds to the terminal age main sequence in our models, as determined by the core hydrogen fraction \citep{2016ApJS..222....8D}. We also discarded stars for which model ages are larger than 14\,Gyr, {  which could be due to improper modeling as we do not expect to have such old stars. Therefore, we prefer to adopt this strict criteria to avoid biases in the subsequent analyses. Removing these stars is both physically motivated and unlikely to affect any analysis. } 

With all these cuts done on the stars where the models converged, we end with a sample of {  38,593} stars. 
Table~\ref{tab:stellarparam} provides $L$, $\log g$, $\teff$, $\feh$,  $\tauc$, $\prot$, and $Ro$ for that sample along with the flag on the provenance of the $\teff$ and $\feh$.

\subsection{The main-sequence sample}
\label{sec:MS}
 In Figure~\ref{Sph_Rom_all}, we show the magnetic activity proxy computed from the {\it Kepler} light curves as a function of the Rossby number, where $\tau_c$ was computed from the YREC models. The Rossby number has also been normalized to the solar value for the YREC models, $Ro_\odot$\,=\ 2.16. 

We clearly see different features in this representation. For solar-like stars of a given spectral type (F, G, and K) on the main sequence, the convective overturn timescale is constant on average with age, which means that we can globally follow the evolution of stars, as they spin down, from low $Ro$ values to high $Ro$ values. 

First, we emphasize that the {\it Kepler} sample is mostly probing the unsaturated regime in terms of magnetic activity. As mentioned in Section~\ref{sec:intro}, \citet{2011ApJ...743...48W} showed that in the X-ray emission, $L_X$-$Ro$ representation, two regimes could be observed: the saturated and the unsaturated regimes. {  \citet{2018MNRAS.479.2351W} found the same behaviour for fully convective M dwarfs, suggesting that similar dynamos operate in partially and fully convective stars. However the transition from saturated to unsaturated regimes is not fully understood.} 
The sample of stars, in the field of view of the nominal \textit{Kepler} mission, are rather old and only a very small sample of stars would be in the saturated regime. Nevertheless, the {\it Kepler} sample with the large number of stars observed allows us to see additional features in the unsaturated regime. While the change of slope below 0.4\,$Ro_\odot$ could be mistakenly interpreted as the saturated regime, \citet{2022ApJ...933..195M} already showed that it was not the case by representing the \citet{2018MNRAS.479.2351W} sample together with the {\it Kepler} sample on the same scale for the $Ro$ \citep[see Figure 16 of][]{2022ApJ...933..195M}. This is further discussed by \citet{2024FrASS..1156379S} and illustrated in Figure~\ref{Sph_Rom_all} where the orange dotted lines are an approximate representation of those regimes. It was also found that in the {\it Kepler}\, sample, the parameter space corresponding to the saturated regime is populated by binaries, particularly tidally-synchronized binaries \citep{2019ApJ...871..174S,2020AJ....160...90A}. 


Looking at the upper envelope of the $\sph$-$Ro$ diagram,  we can see an interesting feature with a decrease of $\sph$ until $Ro_{\rm model}/Ro_\odot$\,$\sim$\,0.3 followed by an increase up to $Ro_{\rm model}/Ro_\odot$\,$\sim$\,0.4. This feature was already mentioned by several works as the {\it kink} or the {\it dip} \citep{2020A&A...635A..43R,2021A&A...652L...2C,2021ApJ...912..127S,2022ApJ...933..195M}. Different interpretations of that dip have been given, in many of them assuming that the low $Ro$ stars were part of the saturated regime. \citet{2019A&A...621A..21R} suggested that this could result from a cancellation effect between active regions and bright faculae, leading to a lower variability. Finally, from that point moving towards higher $Ro$, $S_{\rm ph}$ monotonically decreases when the star evolves. We also see a slight accumulation of points, or a larger number density of stars, at $Ro_{\rm model}/Ro_\odot$\,$\sim$\,0.8 \citep{2023ApJ...948L...6M} also seen in the period-effective temperature diagram \citep{2022ApJ...933..114D}.  


Below 1\,$Ro_{\odot}$, the upper envelope of the diagram has a very sharp edge, as also noted by \citet{2021ApJ...912..127S}. That could correspond to stars that are seen with inclination angle of the rotation axis of 90 degrees, like the Sun, allowing us to fully see the magnetic features (in particular spots) and yielding the maximum values of $\sph$ possible. Moreover stars near that upper edge should correspond to stars observed close to the maximum of their magnetic cycles. We will further discuss this point with specific stars in Sections~\ref{sec:Sunlike} and \ref{sec:specific}.

\subsection{By spectral type}
\label{sec:spectral_type}

\begin{figure*}[!h]
    \centering
    \includegraphics[width=8cm]{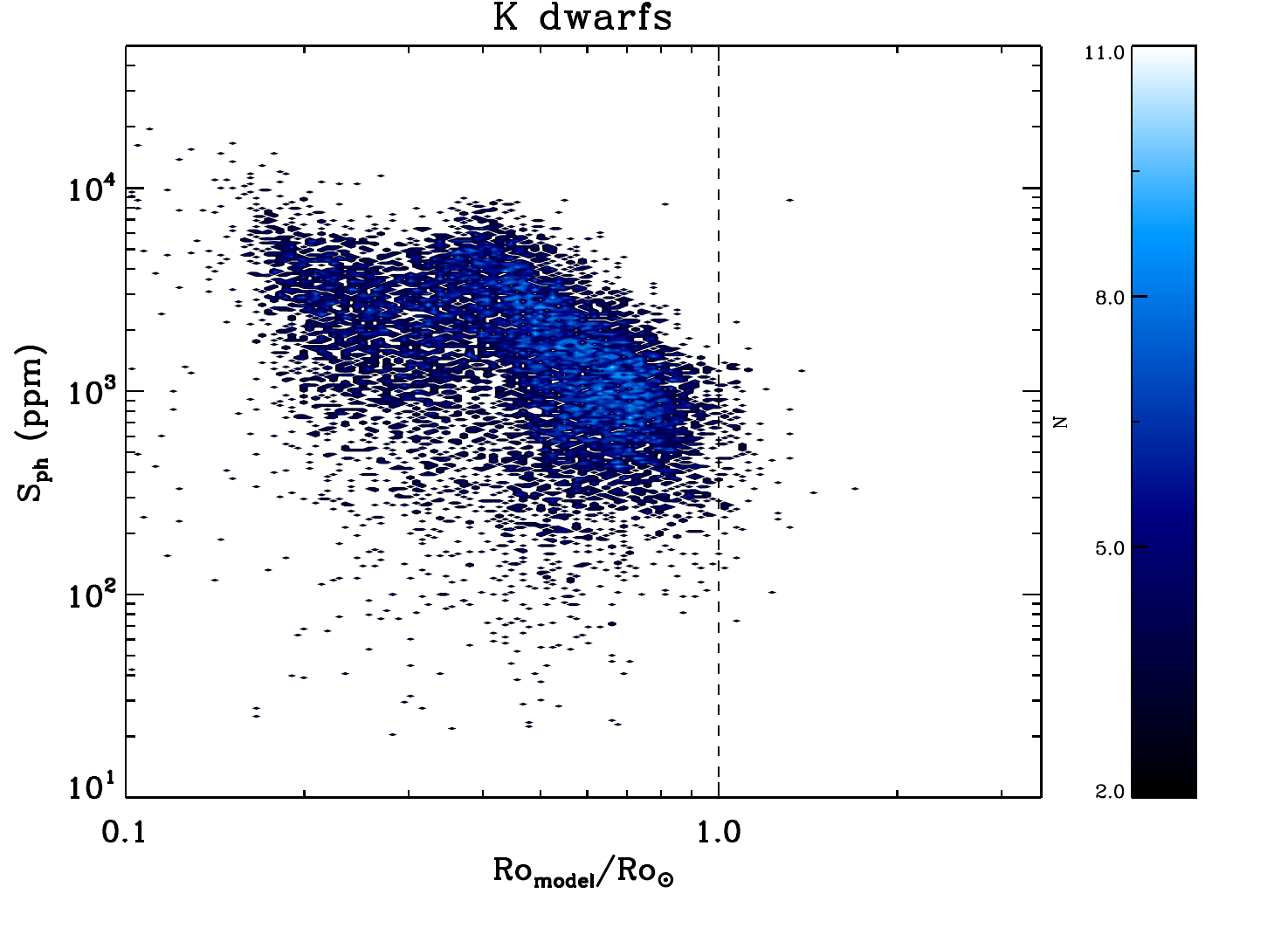}
    \includegraphics[width=8cm]{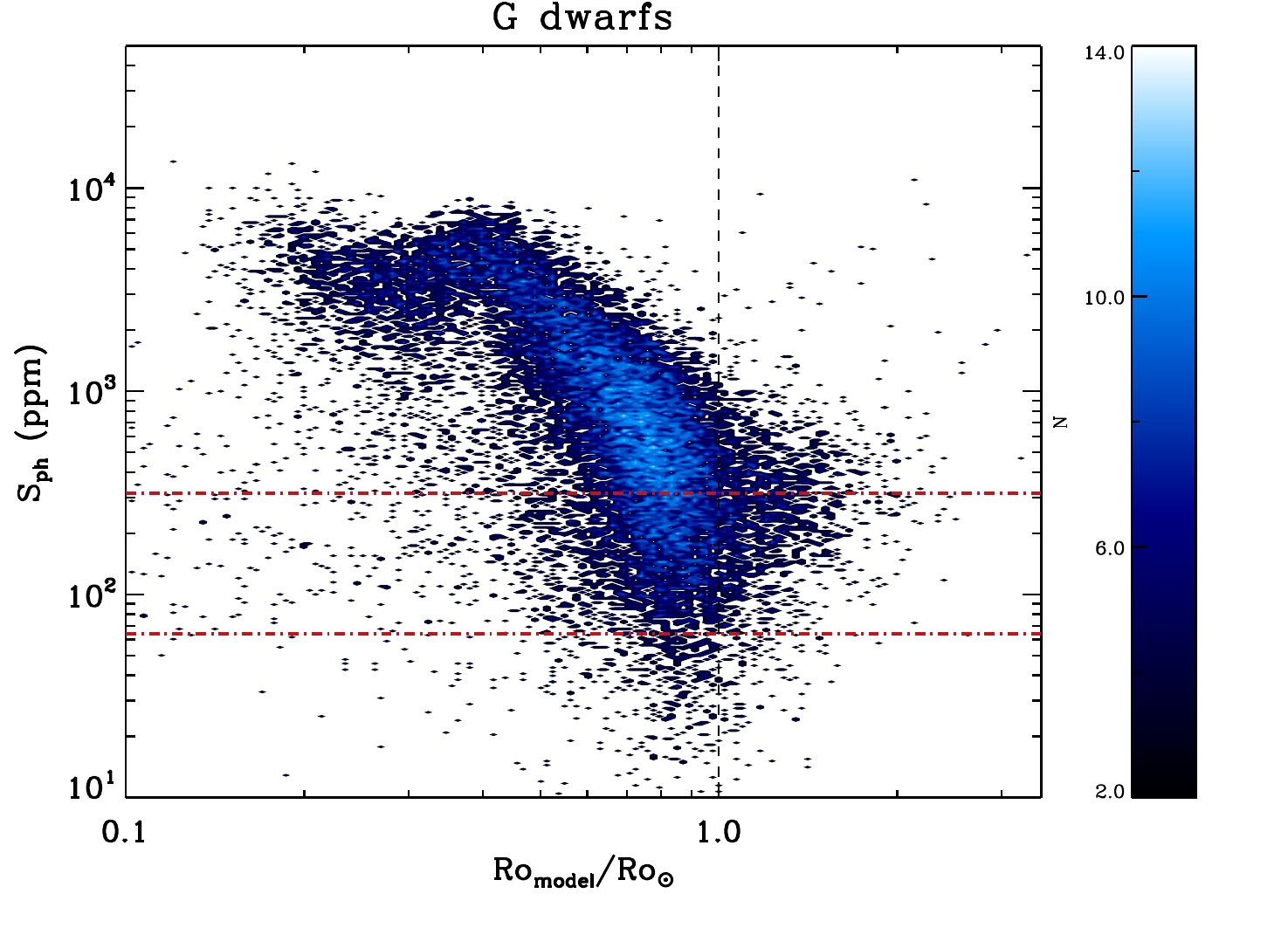}
    \includegraphics[width=8cm]{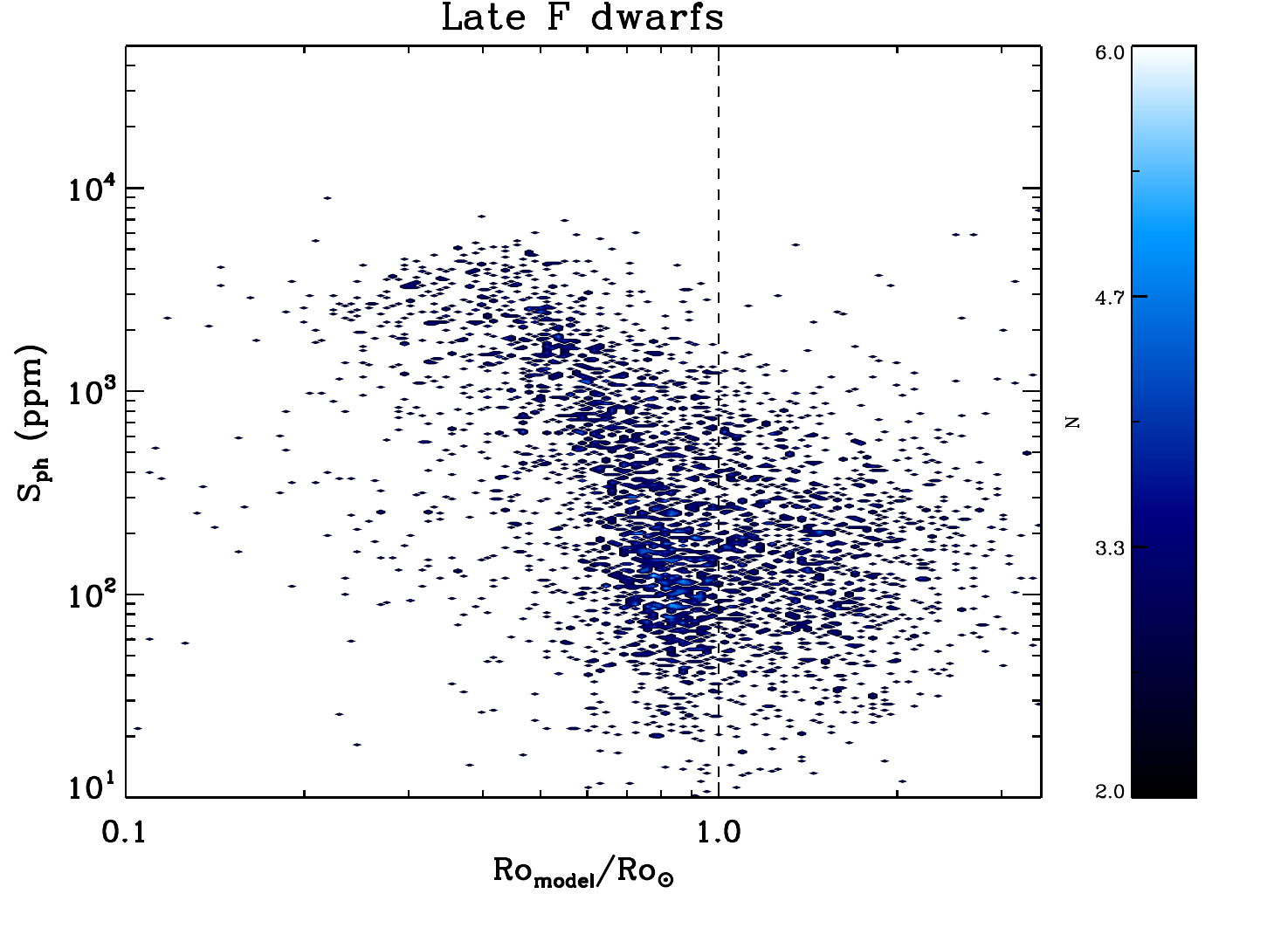}
    \includegraphics[width=8cm]{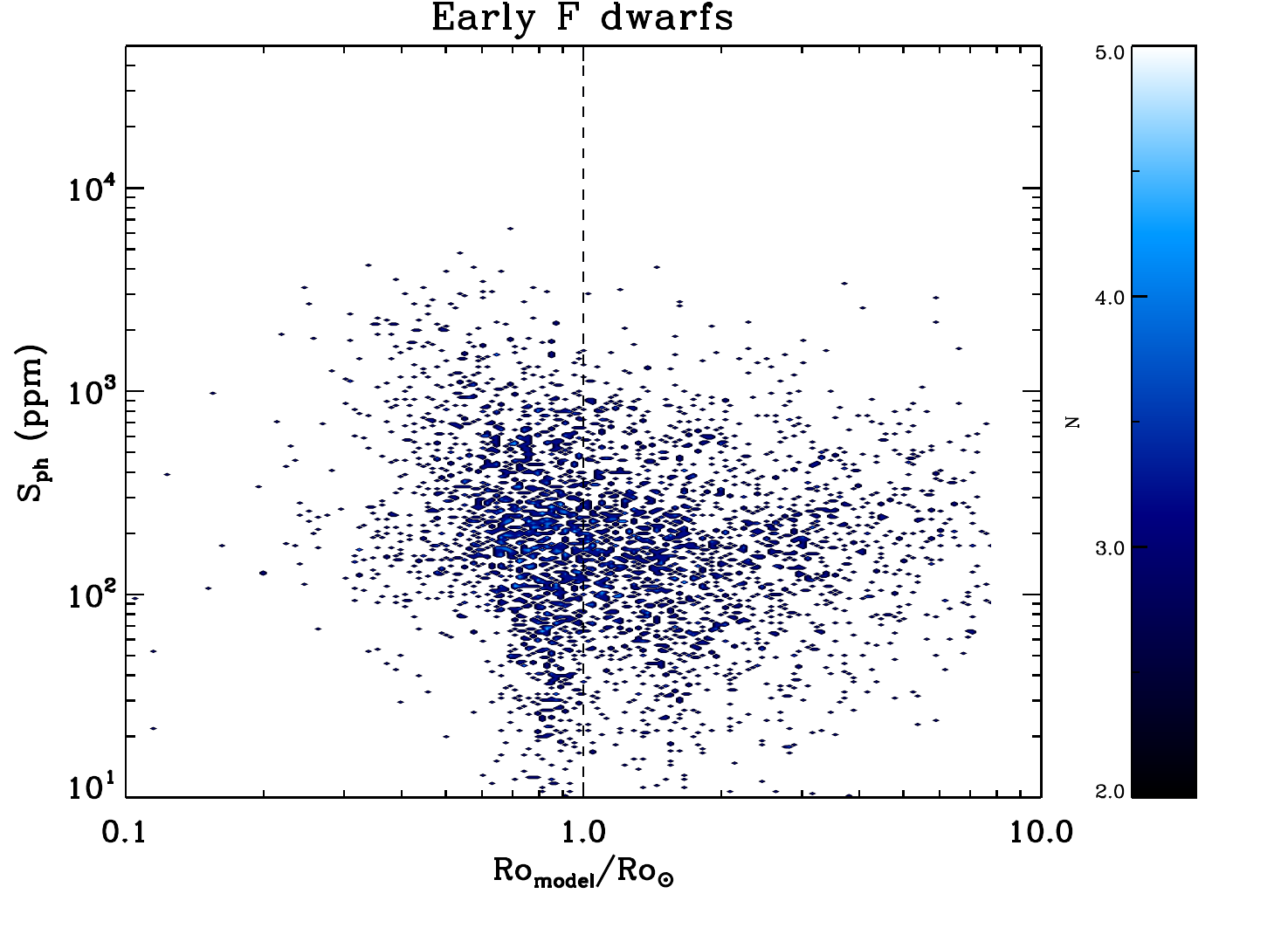}
    \caption{$S_{\rm ph}$ as a function of the model Rossby number normalized to the solar value for different spectral types for all stars color-coded with the number density of stars. In each panel, the vertical dash line corresponds to the solar $Ro$. The horizontal red dash-dot lines for the G dwarfs correspond to the range of $\sph$ for the Sun between minimum and maximum magnetic activity.}
    \label{Sph_Ro_spec}
\end{figure*}

We investigate here the shape and behaviour of the stars in the $\sph$-Ro diagram as a function of the spectral type, with some focus on the aforementioned {\it dip} at low $Ro$. While \citet{2022ApJ...933..195M} also studied in details the shape as a function of mass, the author used a sample of almost 5,000 stars with LAMOST spectra. With our study we increase that number to more than 38,000 stars with different sources for $\teff$ and $\feh$. By comparing the effective temperature from APOGEE, LAMOST and the Gaia-{\it Kepler} catalogs for stars in common, the differences are within 10\%. For the metallicity, the differences are within 0.2\,dex. Though such comparison was done on a few thousands stars, we assume that having atmospheric parameters from different sources should not affect much the general behaviors.

We thus divided our sample into different spectral types as follows \citep{2013ApJS..208....9P}: early F dwarfs are selected with $T_{\rm eff}$ above 6,250\,K, late F dwarfs with $T_{\rm eff}$ between 6,000 and 6,250\,K, G dwarfs with $T_{\rm eff}$ between 5,200 and 6,000\,K, and K dwarfs with $T_{\rm eff}$ between 3,700 and 5,200\,K (see Figure~\ref{Sph_Ro_spec}). 


{  We remind that the sample presented here corresponds to stars with detected rotation periods, which is a fraction of the full sample of stars observed by {\it Kepler}. Indeed, \citet{2021ApJS..255...17S} reported that the detection rate was 51\% for K dwarfs, 31.1\% for G dwarfs, and 29.3\% for F dwarfs. The results that we obtain and that are discussed below cannot be generalized to all the stars but stars with measured rotation periods. }

Different spectral types present different shapes in the $\sph$-$Ro$ diagram. For K dwarfs (top left panel), $\sph$ probes a smaller Rossby number range than G (top right panel) and F dwarfs (bottom panels) that have lower magnetic activity. However, the scatter for a given $Ro$ appears to be smaller for G dwarfs compared to K dwarfs. Similarly to \citet{2022ApJ...933..195M}, we see that the downward slope above 0.4\,$Ro_\odot$ is steeper for G dwarfs than for K dwarfs. The magnetic activity of K dwarfs (and slightly for G dwarfs) seems to keep increasing towards low $Ro$. This suggests that the behavior of magnetic activity of K and G dwarfs is similar but with some slight differences that could be related to the underlying physics of the magnetic activity of the stars. 

 Looking at the dip and the kink for the G and K dwarfs, we can compare them quantitatively. {  To define the location of the dip, we first take bins of size 0.0025 $Ro/Ro_{\odot}$ and compute the 95th percentile of $\sph$ in each bin. We then fit a second order polynomial. The minimum of the polynomial provides the location of the dip.} For the K dwarfs, {  we measure the dip at 0.294\,$\pm$\,0.058\,$Ro_\odot$ compared to 0.286\,$\pm$\,0.077\,$Ro_\odot$ for G dwarfs.} At the location of the dip, {  for K dwarfs, the $\sph$ range (between the 30th percentile and the 95th percentile) is between 1070\,ppm and 4620\,ppm while for G dwarfs $\sph$ ranges between 2000\,ppm and 5370\,ppm.}
 {  We apply the same method to determine the location of the bump or peak.} The peak {  is located 0.396\,$\pm$\,0.092\,$Ro_\odot$ for K dwarfs and at 0.387\,$\pm$\,0.077\,$Ro_\odot$}. For the K dwarfs, at the {  peak}, the magnetic activity proxy is between {  1760\,ppm and 6200\,ppm  compared to 2350\,ppm and 6900\,ppm for G dwarfs}.


Observationally, there is a change of behaviour in the $\sph$-Ro diagram for F dwarfs, which is clear when comparing early and late types.  Early F dwarfs do not populate the Rossby number space associated with the dip, and are instead scattered towards higher $Ro$ values and low $\sph$ values.  For the late F dwarfs, a slight increase in $\sph$ with $Ro$ similar to the G dwarfs can be seen between 0.2 and 0.4\,$Ro_\odot$, although the number of stars in this  $Ro$ range is much smaller. Early F dwarfs have a larger scatter in both $\sph$ and $Ro$, blurring any correlation between the two quantities. Having a much thinner convective zone than the late F dwarfs, they go through an even weaker magnetic braking. This means that they do not have a significant spin down and remain relatively fast during the main sequence \citep{2015ApJ...799L..23M}. As a consequence, the larger $Ro$ is mostly due to very small convective overturn timescale.
While $\prot$ is in general shorter for F dwarfs than for the other spectral types, $\tau_c$ is much longer for G and K dwarfs. Typical $\tau_c$ values for F dwarfs range between 1 and 12\,days while for G and K dwarfs they extend from 7 to 25\,days.  However we note that for the early F dwarfs, caution should be taken due to the very thin convective zone. These trends can also be identified when plotting $\sph$ as a function of $\prot$ instead of $Ro$  \citep[see Figure~7 of][]{2021ApJS..255...17S}. Stars move from the left of the $\sph$-$\prot$ diagram (short $\prot$) towards the right of the $\sph$-$Ro$ diagram (high $Ro$). 
We will discuss these differences further in Section~\ref{sec:discussion}.

\subsection{As a function of mass}

\begin{figure}[h!]
    \centering
    \includegraphics[width=9cm]{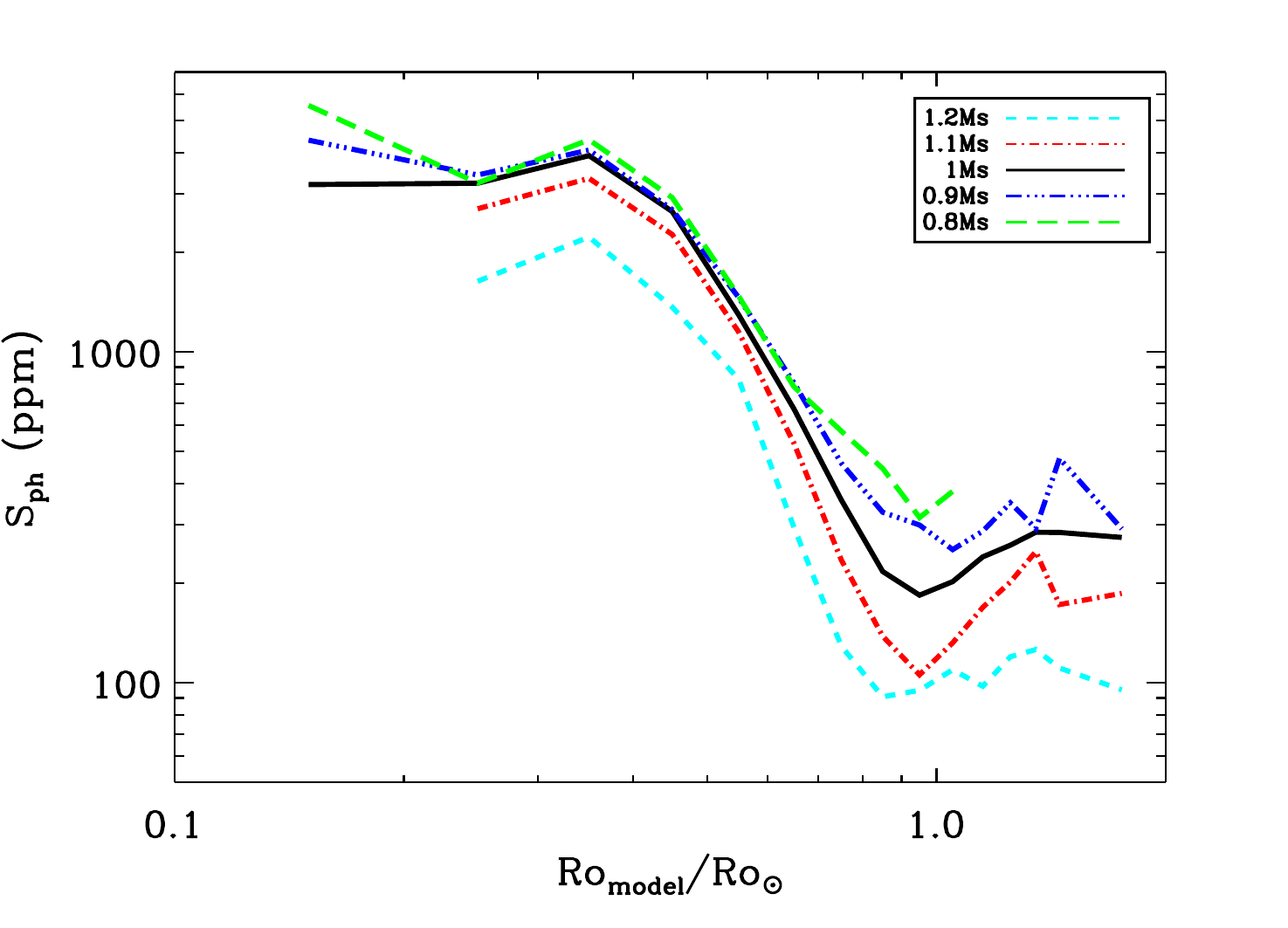}
    \caption{Median $\sph$ in bins of 0.1\,$Ro/Ro_\odot$ as a function of the normalized $Ro$ for masses between 0.8$M_\odot$ and 1.2$M_\odot$.}
    \label{Sph_Ro_Mass_median}
\end{figure}

 We divide our sample into bins of \texttt{kiauhoku} masses of 0.8, 0.9, 1, 1.1, and 1.2\,M$_\odot$ (within 0.1\,M\,$_\odot$) also by selecting them in a given temperature range of 5350, 5550, 5750, 5950, 6160\,K (within 100\,K) to have a better defined sample. {  We applied these cuts in both temperatures and mass in order to remove trends of $\sph$ with metallicity.} By taking the median value of $\sph$ in bins of 0.1\,$Ro/Ro_{\odot}$, we represented them for the different masses in Figure~\ref{Sph_Ro_Mass_median}. For a given $Ro$, $\sph$ decreases as mass increases. 

For stars with masses of 0.8 and 0.9\,M$_\odot$, $\sph$ increases towards low $Ro$. We can see the dip at low $Ro$, and an increase in activity towards high $Ro$ (above 1\,$Ro_{\odot}$). Stars with masses between 0.9 and 1.1\,$M_\odot$ present a notable increase in $\sph$ above the solar Rossby.

For the 1.2\,$M_\odot$ track, given the small number of stars with $Ro$ below 0.3\,$Ro_{\odot}$, we represent only the bins with more than 10 stars. We see a flat behaviour for high $Ro$ as noted for the F dwarfs in the previous section.

\subsection{The Sun and Sun-like stars}
\label{sec:Sunlike}

\begin{figure*}[h!]
    \centering
    \includegraphics[width=12cm]{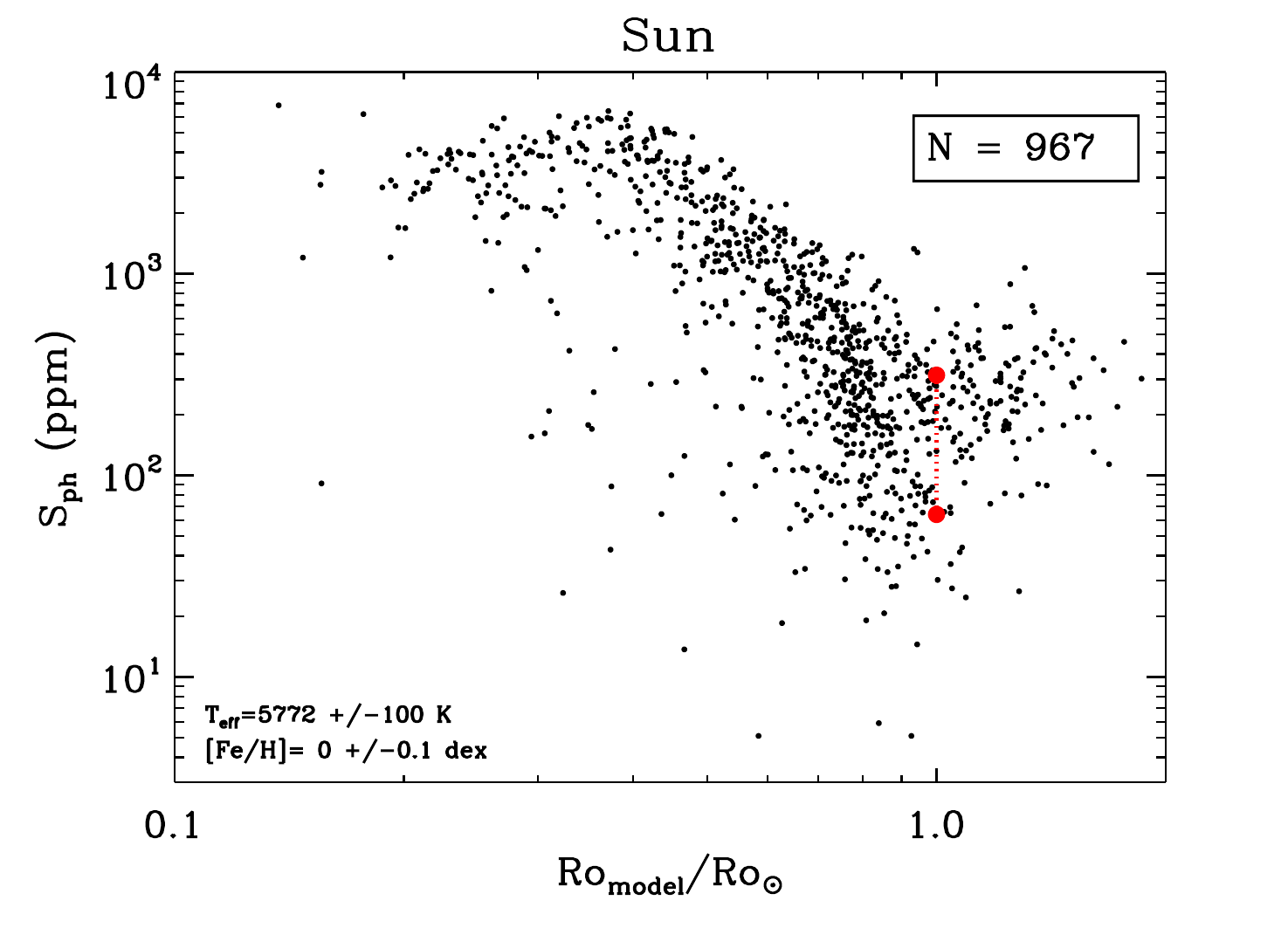}
    \caption{$\sph$ as a function of Rossby number for solar analogs compared to the Sun represented by the red circles for minimum and maximum magnetic activity during a solar cycle. Only stars with spectroscopic values were selected and the number of stars is shown in the top right part of the figure. The ranges of $\teff$ and $\feh$ are given in the bottom left part of the figure and as explained in Section~\ref{sec:Sunlike}.}
    \label{Sph_Ro_spec_Sun}
\end{figure*}

Placing the Sun's magnetic activity in context with that of its siblings is vital for understanding the solar/stellar dynamo. However, the identification of solar twins and analogs can be challenging. Different criteria have been used to select these solar analogs \citep[e.g.][]{2020ApJ...898..173D,2021ApJ...908L..21R}. Here we take stars with atmospheric parameters from spectroscopy to have a better comparison sample. We define the solar analogs as stars with the solar effective temperature $\pm$\,100K and solar metallicity $\pm$\,0.1\,dex. {  We do not make any selection on luminosity or surface gravity to be able to see the evolution with age.} The result is shown in Figure~\ref{Sph_Ro_spec_Sun}. 

For the Sun, using data from the Variability of solar IRradiance and Gravity Oscillations instrument \citep[VIRGO][]{1995SoPh..162..101F}, between minimum and maximum activity, $\sph$ varies from 64 to 310\,ppm \citep{2019FrASS...6...46M}. While comparing these values to all the G dwarfs {  of our sample with measured $\prot$} (top right panel of Figure~\ref{Sph_Ro_spec}), we could say that the Sun is less active than solar-like stars. However, compared to our selection of solar analogs {  with detection of rotation}, we can see that the Sun's activity fits well within the range of values from the solar analogs with close to solar Rossby number (which would correspond to a similar age) as pointed out by \citet{2020arXiv200704416M} and \citet{2023A&A...672A..56S}. 

{  There has been some discussion on whether the Sun would be part of the stars with detected periods by {\it Kepler}. We analyzed more than 20 years of VIRGO data taking subseries of 4 yrs to mimic the {\it Kepler} observations and shifting the center of each box by 1 yr. This gives us 26 subseries for which we ran the rotation pipeline along with the machine learning ROOSTER (see Section~\ref{sec:rot}). Trained on the same sample as the one used for the rotation catalog, we obtained the correct rotation period for 19/26 cases, i.e. 74\%. The training of ROOSTER was done to select the rotation period from the wavelet analysis, which we know can provide sometimes half of the rotation periods. If we change the training to select the rotation period from the Composite Spectrum (the combination of the wavelet power spectrum and the ACF), in 100\% of the cases we retrieve the correct rotation period of the Sun. This analysis shows that the Sun would have a detection probability between 74 and 100\% and subsequently would indeed be part of the sample with rotation detection and with the correct rotation period.}

We also notice that at maximum activity, the Sun is close to the upper edge of the $\sph$-Ro envelope, in agreement with our suggestion from Section~\ref{sec:MS} that stars with a 90\,$^\circ$ inclination angle are close to the upper edge of the $\sph$-$Ro$ diagram. 

We finally note a visible break of slope around the solar Rossby number. Sun-like stars, with larger Rossby values than the Sun, appear to have larger $\sph$ values compared to Sun-like stars around the solar $Ro$. {  Even though we removed all known binaries, we cannot exclude that some remain in our sample, including at high $Ro$. However it seems unlikely that all of them are in binary systems.} We need more observational targets to conclude if this is a real trend, but it seems to go in the direction of \citet{2018ApJ...855L..22B} observations and \citet{2024A&A...684A.156N} numerical results, that will be discussed in Section~\ref{sec:discussion}.

\begin{figure*}[h!]
    \centering
    \includegraphics[width=8cm]{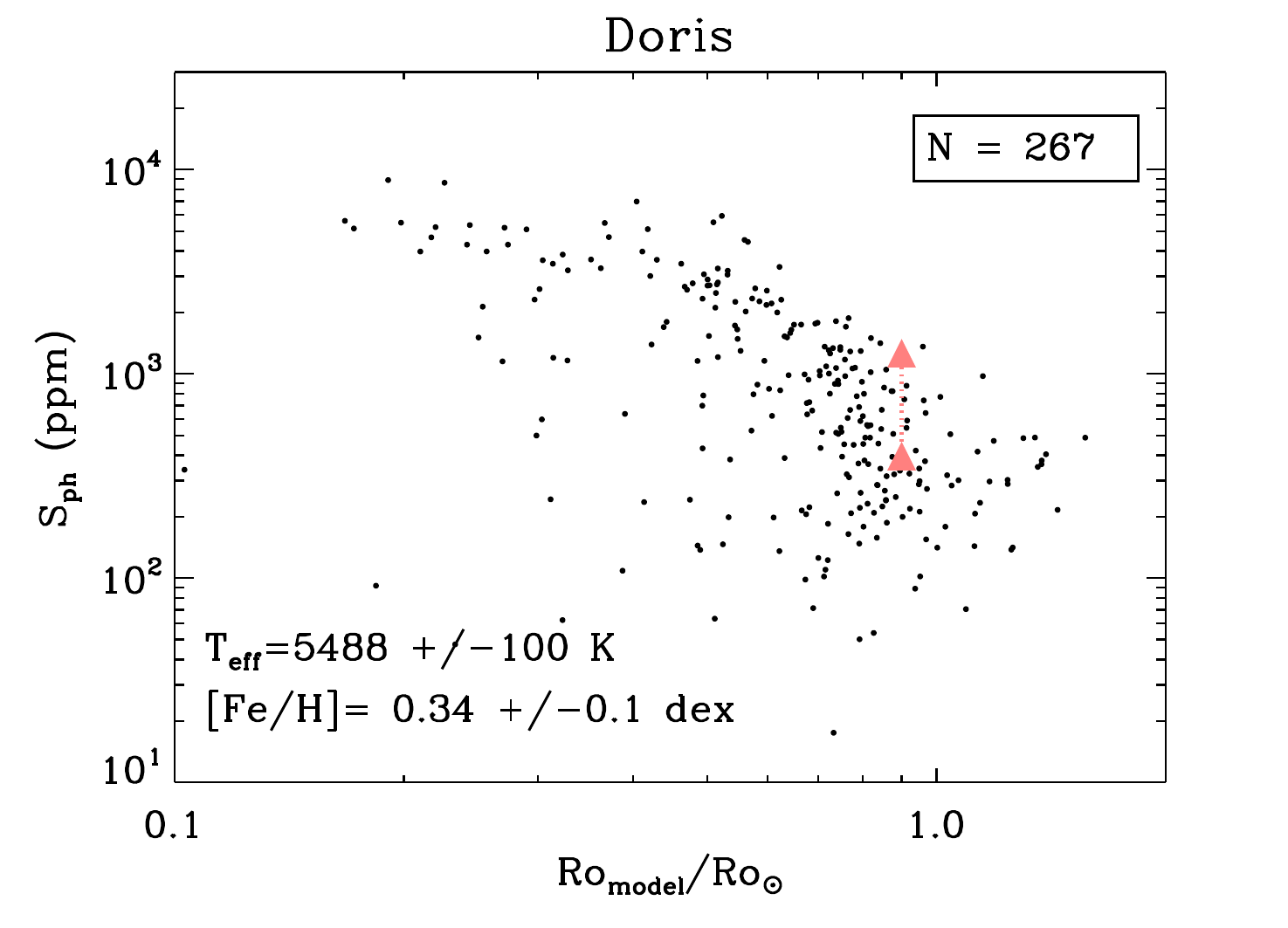}
    \includegraphics[width=8cm]{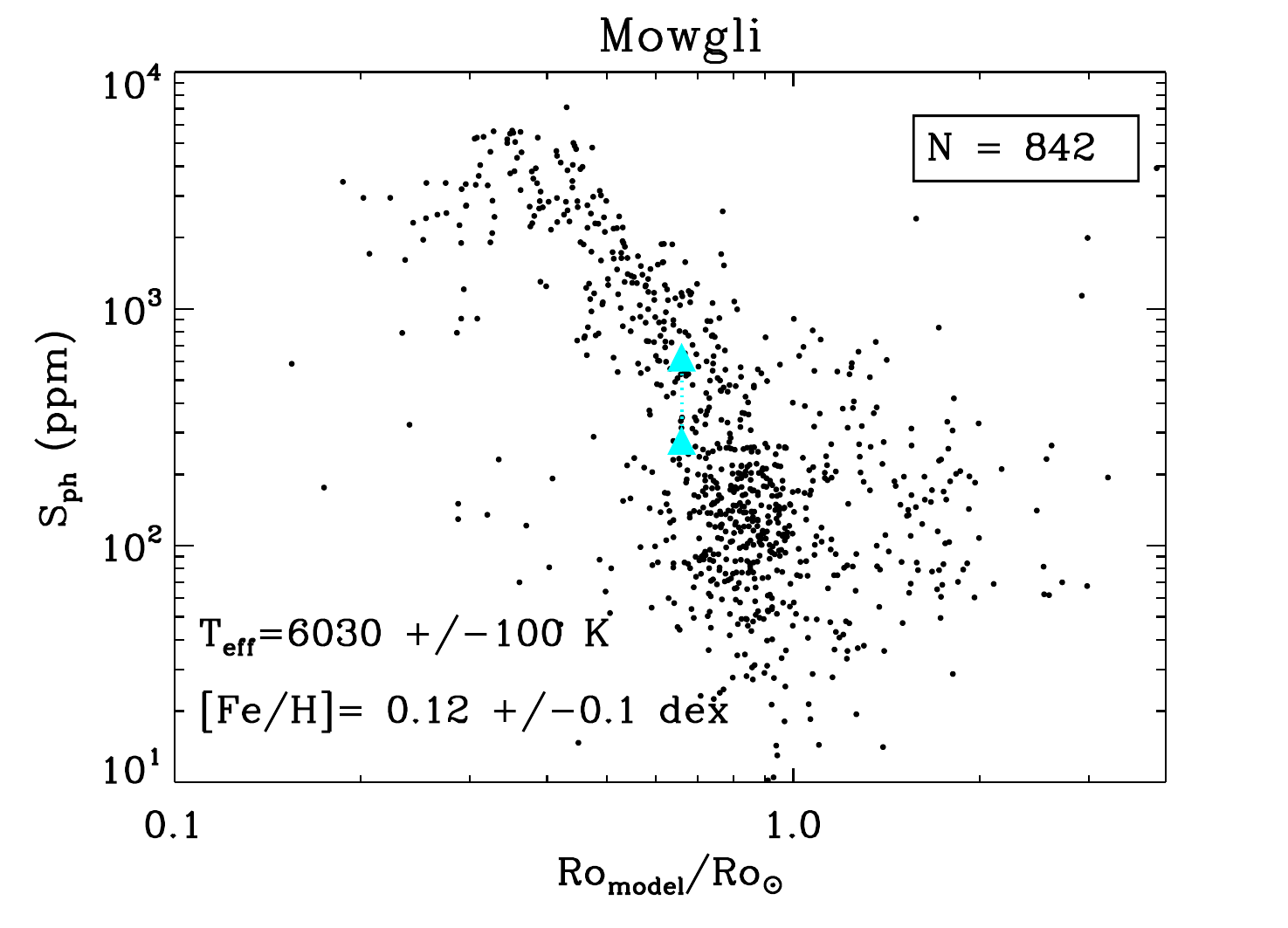}
    \caption{$\sph$ as a function of Rossby number for stars similar (see Section~\ref{sec:specific} for details) to two very well seismically studied {\it Kepler} stars : KIC~8006161 (left panel, pink triangles) and KIC~10644253 (right panel, cyan triangles). The colored symbols represent the range of magnetic activity during the {\it Kepler} observations. Only stars with spectroscopic values were selected and the number of stars is shown in the top right part of the figure. The ranges of $\teff$ and $\feh$ are given in the bottom left part of the figure.}
    \label{Sph_Ro_specialstars}
\end{figure*}

\subsection{Stars with known magnetic cycles}
\label{sec:specific}


Similarly to the previous comparison of the Sun with solar analogs, we study a few {\it Kepler} seismic targets known to have cycles or magnetic activity variability. 

The first one of them is KIC~8006161 (also known as Doris) which is similar to the Sun with $\teff$ of 5488\,$\pm$\,77\,K but metal-rich with 0.34\,$\pm$\,0.1\,dex and a seismic age of 4.57\,$\pm$\,0.36\,Gyr. By combining the {\it Kepler} observations with ground-based spectroscopic observations, \citet{2018ApJ...852...46K} have shown that the higher metallicity was responsible for a shorter and stronger cycle. In the left panel of Figure~\ref{Sph_Ro_specialstars}, we compare the minimum and maximum levels of activity of that star with stars that have similar spectroscopic properties, i.e. $\teff$ within $\pm$\,100\,K and [Fe/H] within $\pm$\,0.1\,dex of Doris' stellar parameters. While we know that during the {\it Kepler} observations we could not see the full cycle (known to be of $\sim7$ years), we probably saw most of the range of magnetic activity of one cycle. Compared to its pairs, KIC~8006161 seems to be close to the upper bound of the $\sph$-$Ro$ diagram. This means that even though the inclination angle derived by seismology is lower than the Sun ($\sim$\,38\,$^\circ$), we are still able to see most of the active latitudes. The upper envelope in the $\sph$-$Ro$ diagram should then correspond to stars for which the inclination angle allows us to observe the majority of the active regions or active latitudes.



The second star with seismic and magnetic activity studies is KIC~10644253 (aka Mowgli). It was shown by \citet{2016A&A...589A.118S} that Mowgli has a magnetic activity going on thanks to the shifts in the frequencies of the acoustic modes as well as with the temporal variation of $\sph$. This star is a young Sun with an effective temperature of 6030\,$\pm$\,60\,K and [Fe/H] of 0.12\,$\pm$\,0.6\,dex \citep{2016A&A...589A.118S} but with a seismic age of 1.6\,$\pm$\,0.25\,Gyr. In the right panel of Figure~\ref{Sph_Ro_specialstars}, Mowgli is compared to stars with similar spectroscopic parameters as done for Doris. It appears to be in the same range of magnetic activity level between minimum and maximum (as observed by {\it Kepler}) as stars with similar $Ro$ number.

\subsection{Metallicity effect}

Previous works based on observations and models showed that magnetic activity depends on the metallicity \citep[e.g.][]{2018ApJ...852...46K,2020ApJ...889..108A,2021ApJ...912..127S,2023MNRAS.524.5781S}. Indeed, metal-rich stars have a deeper convective zone, affecting the convective overturn timescale that is larger. The convective velocities become larger and it leads to higher levels of activity. We looked for such effect by selecting stars with spectroscopic input to ensure that reliable metallicity values were used. We then selected stars similar to the Sun in terms of their effective temperatures in the range 5500\,K-6000\,K and with masses from the \texttt{kiauhoku} model selection between 0.9\,M$_\odot$ and 1.1\,M$_\odot$.  

\begin{figure}[h!]
    \centering
    \includegraphics[width=9cm]{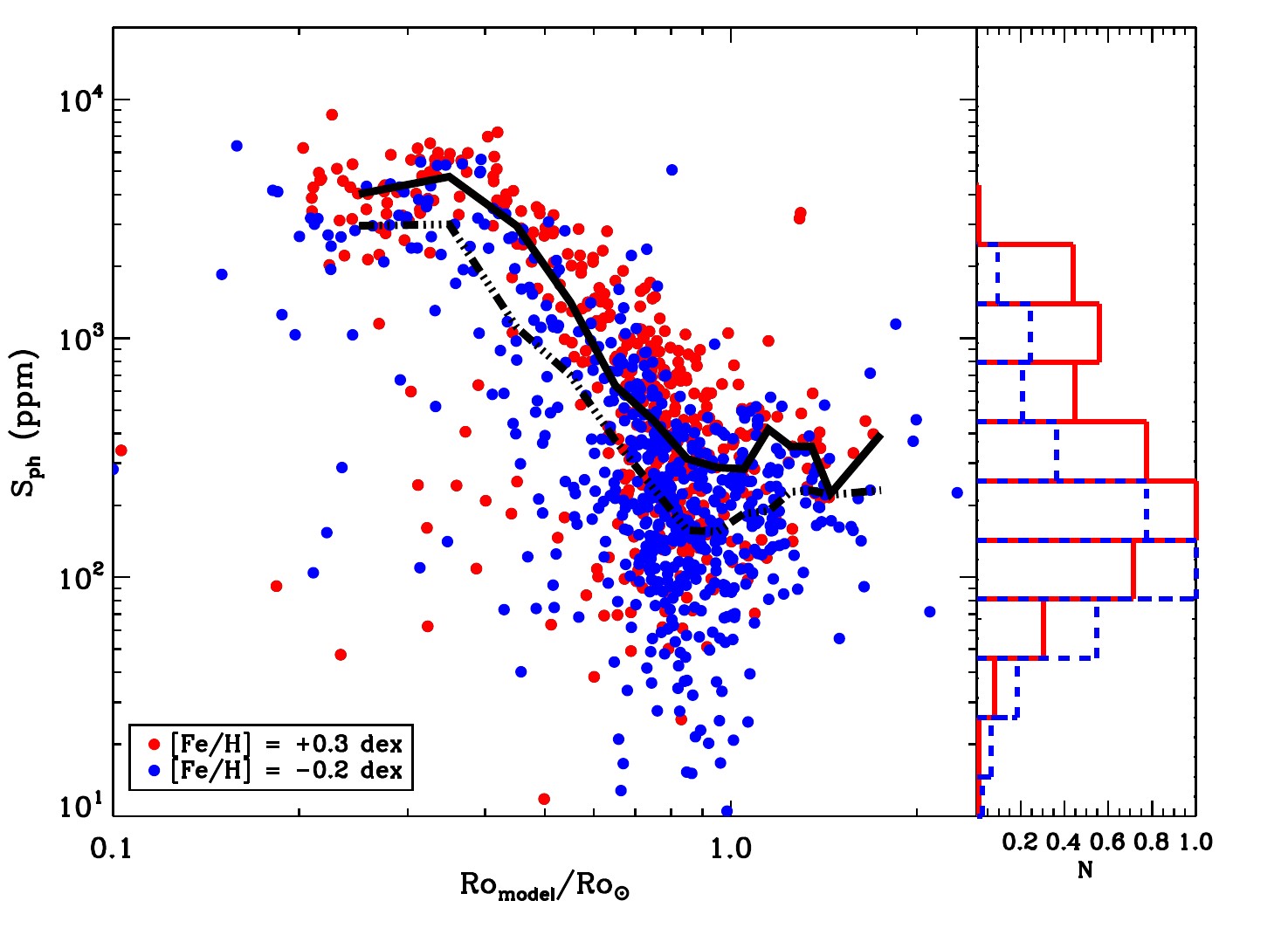}
    \caption{Left panel: $\sph$ as a function of Rossby number for stars with $\teff$ between 5500\,K and 6000\,K and masses between 0.9\,M$_\odot$ and 1.1\,M$_\odot$ for metal-rich (red symbols) and metal-poor (blue symbols) stars. The solid line (resp. triple dot-dash line) represents the median values computed in bins of 0.1\,dex for metal-rich stars (resp. metal-poor stars). Right panel: distribution of the metal-rich (red solid line) and of the metal-poor (blue dash line).}
    \label{Sph_Ro_FeH}
\end{figure}

In Figure~\ref{Sph_Ro_FeH}, we represent $\sph$ as a function of the Rossby number for metal-rich (+0.3\,$\pm$\,0.1\,dex) and metal-poor (-0.2\,$\pm$\,0.1\,dex) stars with a mass between 0.9 and 1.1\,M$_\odot$ {  and the $\teff$ cut mentioned above}, where these values were chosen to have a similar number of stars in both samples. While metal-rich stars (red symbols) seem to lay above metal-poor stars (blue symbols), we also compute the median values of $\sph$ in bins of 0.1 in $Ro/Ro_\odot$, which allows us to quantify the difference between the two samples. On the right panel, we can see that the distribution of the metal-rich stars tends towards higher magnetic activity compared to the metal-poor stars. Similarly to the works aforementioned, we find that metal-rich stars have larger $\sph$ values for a given Rossby number than the metal-poor stars. We also note that there may be a bias due to the fact that we do not see the full magnetic cycles of stars during the 4 years of {\it Kepler} observations.  Another potential bias might stem from the influence of the metallicity on the the length of magnetic cycles. This was exemplified in the comparison between the Sun and a comparable metal-rich star \citep{2018ApJ...852...46K}, where the latter exhibited a shorter cycle. However, since this analysis was conducted solely on a single pair of stars, a broader sample would be necessary to validate this pattern. 

 According to the YREC evolution models, $\tau_c$ varies strongly with temperature and more weakly (but measurably, a factor of 1.25 per dex) with metallicity in a way that is independent of temperature, mass, age, or $\sph$. The difference in $\sph$ at fixed $Ro$ but different metallicities therefore suggests that $\sph$ has some metallicity dependence that is separate from the $Ro$ dependence on metallicity.





\section{Discussion}
\label{sec:discussion}

For our {\it Kepler} targets, the diagram of the magnetic activity proxy as a function of the Rossby number can be divided into four different regimes: at very low $Ro$ (below 0.3\,$Ro_\odot$),  low $Ro$ (between 0.3 and 0.4\,$Ro_\odot$),  intermediate $Ro$ (between 0.4 and 1\,$Ro_\odot$), and high $Ro$ (above 1\,$Ro_\odot$). All the regimes are not observed in all the spectral types and we discuss here some hypotheses on the physical reasons for these differences and for the changes during the evolution of a solar-like star.  

\subsection{Differences between G and K dwarfs for low $Ro$}

While G and K dwarfs show similar behaviors in Figure~\ref{Sph_Ro_spec}, one main difference is the {  level of magnetic activity. We saw in Section~\ref{sec:spectral_type} that} for a given $Ro$ value {  below 0.5\,$Ro_\odot$, $\sph$} is {  smaller} for K dwarfs compared to G dwarfs. {  The scatter in $\sph$ at the dip and the peak is larger by 180\,ppm for K dwarfs than for G dwarfs. This is more than half of the $\sph$ range between minimum and maximum activity for the Sun, i.e. not negligible. This suggests that K dwarfs are less active but more variable than G dwarfs around the dip.} In general, we assume that $\sph$ is mostly sensitive to spots present on the surface of the star. Indeed \cite{2016A&A...589A..46S} and more recently \citet{2024ApJ...963..102L} showed that with the {\it Kepler} bandpass and given the rotation time scale, the detection of rotational modulation is mostly sensitive to the spots. This was also revisited by \cite{2019A&A...629A..42M} with synthetic time series for an even broader range of spectral types. This means that lower values of $\sph$ could be related to different physical or geometrical reasons:

\begin{itemize}
    \item a smaller coverage of spots/active regions on the stellar surface: for  $Ro/Ro_\odot$ below 0.4, G dwarfs could present a narrower range of spot coverage compared to K dwarfs. It does not say anything about the size or number of spots but the general surface coverage. {  This is in agreement with the findings by \citet{2023ApJ...951L..49C} who studied the spot filling factors of G and K dwarfs.}
    \item our magnetic proxy $\sph$ depends on the inclination angle of the rotation axis, the position of the active latitudes, {  and the active longitudes}. Given that dependency, one other hypothesis to explain the difference between K and G dwarfs is that the active latitudes in K dwarfs could probe a larger band than G dwarfs where the spots would appear in a narrower band of latitudes \citep{2018A&A...620A.177I}. 
    \item finally, the length of the magnetic activity cycles could also impact the measurement of the magnetic activity levels. If a star has long cycle periods (longer than 10 years), during the 4 years of the {\it Kepler} observations we would see only part of the cycles and would be biased towards high or low $\sph$, while for a star with shorter cycles, we could measure an average level of magnetic activity. One hypothesis on the smaller scatter for the G dwarfs compared to the K dwarfs is that we capture more often a full cycle for the G dwarfs than for the K dwarfs. {  We discuss it in Appendix~\ref{app:Sun_cycle} by testing the $\sph$ calculation on the Sun.} This also goes in the same direction that K dwarfs being in general slower rotators than G dwarfs: we expect them to have longer cycles as suggested from Mount Wilson observations in Ca H\&K \citep{1995ApJ...438..269B,2002AN....323..357S,2017ApJ...845...79B,2022ApJ...939L..26B}. There might be some hint in the spectroscopic observations \citep{2017ApJ...845...79B}, but we would need a larger sample to confirm it.
\end{itemize}

These are just hypotheses and it could be a combination of some or all of them. Unfortunately we cannot confirm any of them due to the lack of information on the size of spots, inclination angle, and active latitudes. Asteroseismology, also combined with photometric analysis, could provide an answer for some of those properties \citep[e.g.][]{2017A&A...599A...1S,2018A&A...619L...9B,2019MNRAS.485.3857T,2021NatAs...5..707H,2023A&A...680A..27B} and in that respect the future European Space Agency mission PLATO \citep[PLAnetary Transits and Oscillations of stars;][]{2014ExA....38..249R} is promising in order to provide a better overview, as we expect several thousands of solar-like stars very well characterized with seismology \citep{2024A&A...683A..78G}.


\subsection{Change of behavior for F dwarfs at intermediate and high $Ro$}

In Section~\ref{sec:spectral_type}, we observed a change in the $\sph$-$Ro$ diagram when moving towards hotter F dwarfs. 


Being more massive stars, they evolve faster than G and K dwarfs. The lack of  F dwarfs with high activity below 0.3\,$Ro_\odot$, and hence the absence of dip, could be related to the fact that they quickly move to higher $Ro$ rather than an observation bias, since we can detect high $\sph$ values for the less massive stars. This is in agreement with the models by \citet{2020A&A...636A..76S}. However, some observational bias cannot be excluded. Indeed, we could also hypothesize that the manifestation of magnetic activity for those young F dwarfs is different and such that it prevents us from detecting rotation periods with our usual techniques. Such absence of trend is not specific to the $\sph$ proxy for magnetic activity, as \citet{2019ApJS..241...29Y} found a similar behavior for the F dwarfs when studying the flare activity of the {\it Kepler} targets.

From a stellar structure point of view, these stars have a thinner convection zone (a few percent of the stellar radius) compared to K and G dwarfs (more than 20\% of the stellar radius) leading to smaller $\tauc$ and thus higher $Ro$ values and allowing us to probe a not well known regime above the solar $Ro$. While late F dwarfs have their modeled masses peaking around 1.15\,M$_\odot$, early F dwarfs are more massive, around 1.3\,M$_\odot$, close to the Kraft break where the evolution of rotation is drastically different \citep{1967ApJ...150..551K}. This change in the stellar structure will obviously affect how the dynamo operates and could explain the lower $\sph$ values for those stars. Some 3D Magneto-Hydrodynamical simulations started to look into the dynamo in F stars \citep{2013ApJ...777..153A} but more observational constraints are needed as well.  It is also possible that $\tau_c$ from the models is less well constrained in these regimes of very thin convective zones, which could lead to some additional scatter that is not real.

For the high $Ro$ values, there is no clear consensus in the current simulations on what to expect for these stars. 
While the simulations by \citet{2022ApJ...926...21B} are done up to 1.1\,$M_\odot$ stars, when stars go beyond the solar Rossby, the simulations suggest an absence of magnetic cycles. Such behaviour was found with two very different codes, suggesting a robust finding. This threshold also corresponds to the transition in the simulations from solar-like differential rotation to anti-solar differential rotation\footnote{faster rotation at the poles than at the equator}, which is thought to sustain stationary dynamos without cycles in most cases \citep{2022ApJ...926...21B}. In our observations we also notice that stars with $Ro$ above 1\,$Ro_\odot$ have a low $\sph$ that is flat with $Ro$. 
 However, there is no clear consensus on what happens at high $Ro$ as observational constraints are lacking \citep{2022A&A...667A..50N} and other simulations by \citet{2019ApJ...886...21V} lead to some cycle variation.  

Finally, in contrast to F dwarfs, we also note that G dwarfs with high $Ro$ have an increased level of magnetic activity.  \citet{2018ApJ...855L..22B} suggested that stars with anti-solar differential rotation, usually found at high $Ro$ \citep{2022A&A...667A..50N}, have enhanced magnetic activity, which goes in the same direction as some of our observations. Recent findings of \citet{2024A&A...684A.156N} report strengthening of large-scale magnetic topologies at high Rossby number from numerical simulations. {  As mentioned above, at high Rossby numbers, stars could transition to anti-solar differential rotation due to the weakening influence of rotation on convective motions, as was shown by multiple theoretical studies \citep[see][and references therein]{2013A&A...549L...5G,2022ApJ...926...21B}. This changes the sign of the Omega-effect responsible for regenerating the toroidal magnetic field from the poloidal field. Flux emergence then produces active regions that reinforce the pre-existing orientation of the large-scale field. In this case, the dynamo mechanism is therefore no longer cyclic, and instead strengthens the polar magnetic fields until it saturates (either non-linearly by quenching the differential rotation or by mere turbulent diffusion), reaching effectively stronger fields compared to their cyclic analogs. This was demonstrated in detail with 2D axisymmetric mean-field dynamo models by \citet{2020MNRAS.491.3155K} and \citet{2022A&A...667A..50N}, and further confirmed with 3D turbulent models in e.g. \citet{2022ApJ...926...21B}.
}

\subsection{The dip}
\label{sec:dip}

We investigate here in more details the dip observed in the G and K dwarfs around 0.3\,$Ro_\odot$. 


\begin{figure}
    \centering
    \includegraphics[width=\hsize]{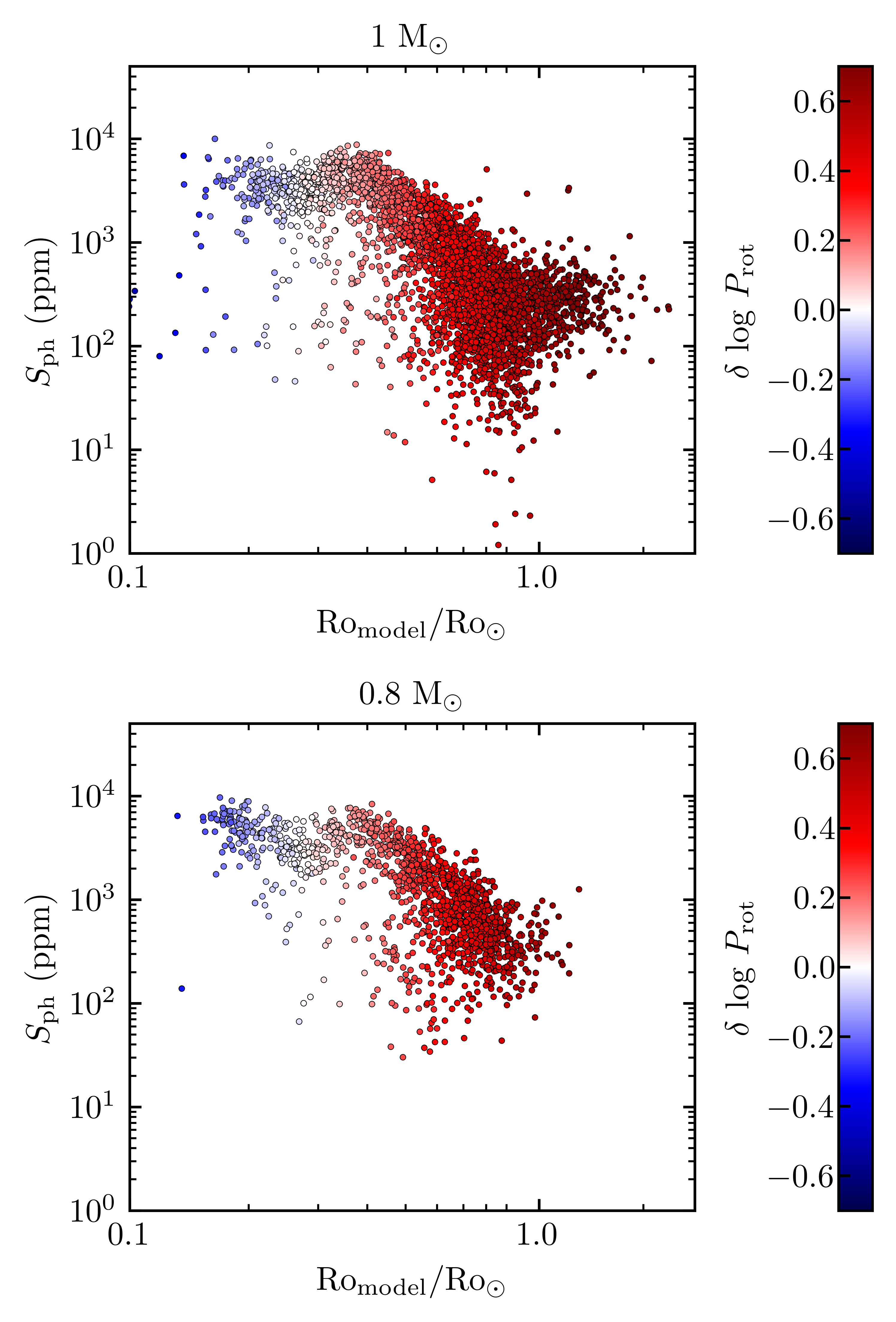}
    \caption{$\sph$ as a function of Rossby for 1$M_\odot$ (top) and 0.8$M_\odot$ (bottom) color-coded by the distance to the intermediate-$\prot$ gap. }
    \label{Sph_Ro_slow_fast}
\end{figure}

We remind that all the stars are in the unsaturated regime described by \citet{2011ApJ...743...48W} where the spin down of the stars follows the Skumanich law. We come back to the $\prot$-$\teff$ diagram of the {\it Kepler} sample, in particular to the intermediate rotation period gap seen for temperatures between 6000 and 4500\,K \citep[e.g.][]{2014ApJS..211...24M,2021ApJ...913...70G,2021ApJS..255...17S}. 
Adopting the location of the intermediate-$\prot$ gap (IPG) by \citet[][following \citeauthor{2023A&A...672A..56S} 2023]{2024FrASS..1156379S}, Figure~\ref{Sph_Ro_slow_fast} shows the $\sph$-Ro diagram color-coded by the distance of each star to the gap, which is represented by the difference between the $\prot$ logarithms. Stars rotating more rapidly than the gap are shown in shades of blue, while stars rotating more slowly than the gap are shown in shades of red. Light colored symbols show stars close to the gap.
We selected stars with masses between 0.9 and 1.1\,M$_\odot$ and $\teff$ between 5500 and 6000\,K (upper panel) and with masses between 0.7 and 0.9\,M$_\odot$ and $\teff$ between 5000 and 5500\,K (lower panel). We clearly see that stars before the gap (or the so-called fast rotation sequence) are generally before the dip while stars after the gap (slow rotation sequence) are after the dip confirming that the surface magnetism goes through some changes. There is even less overlap between both rotation sequences for the $\sim$\,0.8\,M$_\odot$ stars. As suggested by other studies \citep{2021ApJ...913...70G,2022AJ....164..251L}, this could correspond to the {  beginning of the} coupling of the radiative zone with the convection zone \citep{2020A&A...636A..76S}. This is also illustrated in Figure~\ref{Sph_Rom_all_trends}. Since the dip corresponds to the intermediate $\prot$ gap, the coupling should start around that point of the $\sph$-$Ro$ diagram. However it is not clear if the {  decoupling (or the time to be fully coupled)} lasts until the peak at 0.4\,$Ro_\odot$ (point where $\sph$ starts to decrease again). Assuming that the dip corresponds to the core-envelope coupling, the fact that $\sph$ increases just after it suggests that some physical process triggers more magnetic activity before starting to decay again, in agreement with \citet{2023ApJ...951L..49C}. 

We also see that stars very close to the IPG can have a significant drop in $\sph$ down to tens of ppm, in particular for the 0.8\,$M_\odot$ stars. At this stage, with the present dataset, we cannot exclude that this reduction in the surface magnetism is such that the detection of rotation periods is more complicated, and could be partially responsible for the gap in the IGP. {  Another possible explanation for some of those very low $\sph$ stars ($\sph$\,$<$\,500\,ppm) could be a low stellar inclination angle.} 

That dip is not visible in the flare analysis by \citet{2019ApJS..241...29Y}. It could be because they have a smaller sample of stars or that the coupling does not affect the physics of the flares. Based on a different sample, \citet{2024FrASS..1156379S} showed that the X-ray and chromospheric activity also present such dip corresponding to the intermediate $\prot$ gap. Not observing the dip in the flares of the {\it Kepler} targets could be due to the lower sensitivity of flares in the visible than in X-ray or an underestimation of the full flare energy in the visible. 



\section{Summary and Conclusions}
\label{sec:conclusion}

We presented here the study of the evolution of magnetic activity of 38,930 single main-sequence stars observed by {\it Kepler} for 4 continuous years and for which rotation period and the photometric magnetic activity index $\sph$, were computed by \citet{2019ApJS..244...21S, 2021ApJS..255...17S}.

\begin{itemize}
    \item By dividing the sample by spectral types, we found different behaviours. From K to F dwarfs, i.e. with increasing mass, stars are less and less active even accounting for the differences in $Ro$, and this is strongest at high and low $Ro$. More massive stars probe higher values of $Ro$. 
    \item At small $Ro$ below 0.3\,$Ro_{\odot}$ and for a given value, the scatter for K dwarfs is larger than for G dwarfs. This could be due to geometrical effect related to spot coverage or the location of the active latitudes that might depend on the spectral type. Another hypothesis is based on different magnetic activity cycles lengths between K and G dwarfs as it is generally found that K dwarfs are slower rotators than G dwarfs. Hence they should have longer cycles than G dwarfs. 
    \item There is an absence of low $Ro$ for F dwarfs, which can be due to a fast evolution of these more massive stars and also to a larger spin-down time \citep{2015ApJ...799L..23M}. Those hotter stars also probe a quite unknown regime of $Ro$ larger than the solar $Ro$. {  However, we note that the calculation of the convective overturn timescale for those more massive stars should be treated with cautious.}
    \item We found that the upper envelope of the $\sph$-$Ro$ diagram {  corresponds to stars observed close to maximum magnetic activity and with an inclination angle that allows the active regions to be more visible.}
    \item By selecting stars similar to the Sun in terms of their effective temperatures, metallicities, and Rossby numbers we concluded that the magnetic activity of the Sun is similar to that of {  the {\it Kepler} Sun-like stars with measured rotation periods}.
    \item We observed an increase of magnetic activity above the solar $Ro$ for G-type stars, which is also suggested in dynamo models {  where a transition from solar to anti-solar differential rotation can occur \citep{2022ApJ...926...21B,2024A&A...684A.156N}}.
    \item  Using our sample, we found that metal-rich Sun-like stars seem to be more active than metal-poor stars, in agreement with previous works.
    \item We investigated the dip in the $\sph$-$Ro$ diagram. This is an extra feature in the unsaturated regime. We showed that more rapidly rotating stars below the intermediate period gap are located before the dip while stars just after the gap go through an increase of magnetic activity before starting to decrease again. Relating the gap to the core-envelope coupling suggested by \citet{2020A&A...636A..76S}, it would imply that once the coupling starts some physical process begins, which seems to trigger enhanced magnetic activity until the coupling is fully done. 
\end{itemize}

\begin{figure*}
    \centering
    \includegraphics[width=\hsize]{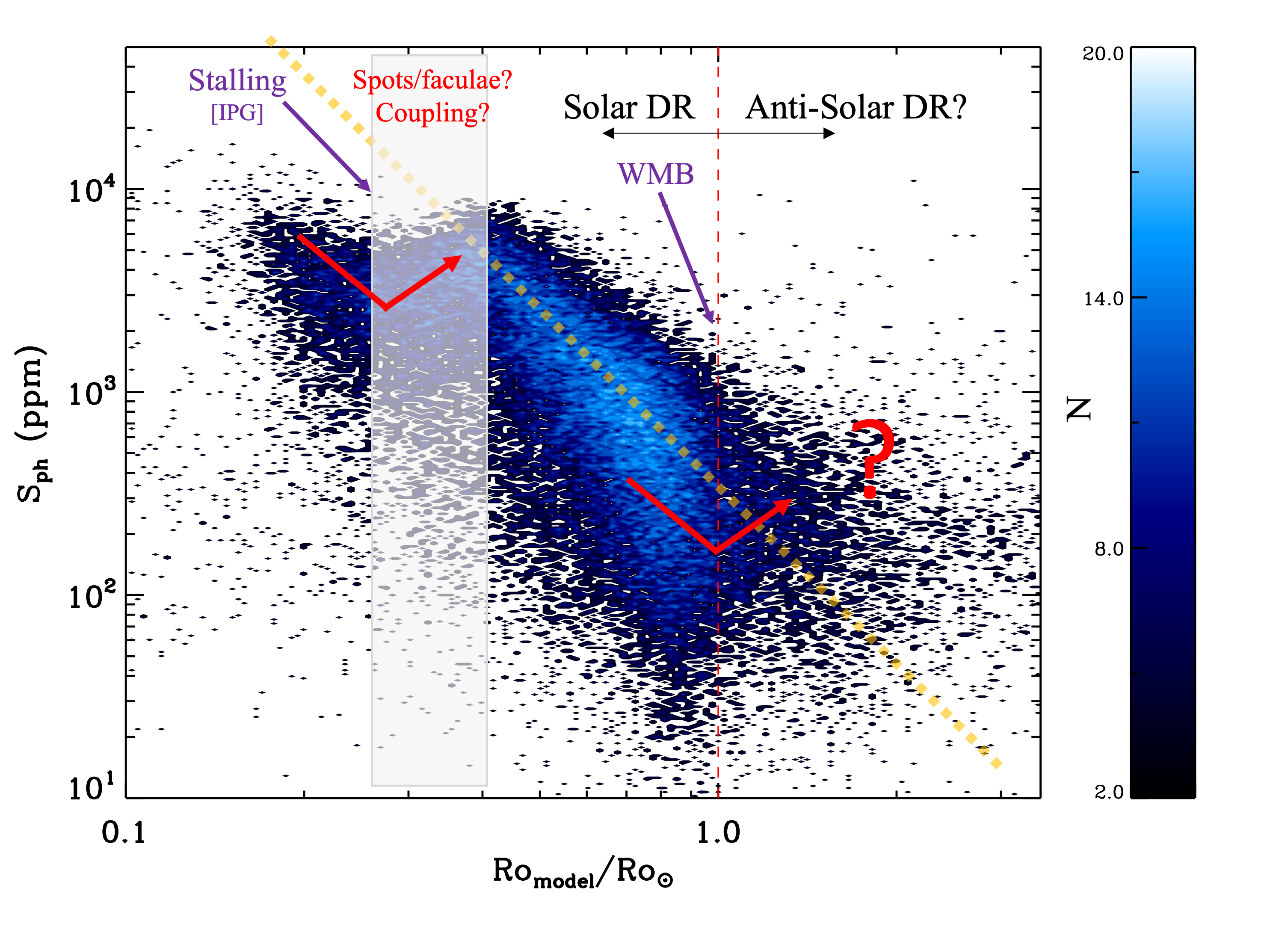}
    \caption{Same as Figure~\ref{Sph_Rom_all} with points discussed in Section~\ref{sec:discussion}. The vertical red dash line represents the solar Rossby number and separates the solar latitudinal differential rotation (DR) regime from the anti-solar differential rotation one as expected from simulations. The location of the stalling (corresponding to the intermediate period gap, IPG) and the weakened magnetic braking (WMB) are shown with the purple arrows. The white semi-opaque area represents the location of the possible coupling between the radiative and convective envelopes discussed in Section~\ref{sec:dip}. The red solid lines are shown to guide the eye on the average evolution of $\sph$. The regime above 1\,$Ro_\odot$ is little known.}
    \label{Sph_Rom_all_trends}
\end{figure*}

 Following our discussion, we summarize in Figure~\ref{Sph_Rom_all_trends} the different theories on the changes occurring during the evolution of solar-like stars in the $\sph$-$Ro$ diagram. We highlight the observed stalling and weakened magnetic braking \citep[WMB;][]{2016Natur.529..181V} as well as the expected changes from solar to anti-solar differential rotation from simulations.

More observations and theoretical development would be needed to confirm some of the theories and hypotheses stated above. Observations with the PLATO mission that will provide several thousands of stars with a precise asteroseismic characterization and combined with magnetic activity monitoring  will likely shed a light on the physics behind these different observations and evolution of magnetic activity. In this framework, revisiting the work by \citet{2019FrASS...6...46M} on the detection of stellar oscillation modes as a function of $\sph$ and $\prot$ by taking into account their intrinsic relationship shown in this work, will be of great importance.
\newline

{  We thank the referee for constructive comments that improved the paper.}
This paper includes data collected by the \emph{Kepler} mission and obtained from the MAST data archive at the Space Telescope Science Institute (STScI). Funding for the \emph{Kepler} mission is provided by the NASA Science Mission Directorate. STScI is operated by the Association of Universities for Research in Astronomy, Inc., under NASA contract NAS 5–26555. We acknowledge
that this research was supported in part by the National Science
Foundation under grant No. NSF PHY-1748958.
S.M.\ acknowledges support by the Spanish Ministry of Science and Innovation with the Ramon y Cajal fellowship number RYC-2015-17697, the grant no. PID2019-107061GB-C66, and through AEI under the Severo Ochoa Centres of Excellence Programme 2020--2023 (CEX2019-000920-S). S.M. and D.G.R. acknowledge support from the Spanish Ministry of Science and Innovation (MICINN) with the grants No. PID2019-107187GB-I00 and PID2023-149439NB-C41. A.R.G.S acknowledges the support from the FCT through national funds and FEDER through COMPETE2020 (UIDB/04434/2020, UIDP/04434/2020 \& 2022.03993.PTDC) and the support from the FCT through the work contract No. 2020.02480.CEECIND/CP1631/CT0001. R.A.G., L.A., A.S. A.S.B, D.B.P, and St.M. acknowledge the support from PLATO CNES grant. A.S., A.S.B., R.A.G. and S.M. acknowledge funding from the Programme National de Plan\'etologie. 
P.G.B. acknowledges support by the Spanish Ministry of Science and Innovation with the \textit{Ram{\'o}n\,y\,Cajal} fellowship number RYC-2021-033137-I and the number MRR4032204. P.G.B., D.G.R., and R.A.G. acknowledge support from the Spanish Ministry of Science and Innovation with the grant no. PID2023-146453NB-100 (\textit{PLAtoSOnG}). S.N.B acknowledges support from PLATO ASI-INAF agreement n.~2015-019-R.1-2018. A.S.B., A.S. and A.J.F acknowledge funding support by ERC Whole Sun \#810218. D.G.R. acknowledges support from the Spanish Ministry of Science and Innovation (MICINN) with the Juan de la Cierva program under contract JDC2022-049054-I. M.H.P. acknowledges support from the Fundaci\'n Occident and the Instituto de Astrof\'sica de Canarias under the Visiting Researcher Programme 2022-2025 agreed between both institutions. St.M. acknowledges sup-
port from the European Research Council (ERC) under the Horizon Europe
programme (Synergy Grant agreement 101071505: 4D-STAR). While partially funded by the European Union, views and opinions expressed are however those of the author only and do not necessarily reflect those of the European Union or the European Research Council. Neither the European Union
nor the granting authority can be held responsible for them.


 \begin{table*}
\small
\centering
 \caption{Input parameters for modeling ($\teff$, $L$, [Fe/H]), resulting stellar fundamental parameters ($\log$\,g, $M$, $Ro$) from \texttt{kiauhoku}, and the rotation periods from \citet{2021ApJS..255...17S}.}
\begin{tabular}{ccccccccc}
  \hline
  \hline
KIC & $T_{\rm eff}$ (K) & $\log L/L_{\odot}$ & [Fe/H] (dex)& $\log\,g$ (dex)&  $M$ (M$_\odot$)&  $Ro$ & $\prot$ &  flag$_{\rm spec}$ \\
\hline
\hline
893033 &  4707\,$\pm$\,  84&   -0.715\,$\pm$\,   0.034&-0.240\,$\pm$\,0.110& 
4.331$^{+0.029}_{-0.026}$& 0.89$^{+0.03}_{-0.02}$& 0.528&   25.9\,$\pm$\,   1.9
& 3\\
893209 &  6051\,$\pm$\, 108&    0.586\,$\pm$\,   0.036& 0.030\,$\pm$\,0.140& 
4.474$^{+0.053}_{-0.052}$& 0.94$^{+0.08}_{-0.07}$& 0.354&    4.6\,$\pm$\,   0.4
& 3\\
893286 &  5297\,$\pm$\, 100&   -0.302\,$\pm$\,   0.039&-0.040\,$\pm$\,0.140& 
4.235$^{+0.037}_{-0.050}$& 1.00$^{+0.04}_{-0.05}$& 0.658&   24.5\,$\pm$\,   2.2
& 3\\
893383 &  5680\,$\pm$\, 103&   -0.152\,$\pm$\,   0.034&-0.170\,$\pm$\,0.130& 
4.409$^{+0.032}_{-0.037}$& 0.89$^{+0.04}_{-0.04}$& 0.758&   21.2\,$\pm$\,   2.6
& 3\\
893559 &  5095\,$\pm$\,  91&   -0.417\,$\pm$\,   0.044&-0.040\,$\pm$\,0.130& 
3.659$^{+0.035}_{-0.045}$& 1.91$^{+0.04}_{-0.04}$& 0.350&   14.7\,$\pm$\,   1.1
& 3\\
1026287 &  4800\,$\pm$\, 233&   -0.524\,$\pm$\,   0.032&-0.130\,$\pm$\,0.220& 
4.587$^{+0.034}_{-0.039}$& 0.75$^{+0.04}_{-0.04}$& 0.606&   27.8\,$\pm$\,   5.2
& 3\\
1026838 &  5952\,$\pm$\,  75&    0.177\,$\pm$\,   0.030& 0.160\,$\pm$\,0.100& 
4.622$^{+0.032}_{-0.034}$& 0.68$^{+0.04}_{-0.04}$& 0.834&   15.6\,$\pm$\,   1.9
& 3\\
1027330 &  4886\,$\pm$\,  81&   -0.459\,$\pm$\,   0.035& 0.160\,$\pm$\,0.120& 
4.037$^{+0.038}_{-0.059}$& 1.23$^{+0.04}_{-0.04}$& 0.639&   31.3\,$\pm$\,   2.6
& 3\\
1027536 &  6017\,$\pm$\,  75&    0.242\,$\pm$\,   0.029& 0.150\,$\pm$\,0.100& 
4.530$^{+0.033}_{-0.041}$& 0.86$^{+0.04}_{-0.04}$& 1.026&   16.4\,$\pm$\,   4.4
& 3\\
1027740 &  5347\,$\pm$\,  94&   -0.201\,$\pm$\,   0.027& 0.030\,$\pm$\,0.140& 
4.513$^{+0.040}_{-0.067}$& 0.91$^{+0.05}_{-0.06}$& 0.735&   26.8\,$\pm$\,   2.7
& 3\\
1027900 &  5893\,$\pm$\, 125&    0.101\,$\pm$\,   0.027&-0.280\,$\pm$\,0.100& 
3.904$^{+0.049}_{-0.088}$& 1.25$^{+0.05}_{-0.07}$& 0.944&   18.8\,$\pm$\,   3.4
& 3\\
...\\
  \hline
 \end{tabular}
 \flushleft {  Notes.} flag$_{\rm spec}$ corresponds to the origin of the atmospheric parameters (0:CFOP, 1:APOGEE, 2:LAMOST, 3:B20). The full table is available online in a machine readable format.
   \label{tab:stellarparam}
\end{table*}

%





\appendix

\section{Comparison of convective overturn timescales}\label{app:tauc_comp}

To compute the convective overturn timescale, some semi-empirical relations have been derived such as the ones from \citet{1984ApJ...279..763N,2011ApJ...741...54C,2011ApJ...743...48W}  where a dependence with color (B-V or V-K) was established.  More recently, \citet{2021A&A...652L...2C} used the Legacy sample of {\it Kepler} targets with detection of solar-like oscillations in order to calibrate the Legacy Rossby number. Indeed those stars had enough signal-to-noise ratio that a precise asteroseismic modeling had been performed, allowing to have detailed information on the deeper layers of the stars.
We can also directly use the models where we look for the best fit to the observables, as explained in Section~\ref{sec:model}. From that model, we have access to the convective overturn timescale.
Finally, we can go back to 3D numerical simulations of rotating stars, where a few works are looking for scaling relations to estimate the Rossby number, and can deduce the convective overturn timescale. We take the {\it fluid} Rossby that compares the the advection term and the Coriolis force in the Navier–Stokes equation \citep{2022ApJ...926...21B,2022A&A...667A..50N}. 

In addition to the way of computing the convective overturn timescale, it is also important to keep in mind where in the star it is computed. In general, it is taken at one or half a pressure scale height above the base of the convection zone but it is not always the case. In the case of a sample with different Rossby definitions/convective depth, the comparison is still possible as long as each Rossby value is normalized to solar value of its definition, but this latter can sometime be challenging to constrain \citep[see for instance,][]{2022A&A...667A..50N}.

\begin{figure*}[!h]
    \centering
    \includegraphics[width=8cm]{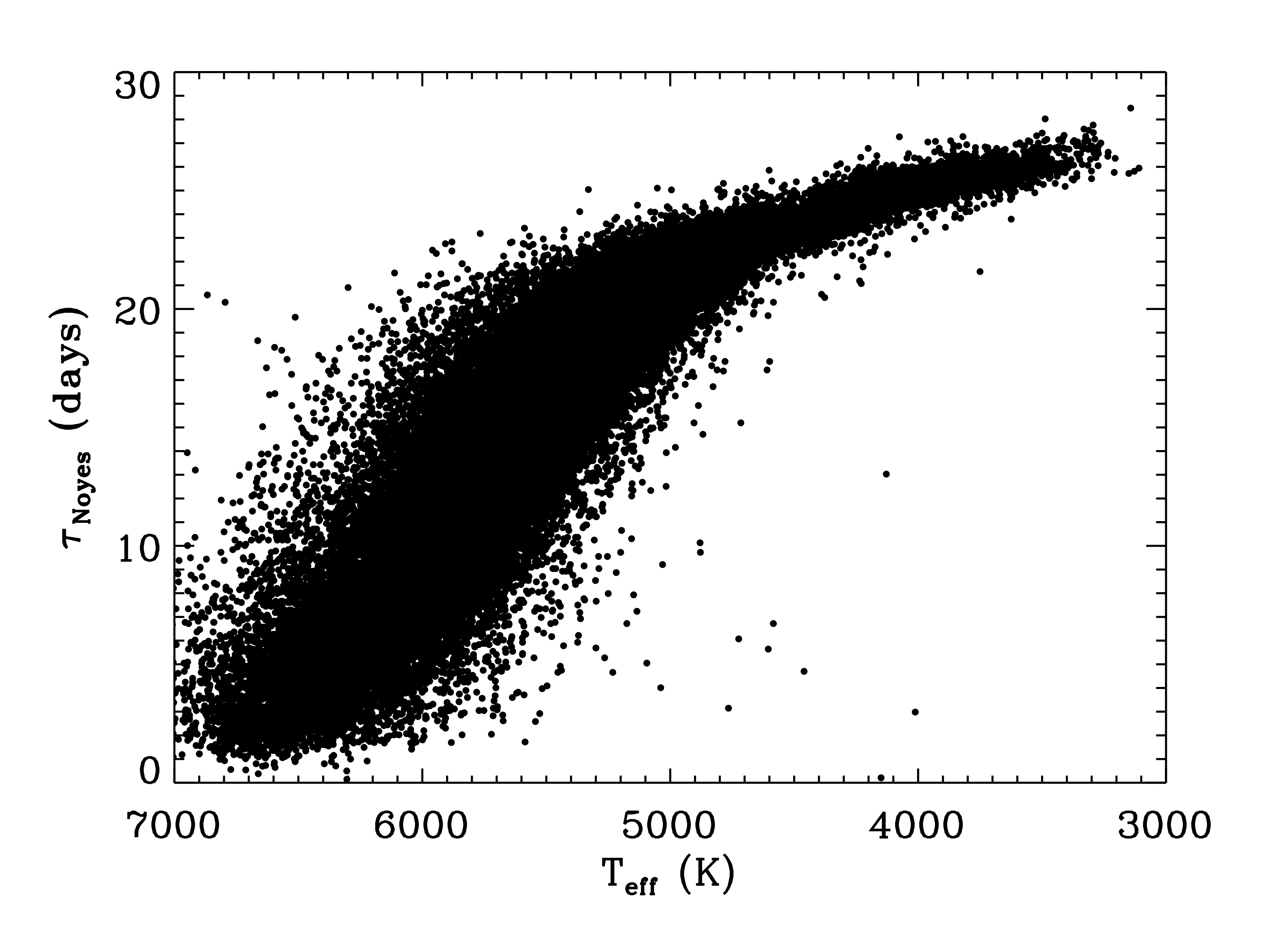}
    \includegraphics[width=8cm]{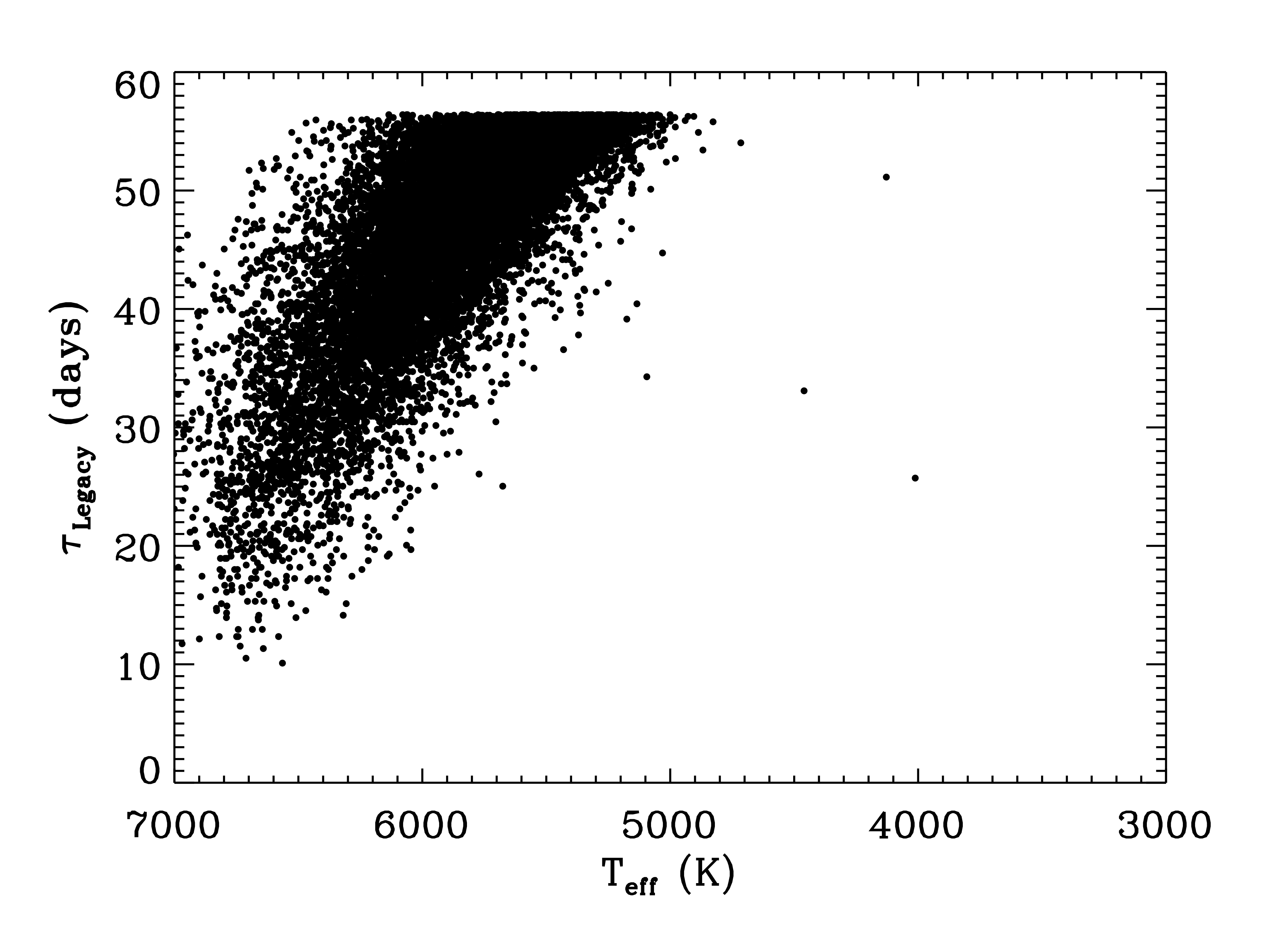}\\
    \includegraphics[width=8cm]{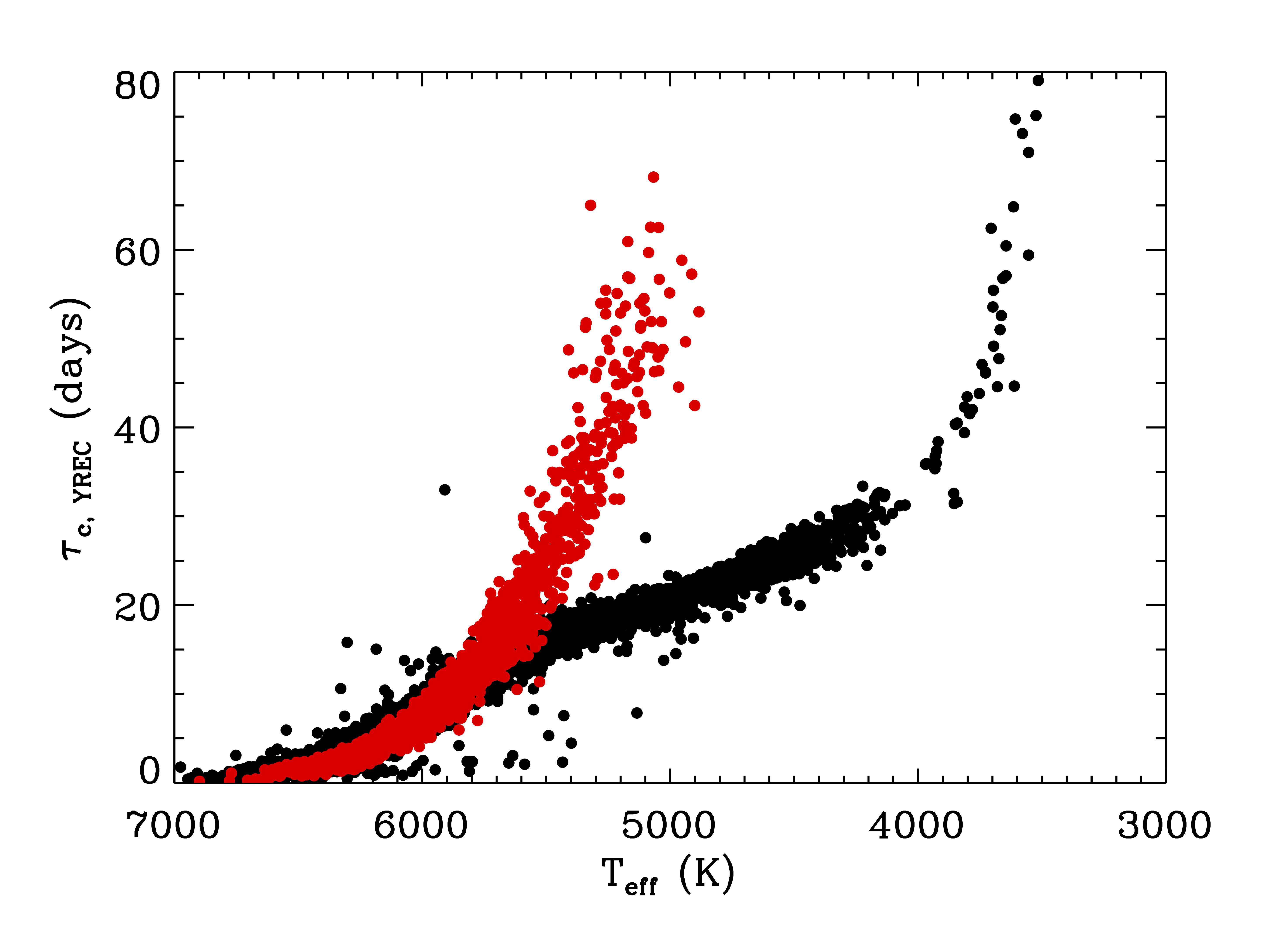}
    \includegraphics[width=8cm]{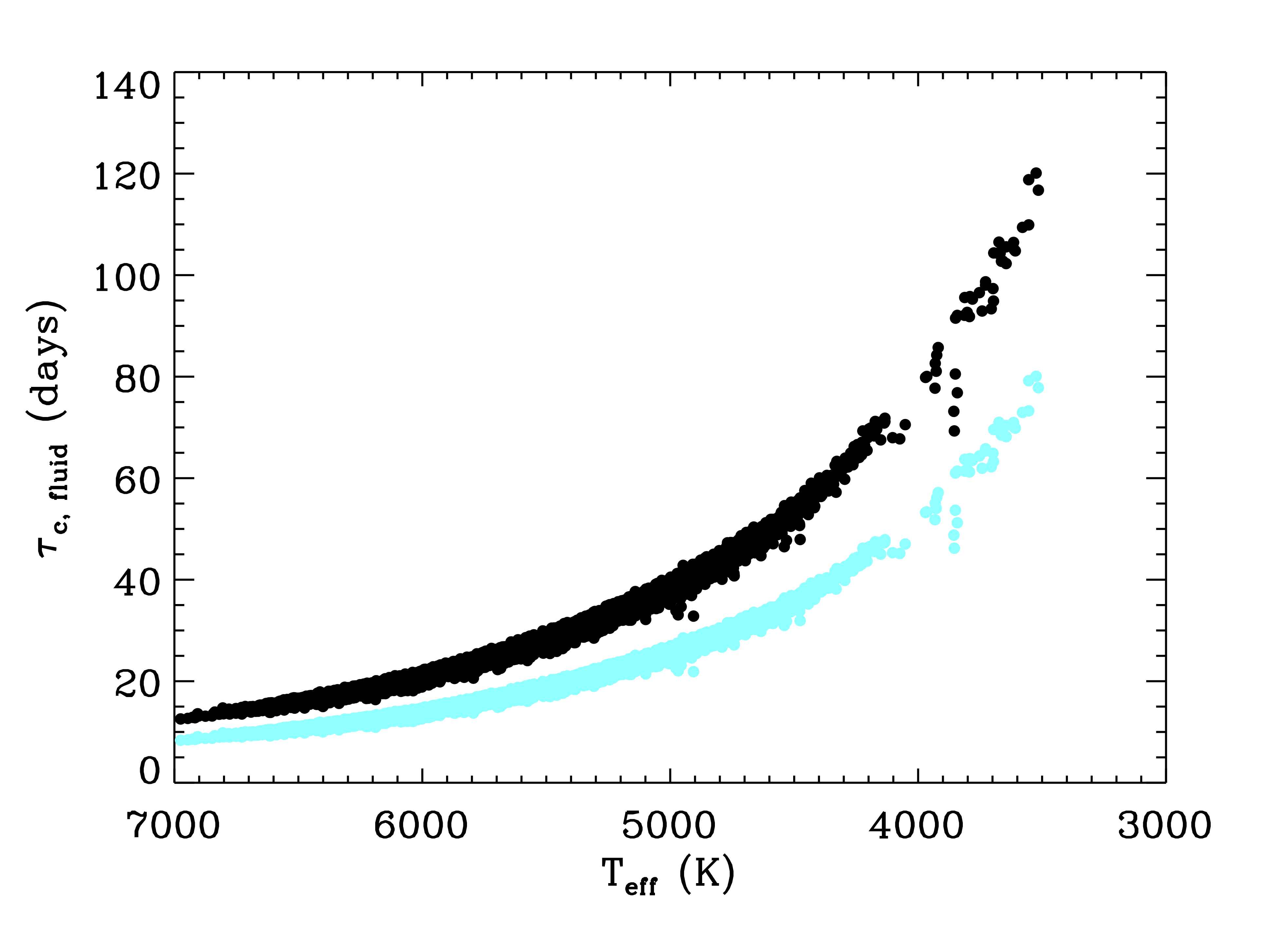}
    \caption{Convective overturn timescale computed with the \citet{1984ApJ...279..763N} relation (top left panel), the Legacy calibration (top right panel), from the best-fit model with YREC (bottom left panel) where subgiants are color-coded in red, and from the fluid computation with simulations (bottom right panel) with two values for the solar Rossby (0.6 in black and 0.9 in cyan) and converted to stellar $Ro$ as explained in Section~\ref{app:tauc_comp}. }
    \label{tauc_comp}
\end{figure*}

In Figure~\ref{tauc_comp}, we show the convective overturn timescale computed by different methods as a function of effective temperature. We clearly see that the relationships are quite different between the different computations. The semi-empirical formula from \citet{1984ApJ...279..763N} was obtained for a sample of targets with chromospheric observations at the Mount Wilson Observatory \citep{1978ApJ...226..379W}. First they used the convective overturn timescale computed from models of the convection by \citet{1979ApJ...231..284G} and combined them with observed rotation periods and the magnetic proxy $R'_{HK}$, finding a tight correlation when using models with the mixing length parameter $\alpha$=2 and fitted a polynomial function between $P_{\rm rot}/\tau_c$ and $R'_{HK}$. After subtracting $P_{\rm rot}$ to the fit, which is $\tau_c$, and representing the values as a function of B-V, they performed a cubic fit leading to the $\tau_c$-(B-V) relation. Note that this calibration was based on the Mount Wilson targets, which were mostly F, G, and K dwarfs. Given that our sample of {\it Kepler} stars also contains late-type stars and subgiants, the relation is not valid in all the domains. We can see in the top left panel of Figure~\ref{tauc_comp} showing $\tau_{\rm Noyes}$ as a function of the effective temperature that for stars cooler than 5,000K the value reaches a kind of plateau, which is not coherent with the fact that those stars should have a deep convective zone, if not fully convective. Similar calibration was also done by \citet{2011ApJ...743...48W} using stars with X-rays observations. 

For the $\tau_c$ calibrated with the {\it Kepler} Legacy sample \citep{2017ApJ...835..173S}, \citet{2021A&A...652L...2C} computed it as the ratio of the depth of the convective zone and the average convective velocity that they relate to the stellar parameters such as mass, radius and luminosity. They then applied it to the stars with the best seismic analysis \citep{2017ApJ...835..173S} where individual modes were fitted \citep{2017ApJ...835..172L}, allowing to have precise information on the stellar interiors, such as the depth of the convective zone and the stellar parameters. With their definition, $\tau_c$ is computed at the base of the convective zone. After performing a polynomial fit of the theoretical convective overturn timescale and the B-V color, they derived a relation between the B-V color and $\tauc$. As the seismic sample had only a small range of B-V values, this implies that the relation is only valid for B-V between 0.38 and 0.75, i.e. 7000\,K\,$>\teff >$\,5000\,K, as shown in the top right panel of Figure~\ref{tauc_comp}.

From the best-fit model computed as explained in Section~\ref{sec:model}, we compute the convective overturn timescale one pressure scale height above the base of the convective zone, as done traditionally. The model $\tau_c$ is shown in the bottom left panel of Figure~\ref{tauc_comp}. Since this calculation is based on models that take into account the evolution of stars, we can see different features that were not in the previous estimations of the convective overturn timescales. For instance, for the cool stars, $\tau_c$ increases with a sharp slope due to the deeper convection zone in those stars. We can also see a departure of another ``branch'' around 6000\,K, corresponding to the subgiants where the convection zone starts to deepen during the more evolved state of the stars. 

Finally the bottom right panel of Figure~\ref{tauc_comp} shows the convective overturn timescale computed from the combination of stellar models and 3D simulations of convection in the middle of the convection zone \citep{2022A&A...667A..50N}. In addition, we convert it to stellar $Ro$ ($\tau_c = P_{rot,*}/Ro_{\rm s} = P_{rot,*} / ( 2.26 \times Ro_{\rm f})$, where $Ro_{\rm f}$ is the fluid Rossby number defined in \citet{2022A&A...667A..50N}), following the result from the comparison between the fluid and stellar $Ro$ by \citet{2024A&A...684A.156N}.

We show it for 2 different values of solar references of 0.6 and 0.9 that correspond to the range of the solar fluid Rossby expected from simulations. The small scatter that we observe is related to the different metallicity of the stars.

The main differences comes from the location where the convective overturn timescale is computed. While the models from \texttt{kiauhoku} compute it at one pressure scale height above the base of the convective zone, the fluid $Ro_{\rm f}$ is computed in the middle of the convection zone, and the Legacy calibration that is at the base of the convection zone. To this we should add the differences on the stellar models used in each method and the way the calibration was done. Since the reference solar value is also different between the different methods and at different depths, this comparison shows the importance to normalize  $\tau_c$ to the solar value.

\section{Impact of the ratio between observation and cycle lengths on $\sph$}\label{app:Sun_cycle}


{  We investigate the impact of the duration of the photometric observations on the $\sph$ calculation depending on the magnetic cycle length, $P_{\rm cyc}$. To do so, we use the red and green channels of the VIRGO photometric observations spanning 27 years (more than 2.5 magnetic cycles). We computed $\sph$ for different observation lengths, $T_{\rm obs}$: 1 to 14 yrs and where we slide the box by 1 yr. For each length selected, we have 27 to 2 independent measurements possible. Given that the cycle length for the Sun is about 11 years, this can be extrapolated to any cycle length with the ratio: R\,=\,$T_{\rm obs}/P_{\rm cyc}$, which we vary from 9 to 127\%. 
If we assume that K dwarfs have longer cycles than G dwarfs \citep[e.g.][]{2017ApJ...845...79B}, K dwarfs would be represented by the cases with R\,$\sim$\,10-30\% and G dwarfs by the cases R\,$\sim$\,40-60\%. 

In the left panel of Figure~\ref{obs_cycle_VIRGO}, we show the variation of $\sph$ as a function of $T_{\rm obs}/P_{\rm cyc}$. On the right panel, we display the minimum and maximum $\sph$ values for each ratio during the first two magnetic cycles. We remind that except for the first ratio value, the points are not independent. We can see how the maximum $\sph$ value decreases with increasing ratio. This is also depicted in the right panel. 

\begin{figure*}[!h]
    \centering
    \includegraphics[width=8cm]{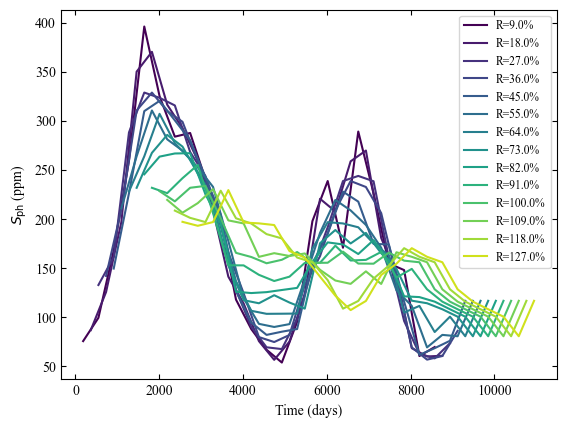}
    \includegraphics[width=8cm]{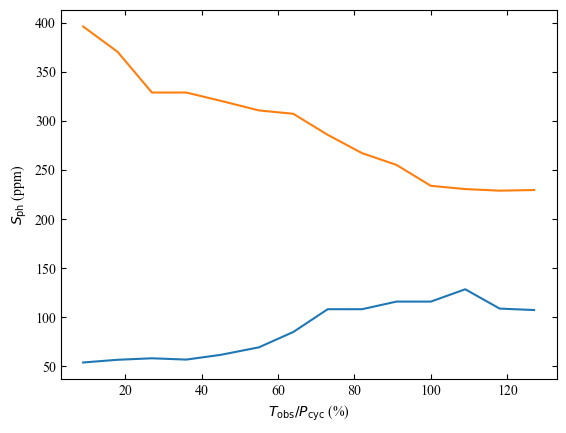}
    \caption{{  Left panel: variation of $\sph$ with the ratio R using the photometric observations of the Sun with VIRGO. Right panel: Minimum (blue) and maximum (orange) $\sph$ for the Sun as a function of the ratio in percentage.}}
\label{obs_cycle_VIRGO}
\end{figure*}

For a 1\,yr observation (or R\,$\sim$\,10\%), we obtain a broad range of $\sph$ values (from 50 to 400\,ppm) depending on when during the cycle the observation is done. When we increase the ratio, the $\sph$ range between minimum and maximum activity slowly decreases (from 100 to 230\,ppm). This agrees with K dwarfs having lower $\sph$ values (when observed at minimum activity) compared to G dwarfs. This also suggests that the K dwarfs should have a larger scatter in $\sph$. However this is assuming that all the stars (K and G) have similar levels of activity at minimum and maximum of the magnetic cycle.}

\bibliography{BIBLIO_sav,sample631}{}

\begin{thebibliography}{}
\expandafter\ifx\csname natexlab\endcsname\relax\def\natexlab#1{#1}\fi
\providecommand{\url}[1]{\href{#1}{#1}}
\providecommand{\dodoi}[1]{doi:~\href{http://doi.org/#1}{\nolinkurl{#1}}}
\providecommand{\doeprint}[1]{\href{http://ascl.net/#1}{\nolinkurl{http://ascl.net/#1}}}
\providecommand{\doarXiv}[1]{\href{https://arxiv.org/abs/#1}{\nolinkurl{https://arxiv.org/abs/#1}}}

\bibitem[{{Ahumada} {et~al.}(2020){Ahumada}, {Allende Prieto}, {Almeida},
  {Anders}, {Anderson}, {Andrews}, {Anguiano}, {Arcodia}, {Armengaud},
  {Aubert}, {Avila}, {Avila-Reese}, {Badenes}, {Balland}, {Barger},
  {Barrera-Ballesteros}, {Basu}, {Bautista}, {Beaton}, {Beers}, {Benavides},
  {Bender}, {Bernardi}, {Bershady}, {Beutler}, {Bidin}, {Bird}, {Bizyaev},
  {Blanc}, {Blanton}, {Boquien}, {Borissova}, {Bovy}, {Brandt}, {Brinkmann},
  {Brownstein}, {Bundy}, {Bureau}, {Burgasser}, {Burtin}, {Cano-D{\'\i}az},
  {Capasso}, {Cappellari}, {Carrera}, {Chabanier}, {Chaplin}, {Chapman},
  {Cherinka}, {Chiappini}, {Doohyun Choi}, {Chojnowski}, {Chung}, {Clerc},
  {Coffey}, {Comerford}, {Comparat}, {da Costa}, {Cousinou}, {Covey}, {Crane},
  {Cunha}, {da Silva Ilha}, {Dai}, {Damsted}, {Darling}, {Davidson}, {Davies},
  {Dawson}, {De}, {de la Macorra}, {De Lee}, {de Andrade Queiroz}, {Deconto
  Machado}, {de la Torre}, {Dell'Agli}, {du Mas des Bourboux},
  {Diamond-Stanic}, {Dillon}, {Donor}, {Drory}, {Duckworth}, {Dwelly},
  {Ebelke}, {Eftekharzadeh}, {Eigenbrot}, {Elsworth}, {Eracleous},
  {Erfanianfar}, {Escoffier}, {Fan}, {Farr}, {Fern{\'a}ndez-Trincado},
  {Feuillet}, {Finoguenov}, {Fofie}, {Fraser-McKelvie}, {Frinchaboy},
  {Fromenteau}, {Fu}, {Galbany}, {Garcia}, {Garc{\'\i}a-Hern{\'a}ndez}, {Garma
  Oehmichen}, {Ge}, {Geimba Maia}, {Geisler}, {Gelfand}, {Goddy},
  {Gonzalez-Perez}, {Grabowski}, {Green}, {Grier}, {Guo}, {Guy}, {Harding},
  {Hasselquist}, {Hawken}, {Hayes}, {Hearty}, {Hekker}, {Hogg}, {Holtzman},
  {Horta}, {Hou}, {Hsieh}, {Huber}, {Hunt}, {Ider Chitham}, {Imig}, {Jaber},
  {Jimenez Angel}, {Johnson}, {Jones}, {J{\"o}nsson}, {Jullo}, {Kim},
  {Kinemuchi}, {Kirkpatrick}, {Kite}, {Klaene}, {Kneib}, {Kollmeier}, {Kong},
  {Kounkel}, {Krishnarao}, {Lacerna}, {Lan}, {Lane}, {Law}, {Le Goff}, {Leung},
  {Lewis}, {Li}, {Lian}, {Lin}, {Long}, {Longa-Pe{\~n}a}, {Lundgren}, {Lyke},
  {Ted Mackereth}, {MacLeod}, {Majewski}, {Manchado}, {Maraston}, {Martini},
  {Masseron}, {Masters}, {Mathur}, {McDermid}, {Merloni}, {Merrifield},
  {M{\'e}sz{\'a}ros}, {Miglio}, {Minniti}, {Minsley}, {Miyaji}, {Mohammad},
  {Mosser}, {Mueller}, {Muna}, {Mu{\~n}oz-Guti{\'e}rrez}, {Myers}, {Nadathur},
  {Nair}, {Nandra}, {do Nascimento}, {Nevin}, {Newman}, {Nidever}, {Nitschelm},
  {Noterdaeme}, {O'Connell}, {Olmstead}, {Oravetz}, {Oravetz}, {Osorio},
  {Pace}, {Padilla}, {Palanque-Delabrouille}, {Palicio}, {Pan}, {Pan},
  {Parker}, {Paviot}, {Peirani}, {Pe{\~n}a Ram{\'r}ez}, {Penny}, {Percival},
  {Perez-Fournon}, {P{\'e}rez-R{\`a}fols}, {Petitjean}, {Pieri},
  {Pinsonneault}, {Poovelil}, {Povick}, {Prakash}, {Price-Whelan}, {Raddick},
  {Raichoor}, {Ray}, {Rembold}, {Rezaie}, {Riffel}, {Riffel}, {Rix}, {Robin},
  {Roman-Lopes}, {Rom{\'a}n-Z{\'u}{\~n}iga}, {Rose}, {Ross}, {Rossi},
  {Rowlands}, {Rubin}, {Salvato}, {S{\'a}nchez}, {S{\'a}nchez-Menguiano},
  {S{\'a}nchez-Gallego}, {Sayres}, {Schaefer}, {Schiavon}, {Schimoia},
  {Schlafly}, {Schlegel}, {Schneider}, {Schultheis}, {Schwope}, {Seo},
  {Serenelli}, {Shafieloo}, {Shamsi}, {Shao}, {Shen}, {Shetrone}, {Shirley},
  {Silva Aguirre}, {Simon}, {Skrutskie}, {Slosar}, {Smethurst}, {Sobeck},
  {Sodi}, {Souto}, {Stark}, {Stassun}, {Steinmetz}, {Stello}, {Stermer},
  {Storchi-Bergmann}, {Streblyanska}, {Stringfellow}, {Stutz}, {Su{\'a}rez},
  {Sun}, {Taghizadeh-Popp}, {Talbot}, {Tayar}, {Thakar}, {Theriault}, {Thomas},
  {Thomas}, {Tinker}, {Tojeiro}, {Toledo}, {Tremonti}, {Troup}, {Tuttle},
  {Unda-Sanzana}, {Valentini}, {Vargas-Gonz{\'a}lez}, {Vargas-Maga{\~n}a},
  {V{\'a}zquez-Mata}, {Vivek}, {Wake}, {Wang}, {Weaver}, {Weijmans}, {Wild},
  {Wilson}, {Wilson}, {Wolthuis}, {Wood-Vasey}, {Yan}, {Yang}, {Y{\`e}che},
  {Zamora}, {Zarrouk}, {Zasowski}, {Zhang}, {Zhao}, {Zhao}, {Zheng}, {Zheng},
  {Zhu}, \& {Zou}}]{2020ApJS..249....3A}
{Ahumada}, R., {Allende Prieto}, C., {Almeida}, A., {et~al.} 2020, \apjs, 249,
  3, \dodoi{10.3847/1538-4365/ab929e}

\bibitem[{{Amard} \& {Matt}(2020)}]{2020ApJ...889..108A}
{Amard}, L., \& {Matt}, S.~P. 2020, \apj, 889, 108,
  \dodoi{10.3847/1538-4357/ab6173}

\bibitem[{{Angus} {et~al.}(2020){Angus}, {Beane}, {Price-Whelan}, {Newton},
  {Curtis}, {Berger}, {van Saders}, {Kiman}, {Foreman-Mackey}, {Lu},
  {Anderson}, \& {Faherty}}]{2020AJ....160...90A}
{Angus}, R., {Beane}, A., {Price-Whelan}, A.~M., {et~al.} 2020, \aj, 160, 90,
  \dodoi{10.3847/1538-3881/ab91b2}

\bibitem[{{Augustson} {et~al.}(2013){Augustson}, {Brun}, \&
  {Toomre}}]{2013ApJ...777..153A}
{Augustson}, K.~C., {Brun}, A.~S., \& {Toomre}, J. 2013, \apj, 777, 153,
  \dodoi{10.1088/0004-637X/777/2/153}

\bibitem[{{Augustson} {et~al.}(2019){Augustson}, {Brun}, \&
  {Toomre}}]{2019ApJ...876...83A}
---. 2019, \apj, 876, 83, \dodoi{10.3847/1538-4357/ab14ea}

\bibitem[{{Bahcall} {et~al.}(2001){Bahcall}, {Pinsonneault}, \&
  {Basu}}]{2001ApJ...555..990B}
{Bahcall}, J.~N., {Pinsonneault}, M.~H., \& {Basu}, S. 2001, \apj, 555, 990,
  \dodoi{10.1086/321493}

\bibitem[{{Baliunas} {et~al.}(1995){Baliunas}, {Donahue}, {Soon}, {Horne},
  {Frazer}, {Woodard-Eklund}, {Bradford}, {Rao}, {Wilson}, {Zhang}, {Bennett},
  {Briggs}, {Carroll}, {Duncan}, {Figueroa}, {Lanning}, {Misch}, {Mueller},
  {Noyes}, {Poppe}, {Porter}, {Robinson}, {Russell}, {Shelton}, {Soyumer},
  {Vaughan}, \& {Whitney}}]{1995ApJ...438..269B}
{Baliunas}, S.~L., {Donahue}, R.~A., {Soon}, W.~H., {et~al.} 1995, \apj, 438,
  269, \dodoi{10.1086/175072}

\bibitem[{{Barnes}(2003)}]{2003ApJ...586L.145B}
{Barnes}, S.~A. 2003, \apjl, 586, L145, \dodoi{10.1086/374681}

\bibitem[{{Basri} \& {Nguyen}(2018)}]{2018ApJ...863..190B}
{Basri}, G., \& {Nguyen}, H.~T. 2018, \apj, 863, 190,
  \dodoi{10.3847/1538-4357/aad3b6}

\bibitem[{{Basri} {et~al.}(2013){Basri}, {Walkowicz}, \&
  {Reiners}}]{2013ApJ...769...37B}
{Basri}, G., {Walkowicz}, L.~M., \& {Reiners}, A. 2013, \apj, 769, 37,
  \dodoi{10.1088/0004-637X/769/1/37}

\bibitem[{{Basri} {et~al.}(2010){Basri}, {Walkowicz}, {Batalha}, {Gilliland},
  {Jenkins}, {Borucki}, {Koch}, {Caldwell}, {Dupree}, {Latham}, {Meibom},
  {Howell}, \& {Brown}}]{2010ApJ...713L.155B}
{Basri}, G., {Walkowicz}, L.~M., {Batalha}, N., {et~al.} 2010, \apjl, 713,
  L155, \dodoi{10.1088/2041-8205/713/2/L155}

\bibitem[{{Basri} {et~al.}(2011){Basri}, {Walkowicz}, {Batalha}, {Gilliland},
  {Jenkins}, {Borucki}, {Koch}, {Caldwell}, {Dupree}, {Latham}, {Marcy},
  {Meibom}, \& {Brown}}]{2011AJ....141...20B}
---. 2011, \aj, 141, 20, \dodoi{10.1088/0004-6256/141/1/20}

\bibitem[{{Bazot} {et~al.}(2018){Bazot}, {Nielsen}, {Mary},
  {Christensen-Dalsgaard}, {Benomar}, {Petit}, {Gizon}, {Sreenivasan}, \&
  {White}}]{2018A&A...619L...9B}
{Bazot}, M., {Nielsen}, M.~B., {Mary}, D., {et~al.} 2018, \aap, 619, L9,
  \dodoi{10.1051/0004-6361/201834251}

\bibitem[{{Beck} {et~al.}(2024){Beck}, {Grossmann}, {Steinwender}, {Schimak},
  {Muntean}, {Vrard}, {Patton}, {Merc}, {Mathur}, {Garcia}, {Pinsonneault},
  {Rowan}, {Gaulme}, {Allende Prieto}, {Arellano-C{\'o}rdova}, {Cao},
  {Corsaro}, {Creevey}, {Hambleton}, {Hanslmeier}, {Holl}, {Johnson}, {Mathis},
  {Godoy-Rivera}, {S{\'\i}mon-D{\'\i}az}, \& {Zinn}}]{2024A&A...682A...7B}
{Beck}, P.~G., {Grossmann}, D.~H., {Steinwender}, L., {et~al.} 2024, \aap, 682,
  A7, \dodoi{10.1051/0004-6361/202346810}

\bibitem[{{Benomar} {et~al.}(2023){Benomar}, {Takata}, {Bazot}, {Sekii},
  {Gizon}, \& {Lu}}]{2023A&A...680A..27B}
{Benomar}, O., {Takata}, M., {Bazot}, M., {et~al.} 2023, \aap, 680, A27,
  \dodoi{10.1051/0004-6361/202347095}

\bibitem[{{Berger} {et~al.}(2020){Berger}, {Huber}, {van Saders}, {Gaidos},
  {Tayar}, \& {Kraus}}]{2020AJ....159..280B}
{Berger}, T.~A., {Huber}, D., {van Saders}, J.~L., {et~al.} 2020, \aj, 159,
  280, \dodoi{10.3847/1538-3881/159/6/280}

\bibitem[{{Bonanno} \& {Corsaro}(2022)}]{2022ApJ...939L..26B}
{Bonanno}, A., \& {Corsaro}, E. 2022, \apjl, 939, L26,
  \dodoi{10.3847/2041-8213/ac9c05}

\bibitem[{{Borucki} {et~al.}(2010){Borucki}, {Koch}, {Basri}, {Batalha},
  {Brown}, {Caldwell}, {Caldwell}, {Christensen-Dalsgaard}, {Cochran},
  {DeVore}, {Dunham}, {Dupree}, {Gautier}, {Geary}, {Gilliland}, {Gould},
  {Howell}, {Jenkins}, {Kondo}, {Latham}, {Marcy}, {Meibom}, {Kjeldsen},
  {Lissauer}, {Monet}, {Morrison}, {Sasselov}, {Tarter}, {Boss}, {Brownlee},
  {Owen}, {Buzasi}, {Charbonneau}, {Doyle}, {Fortney}, {Ford}, {Holman},
  {Seager}, {Steffen}, {Welsh}, {Rowe}, {Anderson}, {Buchhave}, {Ciardi},
  {Walkowicz}, {Sherry}, {Horch}, {Isaacson}, {Everett}, {Fischer}, {Torres},
  {Johnson}, {Endl}, {MacQueen}, {Bryson}, {Dotson}, {Haas}, {Kolodziejczak},
  {Van Cleve}, {Chandrasekaran}, {Twicken}, {Quintana}, {Clarke}, {Allen},
  {Li}, {Wu}, {Tenenbaum}, {Verner}, {Bruhweiler}, {Barnes}, \&
  {Prsa}}]{2010Sci...327..977B}
{Borucki}, W.~J., {Koch}, D., {Basri}, G., {et~al.} 2010, Science, 327, 977,
  \dodoi{10.1126/science.1185402}

\bibitem[{{Brandenburg} \& {Giampapa}(2018)}]{2018ApJ...855L..22B}
{Brandenburg}, A., \& {Giampapa}, M.~S. 2018, \apjl, 855, L22,
  \dodoi{10.3847/2041-8213/aab20a}

\bibitem[{{Brandenburg} {et~al.}(2017){Brandenburg}, {Mathur}, \&
  {Metcalfe}}]{2017ApJ...845...79B}
{Brandenburg}, A., {Mathur}, S., \& {Metcalfe}, T.~S. 2017, \apj, 845, 79,
  \dodoi{10.3847/1538-4357/aa7cfa}

\bibitem[{{Breton} {et~al.}(2021){Breton}, {Santos}, {Bugnet}, {Mathur},
  {Garc{\'\i}a}, \& {Pall{\'e}}}]{2021A&A...647A.125B}
{Breton}, S.~N., {Santos}, A.~R.~G., {Bugnet}, L., {et~al.} 2021, \aap, 647,
  A125, \dodoi{10.1051/0004-6361/202039947}

\bibitem[{{Brown} {et~al.}(2011{\natexlab{a}}){Brown}, {Miesch}, {Browning},
  {Brun}, \& {Toomre}}]{2011ApJ...731...69B}
{Brown}, B.~P., {Miesch}, M.~S., {Browning}, M.~K., {Brun}, A.~S., \& {Toomre},
  J. 2011{\natexlab{a}}, \apj, 731, 69, \dodoi{10.1088/0004-637X/731/1/69}

\bibitem[{{Brown} {et~al.}(2021){Brown}, {Garc{\'\i}a}, {Mathur}, {Metcalfe},
  \& {Santos}}]{2021ApJ...916...66B}
{Brown}, T.~M., {Garc{\'\i}a}, R.~A., {Mathur}, S., {Metcalfe}, T.~S., \&
  {Santos}, {\^A}. R.~G. 2021, \apj, 916, 66, \dodoi{10.3847/1538-4357/ac0635}

\bibitem[{{Brown} {et~al.}(2011{\natexlab{b}}){Brown}, {Latham}, {Everett}, \&
  {Esquerdo}}]{2011AJ....142..112B}
{Brown}, T.~M., {Latham}, D.~W., {Everett}, M.~E., \& {Esquerdo}, G.~A.
  2011{\natexlab{b}}, \aj, 142, 112, \dodoi{10.1088/0004-6256/142/4/112}

\bibitem[{{Brun} \& {Browning}(2017)}]{2017LRSP...14....4B}
{Brun}, A.~S., \& {Browning}, M.~K. 2017, Living Reviews in Solar Physics, 14,
  4, \dodoi{10.1007/s41116-017-0007-8}

\bibitem[{{Brun} {et~al.}(2004){Brun}, {Miesch}, \&
  {Toomre}}]{2004ApJ...614.1073B}
{Brun}, A.~S., {Miesch}, M.~S., \& {Toomre}, J. 2004, \apj, 614, 1073,
  \dodoi{10.1086/423835}

\bibitem[{{Brun} {et~al.}(2022){Brun}, {Strugarek}, {Noraz}, {Perri}, {Varela},
  {Augustson}, {Charbonneau}, \& {Toomre}}]{2022ApJ...926...21B}
{Brun}, A.~S., {Strugarek}, A., {Noraz}, Q., {et~al.} 2022, \apj, 926, 21,
  \dodoi{10.3847/1538-4357/ac469b}

\bibitem[{{Brun} {et~al.}(2017){Brun}, {Strugarek}, {Varela}, {Matt},
  {Augustson}, {Emeriau}, {DoCao}, {Brown}, \& {Toomre}}]{2017ApJ...836..192B}
{Brun}, A.~S., {Strugarek}, A., {Varela}, J., {et~al.} 2017, \apj, 836, 192,
  \dodoi{10.3847/1538-4357/aa5c40}

\bibitem[{{Cao} \& {Pinsonneault}(2022)}]{2022MNRAS.517.2165C}
{Cao}, L., \& {Pinsonneault}, M.~H. 2022, \mnras, 517, 2165,
  \dodoi{10.1093/mnras/stac2706}

\bibitem[{{Cao} {et~al.}(2023){Cao}, {Pinsonneault}, \& {van
  Saders}}]{2023ApJ...951L..49C}
{Cao}, L., {Pinsonneault}, M.~H., \& {van Saders}, J.~L. 2023, \apjl, 951, L49,
  \dodoi{10.3847/2041-8213/acd780}

\bibitem[{{Ceillier} {et~al.}(2017){Ceillier}, {Tayar}, {Mathur}, {Salabert},
  {Garc{\'{\i}}a}, {Stello}, {Pinsonneault}, {van Saders}, {Beck}, \&
  {Bloemen}}]{2017A&A...605A.111C}
{Ceillier}, T., {Tayar}, J., {Mathur}, S., {et~al.} 2017, \aap, 605, A111,
  \dodoi{10.1051/0004-6361/201629884}

\bibitem[{{Charbonnel} {et~al.}(2017){Charbonnel}, {Decressin}, {Lagarde},
  {Gallet}, {Palacios}, {Auri{\`e}re}, {Konstantinova-Antova}, {Mathis},
  {Anderson}, \& {Dintrans}}]{2017A&A...605A.102C}
{Charbonnel}, C., {Decressin}, T., {Lagarde}, N., {et~al.} 2017, \aap, 605,
  A102, \dodoi{10.1051/0004-6361/201526724}

\bibitem[{{Claytor} {et~al.}(2020){Claytor}, {van Saders}, {Santos},
  {Garc{\'\i}a}, {Mathur}, {Tayar}, {Pinsonneault}, \&
  {Shetrone}}]{Claytor2020}
{Claytor}, Z.~R., {van Saders}, J.~L., {Santos}, {\^A}. R.~G., {et~al.} 2020,
  \apj, 888, 43, \dodoi{10.3847/1538-4357/ab5c24}

\bibitem[{{Corsaro} {et~al.}(2021){Corsaro}, {Bonanno}, {Mathur},
  {Garc{\'\i}a}, {Santos}, {Breton}, \& {Khalatyan}}]{2021A&A...652L...2C}
{Corsaro}, E., {Bonanno}, A., {Mathur}, S., {et~al.} 2021, \aap, 652, L2,
  \dodoi{10.1051/0004-6361/202141395}

\bibitem[{{Cranmer} \& {Saar}(2011)}]{2011ApJ...741...54C}
{Cranmer}, S.~R., \& {Saar}, S.~H. 2011, \apj, 741, 54,
  \dodoi{10.1088/0004-637X/741/1/54}

\bibitem[{{David} {et~al.}(2022){David}, {Angus}, {Curtis}, {van Saders},
  {Colman}, {Contardo}, {Lu}, \& {Zinn}}]{2022ApJ...933..114D}
{David}, T.~J., {Angus}, R., {Curtis}, J.~L., {et~al.} 2022, \apj, 933, 114,
  \dodoi{10.3847/1538-4357/ac6dd3}

\bibitem[{{Demarque} {et~al.}(2008){Demarque}, {Guenther}, {Li}, {Mazumdar}, \&
  {Straka}}]{Demarque2008}
{Demarque}, P., {Guenther}, D.~B., {Li}, L.~H., {Mazumdar}, A., \& {Straka},
  C.~W. 2008, \apss, 316, 31, \dodoi{10.1007/s10509-007-9698-y}

\bibitem[{{do Nascimento} {et~al.}(2020){do Nascimento}, {de Almeida},
  {Velloso}, {Anthony}, {Barnes}, {Saar}, {Meibom}, {da Costa}, {Castro},
  {Galarza}, {Lorenzo-Oliveira}, {Beck}, \&
  {Mel{\'e}ndez}}]{2020ApJ...898..173D}
{do Nascimento}, J.~D., J., {de Almeida}, L., {Velloso}, E.~N., {et~al.} 2020,
  \apj, 898, 173, \dodoi{10.3847/1538-4357/ab9c16}

\bibitem[{{Donati} {et~al.}(2023){Donati}, {Lehmann}, {Cristofari},
  {Fouqu{\'e}}, {Moutou}, {Charpentier}, {Ould-Elhkim}, {Carmona}, {Delfosse},
  {Artigau}, {Alencar}, {Cadieux}, {Arnold}, {Petit}, {Morin}, {Forveille},
  {Cloutier}, {Doyon}, {H{\'e}brard}, \& {SLS
  Collaboration}}]{2023MNRAS.525.2015D}
{Donati}, J.~F., {Lehmann}, L.~T., {Cristofari}, P.~I., {et~al.} 2023, \mnras,
  525, 2015, \dodoi{10.1093/mnras/stad2301}

\bibitem[{{Dotter}(2016)}]{2016ApJS..222....8D}
{Dotter}, A. 2016, \apjs, 222, 8, \dodoi{10.3847/0067-0049/222/1/8}

\bibitem[{{Dotter} {et~al.}(2008){Dotter}, {Chaboyer}, {Jevremovi{\'c}},
  {Kostov}, {Baron}, \& {Ferguson}}]{2008ApJS..178...89D}
{Dotter}, A., {Chaboyer}, B., {Jevremovi{\'c}}, D., {et~al.} 2008, \apjs, 178,
  89, \dodoi{10.1086/589654}

\bibitem[{{Duncan} {et~al.}(1991){Duncan}, {Vaughan}, {Wilson}, {Preston},
  {Frazer}, {Lanning}, {Misch}, {Mueller}, {Soyumer}, {Woodard}, {Baliunas},
  {Noyes}, {Hartmann}, {Porter}, {Zwaan}, {Middelkoop}, {Rutten}, \&
  {Mihalas}}]{1991ApJS...76..383D}
{Duncan}, D.~K., {Vaughan}, A.~H., {Wilson}, O.~C., {et~al.} 1991, \apjs, 76,
  383, \dodoi{10.1086/191572}

\bibitem[{{Finley} {et~al.}(2024){Finley}, {Brun}, {Strugarek}, \&
  {Cameron}}]{2024A&A...684A..92F}
{Finley}, A.~J., {Brun}, A.~S., {Strugarek}, A., \& {Cameron}, R. 2024, \aap,
  684, A92, \dodoi{10.1051/0004-6361/202347862}

\bibitem[{{Finley} \& {Matt}(2018)}]{2018ApJ...854...78F}
{Finley}, A.~J., \& {Matt}, S.~P. 2018, \apj, 854, 78,
  \dodoi{10.3847/1538-4357/aaaab5}

\bibitem[{{Fr{\"o}hlich} {et~al.}(1995){Fr{\"o}hlich}, {Romero}, {Roth},
  {Wehrli}, {Andersen}, {Appourchaux}, {Domingo}, {Telljohann}, {Berthomieu},
  {Delache}, {Provost}, {Toutain}, {Crommelynck}, {Chevalier}, {Fichot},
  {D{\"a}ppen}, {Gough}, {Hoeksema}, {Jim{\'e}nez}, {G{\'o}mez}, {Herreros},
  {Cort{\'e}s}, {Jones}, {Pap}, \& {Willson}}]{1995SoPh..162..101F}
{Fr{\"o}hlich}, C., {Romero}, J., {Roth}, H., {et~al.} 1995, \solphys, 162,
  101, \dodoi{10.1007/BF00733428}

\bibitem[{{Furlan} {et~al.}(2018){Furlan}, {Ciardi}, {Cochran}, {Everett},
  {Latham}, {Marcy}, {Buchhave}, {Endl}, {Isaacson}, {Petigura}, {Gautier},
  {Huber}, {Bieryla}, {Borucki}, {Brugamyer}, {Caldwell}, {Cochran}, {Howard},
  {Howell}, {Johnson}, {MacQueen}, {Quinn}, {Robertson}, {Mathur}, \&
  {Batalha}}]{2018ApJ...861..149F}
{Furlan}, E., {Ciardi}, D.~R., {Cochran}, W.~D., {et~al.} 2018, \apj, 861, 149,
  \dodoi{10.3847/1538-4357/aaca34}

\bibitem[{{Gaia Collaboration} {et~al.}(2018){Gaia Collaboration}, {Brown},
  {Vallenari}, {Prusti}, {de Bruijne}, {Babusiaux}, {Bailer-Jones}, {Biermann},
  {Evans}, {Eyer}, {Jansen}, {Jordi}, {Klioner}, {Lammers}, {Lindegren},
  {Luri}, {Mignard}, {Panem}, {Pourbaix}, {Randich}, {Sartoretti}, {Siddiqui},
  {Soubiran}, {van Leeuwen}, {Walton}, {Arenou}, {Bastian}, {Cropper},
  {Drimmel}, {Katz}, {Lattanzi}, {Bakker}, {Cacciari}, {Casta{\~n}eda},
  {Chaoul}, {Cheek}, {De Angeli}, {Fabricius}, {Guerra}, {Holl}, {Masana},
  {Messineo}, {Mowlavi}, {Nienartowicz}, {Panuzzo}, {Portell}, {Riello},
  {Seabroke}, {Tanga}, {Th{\'e}venin}, {Gracia-Abril}, {Comoretto},
  {Garcia-Reinaldos}, {Teyssier}, {Altmann}, {Andrae}, {Audard},
  {Bellas-Velidis}, {Benson}, {Berthier}, {Blomme}, {Burgess}, {Busso},
  {Carry}, {Cellino}, {Clementini}, {Clotet}, {Creevey}, {Davidson}, {De
  Ridder}, {Delchambre}, {Dell'Oro}, {Ducourant},
  {Fern{\'a}ndez-Hern{\'a}ndez}, {Fouesneau}, {Fr{\'e}mat}, {Galluccio},
  {Garc{\'\i}a-Torres}, {Gonz{\'a}lez-N{\'u}{\~n}ez}, {Gonz{\'a}lez-Vidal},
  {Gosset}, {Guy}, {Halbwachs}, {Hambly}, {Harrison}, {Hern{\'a}ndez},
  {Hestroffer}, {Hodgkin}, {Hutton}, {Jasniewicz}, {Jean-Antoine-Piccolo},
  {Jordan}, {Korn}, {Krone-Martins}, {Lanzafame}, {Lebzelter}, {L{\"o}ffler},
  {Manteiga}, {Marrese}, {Mart{\'\i}n-Fleitas}, {Moitinho}, {Mora}, {Muinonen},
  {Osinde}, {Pancino}, {Pauwels}, {Petit}, {Recio-Blanco}, {Richards},
  {Rimoldini}, {Robin}, {Sarro}, {Siopis}, {Smith}, {Sozzetti}, {S{\"u}veges},
  {Torra}, {van Reeven}, {Abbas}, {Abreu Aramburu}, {Accart}, {Aerts},
  {Altavilla}, {{\'A}lvarez}, {Alvarez}, {Alves}, {Anderson}, {Andrei},
  {Anglada Varela}, {Antiche}, {Antoja}, {Arcay}, {Astraatmadja}, {Bach},
  {Baker}, {Balaguer-N{\'u}{\~n}ez}, {Balm}, {Barache}, {Barata}, {Barbato},
  {Barblan}, {Barklem}, {Barrado}, {Barros}, {Barstow}, {Bartholom{\'e}
  Mu{\~n}oz}, {Bassilana}, {Becciani}, {Bellazzini}, {Berihuete}, {Bertone},
  {Bianchi}, {Bienaym{\'e}}, {Blanco-Cuaresma}, {Boch}, {Boeche}, {Bombrun},
  {Borrachero}, {Bossini}, {Bouquillon}, {Bourda}, {Bragaglia}, {Bramante},
  {Breddels}, {Bressan}, {Brouillet}, {Br{\"u}semeister}, {Brugaletta},
  {Bucciarelli}, {Burlacu}, {Busonero}, {Butkevich}, {Buzzi}, {Caffau},
  {Cancelliere}, {Cannizzaro}, {Cantat-Gaudin}, {Carballo}, {Carlucci},
  {Carrasco}, {Casamiquela}, {Castellani}, {Castro-Ginard}, {Charlot},
  {Chemin}, {Chiavassa}, {Cocozza}, {Costigan}, {Cowell}, {Crifo}, {Crosta},
  {Crowley}, {Cuypers}, {Dafonte}, {Damerdji}, {Dapergolas}, {David}, {David},
  {de Laverny}, {De Luise}, {De March}, {de Martino}, {de Souza}, {de Torres},
  {Debosscher}, {del Pozo}, {Delbo}, {Delgado}, {Delgado}, {Di Matteo},
  {Diakite}, {Diener}, {Distefano}, {Dolding}, {Drazinos}, {Dur{\'a}n},
  {Edvardsson}, {Enke}, {Eriksson}, {Esquej}, {Eynard Bontemps}, {Fabre},
  {Fabrizio}, {Faigler}, {Falc{\~a}o}, {Farr{\`a}s Casas}, {Federici},
  {Fedorets}, {Fernique}, {Figueras}, {Filippi}, {Findeisen}, {Fonti},
  {Fraile}, {Fraser}, {Fr{\'e}zouls}, {Gai}, {Galleti}, {Garabato},
  {Garc{\'\i}a-Sedano}, {Garofalo}, {Garralda}, {Gavel}, {Gavras}, {Gerssen},
  {Geyer}, {Giacobbe}, {Gilmore}, {Girona}, {Giuffrida}, {Glass}, {Gomes},
  {Granvik}, {Gueguen}, {Guerrier}, {Guiraud}, {Guti{\'e}rrez-S{\'a}nchez},
  {Haigron}, {Hatzidimitriou}, {Hauser}, {Haywood}, {Heiter}, {Helmi}, {Heu},
  {Hilger}, {Hobbs}, {Hofmann}, {Holland}, {Huckle}, {Hypki}, {Icardi},
  {Jan{\ss}en}, {Jevardat de Fombelle}, {Jonker}, {Juh{\'a}sz}, {Julbe},
  {Karampelas}, {Kewley}, {Klar}, {Kochoska}, {Kohley}, {Kolenberg},
  {Kontizas}, {Kontizas}, {Koposov}, {Kordopatis}, {Kostrzewa-Rutkowska},
  {Koubsky}, {Lambert}, {Lanza}, {Lasne}, {Lavigne}, {Le Fustec}, {Le
  Poncin-Lafitte}, {Lebreton}, {Leccia}, {Leclerc}, {Lecoeur-Taibi},
  {Lenhardt}, {Leroux}, {Liao}, {Licata}, {Lindstr{\o}m}, {Lister}, {Livanou},
  {Lobel}, {L{\'o}pez}, {Managau}, {Mann}, {Mantelet}, {Marchal}, {Marchant},
  {Marconi}, {Marinoni}, {Marschalk{\'o}}, {Marshall}, {Martino}, {Marton},
  {Mary}, {Massari}, {Matijevi{\v{c}}}, {Mazeh}, {McMillan}, {Messina},
  {Michalik}, {Millar}, {Molina}, {Molinaro}, {Moln{\'a}r}, {Montegriffo},
  {Mor}, {Morbidelli}, {Morel}, {Morris}, {Mulone}, {Muraveva}, {Musella},
  {Nelemans}, {Nicastro}, {Noval}, {O'Mullane}, {Ord{\'e}novic},
  {Ord{\'o}{\~n}ez-Blanco}, {Osborne}, {Pagani}, {Pagano}, {Pailler},
  {Palacin}, {Palaversa}, {Panahi}, {Pawlak}, {Piersimoni}, {Pineau}, {Plachy},
  {Plum}, {Poggio}, {Poujoulet}, {Pr{\v{s}}a}, {Pulone}, {Racero}, {Ragaini},
  {Rambaux}, {Ramos-Lerate}, {Regibo}, {Reyl{\'e}}, {Riclet}, {Ripepi}, {Riva},
  {Rivard}, {Rixon}, {Roegiers}, {Roelens}, {Romero-G{\'o}mez}, {Rowell},
  {Royer}, {Ruiz-Dern}, {Sadowski}, {Sagrist{\`a} Sell{\'e}s}, {Sahlmann},
  {Salgado}, {Salguero}, {Sanna}, {Santana-Ros}, {Sarasso}, {Savietto},
  {Schultheis}, {Sciacca}, {Segol}, {Segovia}, {S{\'e}gransan}, {Shih},
  {Siltala}, {Silva}, {Smart}, {Smith}, {Solano}, {Solitro}, {Sordo}, {Soria
  Nieto}, {Souchay}, {Spagna}, {Spoto}, {Stampa}, {Steele},
  {Steidelm{\"u}ller}, {Stephenson}, {Stoev}, {Suess}, {Surdej}, {Szabados},
  {Szegedi-Elek}, {Tapiador}, {Taris}, {Tauran}, {Taylor}, {Teixeira},
  {Terrett}, {Teyssand ier}, {Thuillot}, {Titarenko}, {Torra Clotet}, {Turon},
  {Ulla}, {Utrilla}, {Uzzi}, {Vaillant}, {Valentini}, {Valette}, {van Elteren},
  {Van Hemelryck}, {van Leeuwen}, {Vaschetto}, {Vecchiato}, {Veljanoski},
  {Viala}, {Vicente}, {Vogt}, {von Essen}, {Voss}, {Votruba}, {Voutsinas},
  {Walmsley}, {Weiler}, {Wertz}, {Wevers}, {Wyrzykowski}, {Yoldas},
  {{\v{Z}}erjal}, {Ziaeepour}, {Zorec}, {Zschocke}, {Zucker}, {Zurbach}, \&
  {Zwitter}}]{2018A&A...616A...1G}
{Gaia Collaboration}, {Brown}, A.~G.~A., {Vallenari}, A., {et~al.} 2018, \aap,
  616, A1, \dodoi{10.1051/0004-6361/201833051}

\bibitem[{{Gallet} \& {Bouvier}(2013)}]{2013A&A...556A..36G}
{Gallet}, F., \& {Bouvier}, J. 2013, \aap, 556, A36,
  \dodoi{10.1051/0004-6361/201321302}

\bibitem[{{Garc{\'{\i}}a} {et~al.}(2010){Garc{\'{\i}}a}, {Mathur}, {Salabert},
  {Ballot}, {R{\'e}gulo}, {Metcalfe}, \& {Baglin}}]{2010Sci...329.1032G}
{Garc{\'{\i}}a}, R.~A., {Mathur}, S., {Salabert}, D., {et~al.} 2010, Science,
  329, 1032, \dodoi{10.1126/science.1191064}

\bibitem[{{Garc{\'{\i}}a} {et~al.}(2014){Garc{\'{\i}}a}, {Ceillier},
  {Salabert}, {Mathur}, {van Saders}, {Pinsonneault}, {Ballot}, {Beck},
  {Bloemen}, {Campante}, {Davies}, {do Nascimento}, {Mathis}, {Metcalfe},
  {Nielsen}, {Su{\'a}rez}, {Chaplin}, {Jim{\'e}nez}, \&
  {Karoff}}]{2014A&A...572A..34G}
{Garc{\'{\i}}a}, R.~A., {Ceillier}, T., {Salabert}, D., {et~al.} 2014, \aap,
  572, A34, \dodoi{10.1051/0004-6361/201423888}

\bibitem[{{Gastine} {et~al.}(2013){Gastine}, {Morin}, {Duarte}, {Reiners},
  {Christensen}, \& {Wicht}}]{2013A&A...549L...5G}
{Gastine}, T., {Morin}, J., {Duarte}, L., {et~al.} 2013, \aap, 549, L5,
  \dodoi{10.1051/0004-6361/201220317}

\bibitem[{{Gilman}(1979)}]{1979ApJ...231..284G}
{Gilman}, P.~A. 1979, \apj, 231, 284, \dodoi{10.1086/157191}

\bibitem[{{Godoy-Rivera} {et~al.}(2025){Godoy-Rivera}, {Mathur}, {Garc{\'\i}a},
  {Pinsonneault}, {Santos}, {Beck}, {Grossmann}, {Schimak}, {Bedell}, {Merc},
  \& {Escorza}}]{2025arXiv250118719G}
{Godoy-Rivera}, D., {Mathur}, S., {Garc{\'\i}a}, R.~A., {et~al.} 2025, arXiv
  e-prints, arXiv:2501.18719, \dodoi{10.48550/arXiv.2501.18719}

\bibitem[{{Gordon} {et~al.}(2021){Gordon}, {Davenport}, {Angus},
  {Foreman-Mackey}, {Agol}, {Covey}, {Ag{\"u}eros}, \&
  {Kipping}}]{2021ApJ...913...70G}
{Gordon}, T.~A., {Davenport}, J. R.~A., {Angus}, R., {et~al.} 2021, \apj, 913,
  70, \dodoi{10.3847/1538-4357/abf63e}

\bibitem[{{Goupil} {et~al.}(2024){Goupil}, {Catala}, {Samadi}, {Belkacem},
  {Ouazzani}, {Reese}, {Appourchaux}, {Mathur}, {Cabrera}, {B{\"o}rner},
  {Paproth}, {Moedas}, {Verma}, {Lebreton}, {Deal}, {Ballot}, {Chaplin},
  {Christensen-Dalsgaard}, {Cunha}, {Lanza}, {Miglio}, {Morel}, {Serenelli},
  {Mosser}, {Creevey}, {Moya}, {Garcia}, {Nielsen}, \&
  {Hatt}}]{2024A&A...683A..78G}
{Goupil}, M.~J., {Catala}, C., {Samadi}, R., {et~al.} 2024, \aap, 683, A78,
  \dodoi{10.1051/0004-6361/202348111}

\bibitem[{{Hall} {et~al.}(2021){Hall}, {Davies}, {van Saders}, {Nielsen},
  {Lund}, {Chaplin}, {Garc{\'\i}a}, {Amard}, {Breimann}, {Khan}, {See}, \&
  {Tayar}}]{2021NatAs...5..707H}
{Hall}, O.~J., {Davies}, G.~R., {van Saders}, J., {et~al.} 2021, Nature
  Astronomy, 5, 707, \dodoi{10.1038/s41550-021-01335-x}

\bibitem[{{Holl} {et~al.}(2022){Holl}, {Sozzetti}, {Sahlmann}, {Giacobbe},
  {S{\'e}gransan}, {Unger}, {Delisle}, {Barbato}, {Lattanzi}, {Morbidelli}, \&
  {Sosnowska}}]{2022arXiv220605439H}
{Holl}, B., {Sozzetti}, A., {Sahlmann}, J., {et~al.} 2022, arXiv e-prints,
  arXiv:2206.05439.
\newblock \doarXiv{2206.05439}

\bibitem[{{Huber} {et~al.}(2014){Huber}, {Silva Aguirre}, {Matthews},
  {Pinsonneault}, {Gaidos}, {Garc{\'{\i}}a}, {Hekker}, {Mathur}, {Mosser},
  {Torres}, {Bastien}, {Basu}, {Bedding}, {Chaplin}, {Demory}, {Fleming},
  {Guo}, {Mann}, {Rowe}, {Serenelli}, {Smith}, \&
  {Stello}}]{2014ApJS..211....2H}
{Huber}, D., {Silva Aguirre}, V., {Matthews}, J.~M., {et~al.} 2014, \apjs, 211,
  2, \dodoi{10.1088/0067-0049/211/1/2}

\bibitem[{{I{\c{s}}{\i}k} {et~al.}(2018){I{\c{s}}{\i}k}, {Solanki}, {Krivova},
  \& {Shapiro}}]{2018A&A...620A.177I}
{I{\c{s}}{\i}k}, E., {Solanki}, S.~K., {Krivova}, N.~A., \& {Shapiro}, A.~I.
  2018, \aap, 620, A177, \dodoi{10.1051/0004-6361/201833393}

\bibitem[{{Jouve} \& {Brun}(2007)}]{2007A&A...474..239J}
{Jouve}, L., \& {Brun}, A.~S. 2007, \aap, 474, 239,
  \dodoi{10.1051/0004-6361:20077070}

\bibitem[{{Jouve} {et~al.}(2008){Jouve}, {Brun}, {Arlt}, {Brandenburg},
  {Dikpati}, {Bonanno}, {K{\"a}pyl{\"a}}, {Moss}, {Rempel}, {Gilman}, {Korpi},
  \& {Kosovichev}}]{2008A&A...483..949J}
{Jouve}, L., {Brun}, A.~S., {Arlt}, R., {et~al.} 2008, \aap, 483, 949,
  \dodoi{10.1051/0004-6361:20078351}

\bibitem[{{K{\"a}pyl{\"a}}(2022)}]{2022ApJ...931L..17K}
{K{\"a}pyl{\"a}}, P.~J. 2022, \apjl, 931, L17, \dodoi{10.3847/2041-8213/ac6e6b}

\bibitem[{{Karak} {et~al.}(2020){Karak}, {Tomar}, \&
  {Vashishth}}]{2020MNRAS.491.3155K}
{Karak}, B.~B., {Tomar}, A., \& {Vashishth}, V. 2020, \mnras, 491, 3155,
  \dodoi{10.1093/mnras/stz3220}

\bibitem[{{Karoff} {et~al.}(2018){Karoff}, {Metcalfe}, {Santos}, {Montet},
  {Isaacson}, {Witzke}, {Shapiro}, {Mathur}, {Davies}, {Lund}, {Garcia},
  {Brun}, {Salabert}, {Avelino}, {van Saders}, {Egeland}, {Cunha}, {Campante},
  {Chaplin}, {Krivova}, {Solanki}, {Stritzinger}, \&
  {Knudsen}}]{2018ApJ...852...46K}
{Karoff}, C., {Metcalfe}, T.~S., {Santos}, {\^A}.~R.~G., {et~al.} 2018, \apj,
  852, 46, \dodoi{10.3847/1538-4357/aaa026}

\bibitem[{{Kawaler}(1988)}]{1988ApJ...333..236K}
{Kawaler}, S.~D. 1988, \apj, 333, 236, \dodoi{10.1086/166740}

\bibitem[{{Kraft}(1967)}]{1967ApJ...150..551K}
{Kraft}, R.~P. 1967, \apj, 150, 551, \dodoi{10.1086/149359}

\bibitem[{{Landin} {et~al.}(2023){Landin}, {Mendes}, {Vaz}, \&
  {Alencar}}]{2023MNRAS.519.5304L}
{Landin}, N.~R., {Mendes}, L.~T.~S., {Vaz}, L.~P.~R., \& {Alencar}, S.~H.~P.
  2023, \mnras, 519, 5304, \dodoi{10.1093/mnras/stac3823}

\bibitem[{{Li} \& {Basri}(2024)}]{2024ApJ...963..102L}
{Li}, C., \& {Basri}, G. 2024, \apj, 963, 102, \dodoi{10.3847/1538-4357/ad1e59}

\bibitem[{Liu {et~al.}(2007)Liu, Liang, \& Weisberg}]{liu2007}
Liu, Y., Liang, X., \& Weisberg, R. 2007, Atmos. and Ocean Tech., 24, 2093

\bibitem[{{Lu} {et~al.}(2022){Lu}, {Curtis}, {Angus}, {David}, \&
  {Hattori}}]{2022AJ....164..251L}
{Lu}, Y.~L., {Curtis}, J.~L., {Angus}, R., {David}, T.~J., \& {Hattori}, S.
  2022, \aj, 164, 251, \dodoi{10.3847/1538-3881/ac9bee}

\bibitem[{{Lund} {et~al.}(2017){Lund}, {Silva Aguirre}, {Davies}, {Chaplin},
  {Christensen-Dalsgaard}, {Houdek}, {White}, {Bedding}, {Ball}, {Huber},
  {Antia}, {Lebreton}, {Latham}, {Handberg}, {Verma}, {Basu}, {Casagrande},
  {Justesen}, {Kjeldsen}, \& {Mosumgaard}}]{2017ApJ...835..172L}
{Lund}, M.~N., {Silva Aguirre}, V., {Davies}, G.~R., {et~al.} 2017, \apj, 835,
  172, \dodoi{10.3847/1538-4357/835/2/172}

\bibitem[{{Marsden} {et~al.}(2014){Marsden}, {Petit}, {Jeffers}, {Morin},
  {Fares}, {Reiners}, {do Nascimento}, {Auri{\`e}re}, {Bouvier}, {Carter},
  {Catala}, {Dintrans}, {Donati}, {Gastine}, {Jardine}, {Konstantinova-Antova},
  {Lanoux}, {Ligni{\`e}res}, {Morgenthaler}, {Ram{\`\i}rez-V{\`e}lez},
  {Th{\'e}ado}, {Van Grootel}, \& {BCool Collaboration}}]{2014MNRAS.444.3517M}
{Marsden}, S.~C., {Petit}, P., {Jeffers}, S.~V., {et~al.} 2014, \mnras, 444,
  3517, \dodoi{10.1093/mnras/stu1663}

\bibitem[{{Masuda}(2022)}]{2022ApJ...933..195M}
{Masuda}, K. 2022, \apj, 933, 195, \dodoi{10.3847/1538-4357/ac7527}

\bibitem[{{Mathur} {et~al.}(2019){Mathur}, {Garc{\'\i}a}, {Bugnet}, {Santos},
  {Santiago}, \& {Beck}}]{2019FrASS...6...46M}
{Mathur}, S., {Garc{\'\i}a}, R.~A., {Bugnet}, L., {et~al.} 2019, Frontiers in
  Astronomy and Space Sciences, 6, 46, \dodoi{10.3389/fspas.2019.00046}

\bibitem[{{Mathur} {et~al.}(2014{\natexlab{a}}){Mathur}, {Salabert},
  {Garc{\'{\i}}a}, \& {Ceillier}}]{2014JSWSC...4A..15M}
{Mathur}, S., {Salabert}, D., {Garc{\'{\i}}a}, R.~A., \& {Ceillier}, T.
  2014{\natexlab{a}}, Journal of Space Weather and Space Climate, 4, A15,
  \dodoi{10.1051/swsc/2014011}

\bibitem[{{Mathur} {et~al.}(2010){Mathur}, {Garc{\'{\i}}a}, {R{\'e}gulo},
  {Creevey}, {Ballot}, {Salabert}, {Arentoft}, {Quirion}, {Chaplin}, \&
  {Kjeldsen}}]{2010A&A...511A..46M}
{Mathur}, S., {Garc{\'{\i}}a}, R.~A., {R{\'e}gulo}, C., {et~al.} 2010, \aap,
  511, A46, \dodoi{10.1051/0004-6361/200913266}

\bibitem[{{Mathur} {et~al.}(2014{\natexlab{b}}){Mathur}, {Garc{\'{\i}}a},
  {Ballot}, {Ceillier}, {Salabert}, {Metcalfe}, {R{\'e}gulo}, {Jim{\'e}nez}, \&
  {Bloemen}}]{2014A&A...562A.124M}
{Mathur}, S., {Garc{\'{\i}}a}, R.~A., {Ballot}, J., {et~al.}
  2014{\natexlab{b}}, \aap, 562, A124, \dodoi{10.1051/0004-6361/201322707}

\bibitem[{{Mathur} {et~al.}(2017){Mathur}, {Huber}, {Batalha}, {Ciardi},
  {Bastien}, {Bieryla}, {Buchhave}, {Cochran}, {Endl}, {Esquerdo}, {Furlan},
  {Howard}, {Howell}, {Isaacson}, {Latham}, {MacQueen}, \&
  {Silva}}]{2017ApJS..229...30M}
{Mathur}, S., {Huber}, D., {Batalha}, N.~M., {et~al.} 2017, \apjs, 229, 30,
  \dodoi{10.3847/1538-4365/229/2/30}

\bibitem[{{Mathur} {et~al.}(2023){Mathur}, {Claytor}, {Santos}, {Garc{\'\i}a},
  {Amard}, {Bugnet}, {Corsaro}, {Bonanno}, {Breton}, {Godoy-Rivera},
  {Pinsonneault}, \& {van Saders}}]{2023ApJ...952..131M}
{Mathur}, S., {Claytor}, Z.~R., {Santos}, {\^A}. R.~G., {et~al.} 2023, \apj,
  952, 131, \dodoi{10.3847/1538-4357/acd118}

\bibitem[{{Matt} {et~al.}(2015){Matt}, {Brun}, {Baraffe}, {Bouvier}, \&
  {Chabrier}}]{2015ApJ...799L..23M}
{Matt}, S.~P., {Brun}, A.~S., {Baraffe}, I., {Bouvier}, J., \& {Chabrier}, G.
  2015, \apjl, 799, L23, \dodoi{10.1088/2041-8205/799/2/L23}

\bibitem[{{McQuillan} {et~al.}(2013){McQuillan}, {Aigrain}, \&
  {Mazeh}}]{2013MNRAS.432.1203M}
{McQuillan}, A., {Aigrain}, S., \& {Mazeh}, T. 2013, \mnras, 432, 1203,
  \dodoi{10.1093/mnras/stt536}

\bibitem[{{McQuillan} {et~al.}(2014){McQuillan}, {Mazeh}, \&
  {Aigrain}}]{2014ApJS..211...24M}
{McQuillan}, A., {Mazeh}, T., \& {Aigrain}, S. 2014, \apjs, 211, 24,
  \dodoi{10.1088/0067-0049/211/2/24}

\bibitem[{{Metcalfe} \& {van Saders}(2020)}]{2020arXiv200704416M}
{Metcalfe}, T.~S., \& {van Saders}, J. 2020, arXiv e-prints, arXiv:2007.04416,
  \dodoi{10.48550/arXiv.2007.04416}

\bibitem[{{Metcalfe} {et~al.}(2023){Metcalfe}, {Strassmeier}, {Ilyin}, {van
  Saders}, {Ayres}, {Finley}, {Kochukhov}, {Petit}, {See}, {Stassun},
  {Jeffers}, {Marsden}, {Morin}, \& {Vidotto}}]{2023ApJ...948L...6M}
{Metcalfe}, T.~S., {Strassmeier}, K.~G., {Ilyin}, I.~V., {et~al.} 2023, \apjl,
  948, L6, \dodoi{10.3847/2041-8213/acce38}

\bibitem[{{Meunier} \& {Lagrange}(2019)}]{2019A&A...629A..42M}
{Meunier}, N., \& {Lagrange}, A.~M. 2019, \aap, 629, A42,
  \dodoi{10.1051/0004-6361/201935651}

\bibitem[{{Nandy}(2021)}]{2021SoPh..296...54N}
{Nandy}, D. 2021, \solphys, 296, 54, \dodoi{10.1007/s11207-021-01797-2}

\bibitem[{{Noraz} {et~al.}(2022){Noraz}, {Breton}, {Brun}, {Garc{\'\i}a},
  {Strugarek}, {Santos}, {Mathur}, \& {Amard}}]{2022A&A...667A..50N}
{Noraz}, Q., {Breton}, S.~N., {Brun}, A.~S., {et~al.} 2022, \aap, 667, A50,
  \dodoi{10.1051/0004-6361/202243890}

\bibitem[{{Noraz} {et~al.}(2024){Noraz}, {Brun}, \&
  {Strugarek}}]{2024A&A...684A.156N}
{Noraz}, Q., {Brun}, A.~S., \& {Strugarek}, A. 2024, \aap, 684, A156,
  \dodoi{10.1051/0004-6361/202347939}

\bibitem[{{Noyes} {et~al.}(1984){Noyes}, {Hartmann}, {Baliunas}, {Duncan}, \&
  {Vaughan}}]{1984ApJ...279..763N}
{Noyes}, R.~W., {Hartmann}, L.~W., {Baliunas}, S.~L., {Duncan}, D.~K., \&
  {Vaughan}, A.~H. 1984, \apj, 279, 763, \dodoi{10.1086/161945}

\bibitem[{{Pecaut} \& {Mamajek}(2013)}]{2013ApJS..208....9P}
{Pecaut}, M.~J., \& {Mamajek}, E.~E. 2013, \apjs, 208, 9,
  \dodoi{10.1088/0067-0049/208/1/9}

\bibitem[{{Pinsonneault} {et~al.}(1990){Pinsonneault}, {Kawaler}, \&
  {Demarque}}]{1990ApJS...74..501P}
{Pinsonneault}, M.~H., {Kawaler}, S.~D., \& {Demarque}, P. 1990, \apjs, 74,
  501, \dodoi{10.1086/191507}

\bibitem[{{Pinsonneault} {et~al.}(1989){Pinsonneault}, {Kawaler}, {Sofia}, \&
  {Demarque}}]{1989ApJ...338..424P}
{Pinsonneault}, M.~H., {Kawaler}, S.~D., {Sofia}, S., \& {Demarque}, P. 1989,
  \apj, 338, 424, \dodoi{10.1086/167210}

\bibitem[{{Pourbaix} {et~al.}(2004){Pourbaix}, {Tokovinin}, {Batten}, {Fekel},
  {Hartkopf}, {Levato}, {Morrell}, {Torres}, \& {Udry}}]{Pourbaix2004}
{Pourbaix}, D., {Tokovinin}, A.~A., {Batten}, A.~H., {et~al.} 2004, \aap, 424,
  727, \dodoi{10.1051/0004-6361:20041213}

\bibitem[{{Rauer} {et~al.}(2014){Rauer}, {Catala}, {Aerts}, {Appourchaux},
  {Benz}, {Brandeker}, {Christensen-Dalsgaard}, {Deleuil}, {Gizon}, {Goupil},
  {G{\"u}del}, {Janot-Pacheco}, {Mas-Hesse}, {Pagano}, {Piotto}, {Pollacco},
  {Santos}, {Smith}, {Su{\'a}rez}, {Szab{\'o}}, {Udry}, {Adibekyan}, {Alibert},
  {Almenara}, {Amaro-Seoane}, {Eiff}, {Asplund}, {Antonello}, {Barnes},
  {Baudin}, {Belkacem}, {Bergemann}, {Bihain}, {Birch}, {Bonfils}, {Boisse},
  {Bonomo}, {Borsa}, {Brand{\~a}o}, {Brocato}, {Brun}, {Burleigh}, {Burston},
  {Cabrera}, {Cassisi}, {Chaplin}, {Charpinet}, {Chiappini}, {Church},
  {Csizmadia}, {Cunha}, {Damasso}, {Davies}, {Deeg}, {D{\'{\i}}az}, {Dreizler},
  {Dreyer}, {Eggenberger}, {Ehrenreich}, {Eigm{\"u}ller}, {Erikson}, {Farmer},
  {Feltzing}, {de Oliveira Fialho}, {Figueira}, {Forveille}, {Fridlund},
  {Garc{\'{\i}}a}, {Giommi}, {Giuffrida}, {Godolt}, {Gomes da Silva},
  {Granzer}, {Grenfell}, {Grotsch-Noels}, {G{\"u}nther}, {Haswell}, {Hatzes},
  {H{\'e}brard}, {Hekker}, {Helled}, {Heng}, {Jenkins}, {Johansen},
  {Khodachenko}, {Kislyakova}, {Kley}, {Kolb}, {Krivova}, {Kupka}, {Lammer},
  {Lanza}, {Lebreton}, {Magrin}, {Marcos-Arenal}, {Marrese}, {Marques},
  {Martins}, {Mathis}, {Mathur}, {Messina}, {Miglio}, {Montalban}, {Montalto},
  {Monteiro}, {Moradi}, {Moravveji}, {Mordasini}, {Morel}, {Mortier},
  {Nascimbeni}, {Nelson}, {Nielsen}, {Noack}, {Norton}, {Ofir}, {Oshagh},
  {Ouazzani}, {P{\'a}pics}, {Parro}, {Petit}, {Plez}, {Poretti}, {Quirrenbach},
  {Ragazzoni}, {Raimondo}, {Rainer}, {Reese}, {Redmer}, {Reffert},
  {Rojas-Ayala}, {Roxburgh}, {Salmon}, {Santerne}, {Schneider}, {Schou},
  {Schuh}, {Schunker}, {Silva-Valio}, {Silvotti}, {Skillen}, {Snellen}, {Sohl},
  {Sousa}, {Sozzetti}, {Stello}, {Strassmeier}, {{\v S}vanda}, {Szab{\'o}},
  {Tkachenko}, {Valencia}, {Van Grootel}, {Vauclair}, {Ventura}, {Wagner},
  {Walton}, {Weingrill}, {Werner}, {Wheatley}, \&
  {Zwintz}}]{2014ExA....38..249R}
{Rauer}, H., {Catala}, C., {Aerts}, C., {et~al.} 2014, Experimental Astronomy,
  38, 249, \dodoi{10.1007/s10686-014-9383-4}

\bibitem[{{Reinhold} {et~al.}(2019){Reinhold}, {Bell}, {Kuszlewicz}, {Hekker},
  \& {Shapiro}}]{2019A&A...621A..21R}
{Reinhold}, T., {Bell}, K.~J., {Kuszlewicz}, J., {Hekker}, S., \& {Shapiro},
  A.~I. 2019, \aap, 621, A21, \dodoi{10.1051/0004-6361/201833754}

\bibitem[{{Reinhold} \& {Hekker}(2020)}]{2020A&A...635A..43R}
{Reinhold}, T., \& {Hekker}, S. 2020, \aap, 635, A43,
  \dodoi{10.1051/0004-6361/201936887}

\bibitem[{{Reinhold} {et~al.}(2021){Reinhold}, {Shapiro}, {Witzke},
  {N{\`e}mec}, {I{\c{s}}{\i}k}, \& {Solanki}}]{2021ApJ...908L..21R}
{Reinhold}, T., {Shapiro}, A.~I., {Witzke}, V., {et~al.} 2021, \apjl, 908, L21,
  \dodoi{10.3847/2041-8213/abde46}

\bibitem[{{R{\'e}ville} {et~al.}(2015){R{\'e}ville}, {Brun}, {Matt},
  {Strugarek}, \& {Pinto}}]{2015ApJ...798..116R}
{R{\'e}ville}, V., {Brun}, A.~S., {Matt}, S.~P., {Strugarek}, A., \& {Pinto},
  R.~F. 2015, \apj, 798, 116, \dodoi{10.1088/0004-637X/798/2/116}

\bibitem[{{Rimoldini} {et~al.}(2023){Rimoldini}, {Holl}, {Gavras}, {Audard},
  {De Ridder}, {Mowlavi}, {Nienartowicz}, {Jevardat de Fombelle},
  {Lecoeur-Ta{\"\i}bi}, {Karbevska}, {Evans}, {{\'A}brah{\'a}m}, {Carnerero},
  {Clementini}, {Distefano}, {Garofalo}, {Garc{\'\i}a-Lario}, {Gomel},
  {Klioner}, {Kruszy{\'n}ska}, {Lanzafame}, {Lebzelter}, {Marton}, {Mazeh},
  {Molinaro}, {Panahi}, {Raiteri}, {Ripepi}, {Szabados}, {Teyssier},
  {Trabucchi}, {Wyrzykowski}, {Zucker}, \& {Eyer}}]{2023A&A...674A..14R}
{Rimoldini}, L., {Holl}, B., {Gavras}, P., {et~al.} 2023, \aap, 674, A14,
  \dodoi{10.1051/0004-6361/202245591}

\bibitem[{{Saar} \& {Brandenburg}(2002)}]{2002AN....323..357S}
{Saar}, S.~H., \& {Brandenburg}, A. 2002, Astronomische Nachrichten, 323, 357,
  \dodoi{10.1002/1521-3994(200208)323:3/4<357::AID-ASNA357>3.0.CO;2-I}

\bibitem[{{Salabert} {et~al.}(2017){Salabert}, {Garc{\'{\i}}a}, {Jim{\'e}nez},
  {Bertello}, {Corsaro}, \& {Pall{\'e}}}]{2017A&A...608A..87S}
{Salabert}, D., {Garc{\'{\i}}a}, R.~A., {Jim{\'e}nez}, A., {et~al.} 2017, \aap,
  608, A87, \dodoi{10.1051/0004-6361/201731560}

\bibitem[{{Salabert} {et~al.}(2016{\natexlab{a}}){Salabert}, {Garc{\'{\i}}a},
  {Beck}, {Egeland}, {Pall{\'e}}, {Mathur}, {Metcalfe}, {do Nascimento},
  {Ceillier}, {Andersen}, \& {Trivi{\~n}o Hage}}]{2016A&A...596A..31S}
{Salabert}, D., {Garc{\'{\i}}a}, R.~A., {Beck}, P.~G., {et~al.}
  2016{\natexlab{a}}, \aap, 596, A31, \dodoi{10.1051/0004-6361/201628583}

\bibitem[{{Salabert} {et~al.}(2016{\natexlab{b}}){Salabert}, {R{\'e}gulo},
  {Garc{\'{\i}}a}, {Beck}, {Ballot}, {Creevey}, {P{\'e}rez Hern{\'a}ndez}, {do
  Nascimento}, {Corsaro}, {Egeland}, {Mathur}, {Metcalfe}, {Bigot}, {Ceillier},
  \& {Pall{\'e}}}]{2016A&A...589A.118S}
{Salabert}, D., {R{\'e}gulo}, C., {Garc{\'{\i}}a}, R.~A., {et~al.}
  2016{\natexlab{b}}, \aap, 589, A118, \dodoi{10.1051/0004-6361/201527978}

\bibitem[{{Santos} {et~al.}(2021){Santos}, {Breton}, {Mathur}, \&
  {Garc{\'\i}a}}]{2021ApJS..255...17S}
{Santos}, A.~R.~G., {Breton}, S.~N., {Mathur}, S., \& {Garc{\'\i}a}, R.~A.
  2021, \apjs, 255, 17, \dodoi{10.3847/1538-4365/ac033f}

\bibitem[{{Santos} {et~al.}(2017){Santos}, {Cunha}, {Avelino}, {Garc{\'{\i}}a},
  \& {Mathur}}]{2017A&A...599A...1S}
{Santos}, A.~R.~G., {Cunha}, M.~S., {Avelino}, P.~P., {Garc{\'{\i}}a}, R.~A.,
  \& {Mathur}, S. 2017, \aap, 599, A1, \dodoi{10.1051/0004-6361/201629923}

\bibitem[{{Santos} {et~al.}(2019){Santos}, {Garc{\'\i}a}, {Mathur}, {Bugnet},
  {van Saders}, {Metcalfe}, {Simonian}, \&
  {Pinsonneault}}]{2019ApJS..244...21S}
{Santos}, A.~R.~G., {Garc{\'\i}a}, R.~A., {Mathur}, S., {et~al.} 2019, \apjs,
  244, 21, \dodoi{10.3847/1538-4365/ab3b56}

\bibitem[{{Santos} {et~al.}(2024){Santos}, {Godoy-Rivera}, {Finley}, {Mathur},
  {Garc{\'\i}a}, {Breton}, \& {Broomhall}}]{2024FrASS..1156379S}
{Santos}, {\^A}. R.~G., {Godoy-Rivera}, D., {Finley}, A.~J., {et~al.} 2024,
  Frontiers in Astronomy and Space Sciences, 11, 1356379,
  \dodoi{10.3389/fspas.2024.1356379}

\bibitem[{{Santos} {et~al.}(2023){Santos}, {Mathur}, {Garc{\'\i}a},
  {Broomhall}, {Egeland}, {Jim{\'e}nez}, {Godoy-Rivera}, {Breton}, {Claytor},
  {Metcalfe}, {Cunha}, \& {Amard}}]{2023A&A...672A..56S}
{Santos}, A.~R.~G., {Mathur}, S., {Garc{\'\i}a}, R.~A., {et~al.} 2023, \aap,
  672, A56, \dodoi{10.1051/0004-6361/202245430}

\bibitem[{{See} {et~al.}(2023){See}, {Roquette}, {Amard}, \&
  {Matt}}]{2023MNRAS.524.5781S}
{See}, V., {Roquette}, J., {Amard}, L., \& {Matt}, S. 2023, \mnras, 524, 5781,
  \dodoi{10.1093/mnras/stad2020}

\bibitem[{{See} {et~al.}(2021){See}, {Roquette}, {Amard}, \&
  {Matt}}]{2021ApJ...912..127S}
{See}, V., {Roquette}, J., {Amard}, L., \& {Matt}, S.~P. 2021, \apj, 912, 127,
  \dodoi{10.3847/1538-4357/abed47}

\bibitem[{{See} {et~al.}(2019){See}, {Matt}, {Folsom}, {Boro Saikia}, {Donati},
  {Fares}, {Finley}, {H{\'e}brard}, {Jardine}, {Jeffers}, {Lehmann}, {Marsden},
  {Mengel}, {Morin}, {Petit}, {Vidotto}, {Waite}, \& {BCool
  Collaboration}}]{2019ApJ...876..118S}
{See}, V., {Matt}, S.~P., {Folsom}, C.~P., {et~al.} 2019, \apj, 876, 118,
  \dodoi{10.3847/1538-4357/ab1096}

\bibitem[{{Shapiro} {et~al.}(2016){Shapiro}, {Solanki}, {Krivova}, {Yeo}, \&
  {Schmutz}}]{2016A&A...589A..46S}
{Shapiro}, A.~I., {Solanki}, S.~K., {Krivova}, N.~A., {Yeo}, K.~L., \&
  {Schmutz}, W.~K. 2016, \aap, 589, A46, \dodoi{10.1051/0004-6361/201527527}

\bibitem[{{Silva Aguirre} {et~al.}(2017){Silva Aguirre}, {Lund}, {Antia},
  {Ball}, {Basu}, {Christensen-Dalsgaard}, {Lebreton}, {Reese}, {Verma},
  {Casagrande}, {Justesen}, {Mosumgaard}, {Chaplin}, {Bedding}, {Davies},
  {Handberg}, {Houdek}, {Huber}, {Kjeldsen}, {Latham}, {White}, {Coelho},
  {Miglio}, \& {Rendle}}]{2017ApJ...835..173S}
{Silva Aguirre}, V., {Lund}, M.~N., {Antia}, H.~M., {et~al.} 2017, \apj, 835,
  173, \dodoi{10.3847/1538-4357/835/2/173}

\bibitem[{{Simonian} {et~al.}(2019){Simonian}, {Pinsonneault}, \&
  {Terndrup}}]{2019ApJ...871..174S}
{Simonian}, G. V.~A., {Pinsonneault}, M.~H., \& {Terndrup}, D.~M. 2019, \apj,
  871, 174, \dodoi{10.3847/1538-4357/aaf97c}

\bibitem[{{Simonian} {et~al.}(2020){Simonian}, {Pinsonneault}, {Terndrup}, \&
  {van Saders}}]{2020ApJ...898...76S}
{Simonian}, G. V.~A., {Pinsonneault}, M.~H., {Terndrup}, D.~M., \& {van
  Saders}, J.~L. 2020, \apj, 898, 76, \dodoi{10.3847/1538-4357/ab9a43}

\bibitem[{{Skumanich}(1972)}]{1972ApJ...171..565S}
{Skumanich}, A. 1972, \apj, 171, 565, \dodoi{10.1086/151310}

\bibitem[{{Somers} {et~al.}(2017){Somers}, {Stauffer}, {Rebull}, {Cody}, \&
  {Pinsonneault}}]{2017ApJ...850..134S}
{Somers}, G., {Stauffer}, J., {Rebull}, L., {Cody}, A.~M., \& {Pinsonneault},
  M. 2017, \apj, 850, 134, \dodoi{10.3847/1538-4357/aa93ed}

\bibitem[{{Sowmya} {et~al.}(2021){Sowmya}, {Shapiro}, {Witzke}, {N{\`e}mec},
  {Chatzistergos}, {Yeo}, {Krivova}, \& {Solanki}}]{2021ApJ...914...21S}
{Sowmya}, K., {Shapiro}, A.~I., {Witzke}, V., {et~al.} 2021, \apj, 914, 21,
  \dodoi{10.3847/1538-4357/abf247}

\bibitem[{{Spada} \& {Lanzafame}(2020)}]{2020A&A...636A..76S}
{Spada}, F., \& {Lanzafame}, A.~C. 2020, \aap, 636, A76,
  \dodoi{10.1051/0004-6361/201936384}

\bibitem[{{Strugarek} {et~al.}(2018){Strugarek}, {Beaudoin}, {Charbonneau}, \&
  {Brun}}]{2018ApJ...863...35S}
{Strugarek}, A., {Beaudoin}, P., {Charbonneau}, P., \& {Brun}, A.~S. 2018,
  \apj, 863, 35, \dodoi{10.3847/1538-4357/aacf9e}

\bibitem[{{Strugarek} {et~al.}(2017){Strugarek}, {Beaudoin}, {Charbonneau},
  {Brun}, \& {do Nascimento}}]{2017Sci...357..185S}
{Strugarek}, A., {Beaudoin}, P., {Charbonneau}, P., {Brun}, A.~S., \& {do
  Nascimento}, J.~D. 2017, Science, 357, 185, \dodoi{10.1126/science.aal3999}

\bibitem[{{Thomas} {et~al.}(2019){Thomas}, {Chaplin}, {Davies}, {Howe},
  {Santos}, {Elsworth}, {Miglio}, {Campante}, \& {Cunha}}]{2019MNRAS.485.3857T}
{Thomas}, A. E.~L., {Chaplin}, W.~J., {Davies}, G.~R., {et~al.} 2019, \mnras,
  485, 3857, \dodoi{10.1093/mnras/stz672}

\bibitem[{{Torrence} \& {Compo}(1998)}]{1998BAMS...79...61T}
{Torrence}, C., \& {Compo}, G.~P. 1998, Bulletin of the American Meteorological
  Society, 79, 61, \dodoi{10.1175/1520-0477(1998)079}

\bibitem[{{van Saders} {et~al.}(2016){van Saders}, {Ceillier}, {Metcalfe},
  {Aguirre}, {Pinsonneault}, {Garc{\'{\i}}a}, {Mathur}, \&
  {Davies}}]{2016Natur.529..181V}
{van Saders}, J.~L., {Ceillier}, T., {Metcalfe}, T.~S., {et~al.} 2016, \nat,
  529, 181, \dodoi{10.1038/nature16168}

\bibitem[{{van Saders} \& {Pinsonneault}(2013)}]{vanSaders2013}
{van Saders}, J.~L., \& {Pinsonneault}, M.~H. 2013, \apj, 776, 67,
  \dodoi{10.1088/0004-637X/776/2/67}

\bibitem[{{Vidotto} {et~al.}(2014){Vidotto}, {Gregory}, {Jardine}, {Donati},
  {Petit}, {Morin}, {Folsom}, {Bouvier}, {Cameron}, {Hussain}, {Marsden},
  {Waite}, {Fares}, {Jeffers}, \& {do Nascimento}}]{2014MNRAS.441.2361V}
{Vidotto}, A.~A., {Gregory}, S.~G., {Jardine}, M., {et~al.} 2014, \mnras, 441,
  2361, \dodoi{10.1093/mnras/stu728}

\bibitem[{{Viviani} {et~al.}(2019){Viviani}, {K{\"a}pyl{\"a}}, {Warnecke},
  {K{\"a}pyl{\"a}}, \& {Rheinhardt}}]{2019ApJ...886...21V}
{Viviani}, M., {K{\"a}pyl{\"a}}, M.~J., {Warnecke}, J., {K{\"a}pyl{\"a}},
  P.~J., \& {Rheinhardt}, M. 2019, \apj, 886, 21,
  \dodoi{10.3847/1538-4357/ab3e07}

\bibitem[{{Wilson}(1978)}]{1978ApJ...226..379W}
{Wilson}, O.~C. 1978, \apj, 226, 379, \dodoi{10.1086/156618}

\bibitem[{{Wright} {et~al.}(2011){Wright}, {Drake}, {Mamajek}, \&
  {Henry}}]{2011ApJ...743...48W}
{Wright}, N.~J., {Drake}, J.~J., {Mamajek}, E.~E., \& {Henry}, G.~W. 2011,
  \apj, 743, 48, \dodoi{10.1088/0004-637X/743/1/48}

\bibitem[{{Wright} {et~al.}(2018){Wright}, {Newton}, {Williams}, {Drake}, \&
  {Yadav}}]{2018MNRAS.479.2351W}
{Wright}, N.~J., {Newton}, E.~R., {Williams}, P. K.~G., {Drake}, J.~J., \&
  {Yadav}, R.~K. 2018, \mnras, 479, 2351, \dodoi{10.1093/mnras/sty1670}

\bibitem[{{Yang} \& {Liu}(2019)}]{2019ApJS..241...29Y}
{Yang}, H., \& {Liu}, J. 2019, \apjs, 241, 29, \dodoi{10.3847/1538-4365/ab0d28}

\bibitem[{{Zhao} {et~al.}(2012){Zhao}, {Zhao}, {Chu}, {Jing}, \&
  {Deng}}]{2012RAA....12..723Z}
{Zhao}, G., {Zhao}, Y.-H., {Chu}, Y.-Q., {Jing}, Y.-P., \& {Deng}, L.-C. 2012,
  Research in Astronomy and Astrophysics, 12, 723,
  \dodoi{10.1088/1674-4527/12/7/002}

\bibitem[{{Zong} {et~al.}(2020){Zong}, {Fu}, {De Cat}, {Wang}, {Shi}, {Luo},
  {Zhang}, {Frasca}, {Molenda-{\.Z}akowicz}, {Gray}, {Corbally}, {Catanzaro},
  {Cang}, {Wang}, {Chen}, {Hou}, {Liu}, {Niu}, {Pan}, {Tian}, {Yan}, {Zhang},
  \& {Zuo}}]{2020ApJS..251...15Z}
{Zong}, W., {Fu}, J.-N., {De Cat}, P., {et~al.} 2020, \apjs, 251, 15,
  \dodoi{10.3847/1538-4365/abbb2d}

\end{thebibliography}
\bibliographystyle{aasjournal}



\end{document}